\documentclass[12pt]{article}
\usepackage {amsmath, amssymb, amsthm}
\usepackage {amsfonts,xcolor,mathtools,extarrows,bbm,mathrsfs,dutchcal}
\usepackage {tikz}
\usetikzlibrary{calc}
\usepackage[siunitx]{circuitikz}
\usetikzlibrary{decorations.markings,decorations.pathreplacing, decorations.pathmorphing, calligraphy,positioning}
 \usepackage{hyperref}
\hypersetup{
    colorlinks = true,
    citecolor = red,
}
\usepackage [margin=1.5in] {geometry}

\theoremstyle {definition}
\newtheorem {defn} {Definition}

\theoremstyle {plain}

\newtheorem {lemma}{Lemma}
\newtheorem {prop}{Proposition}
\newtheorem {thm}{Theorem}

\newcommand{\C}{{\mathcal C}}
\newcommand{\Cx}{{\mathbb C}}

\newcommand{\G}{{\mathcal G}}
\newcommand{\be}{\begin{equation}}
\newcommand{\ee}{\end{equation}} 

\newcommand{\eps}{\epsilon}

\newcommand{\R}{{\mathbb R}}
\newcommand{\Z}{{\mathbb Z}}

\newcommand{\E}{{\mathbb E}}

\newcommand{\SL}{\mathrm{SL}}
\newcommand{\GL}{\text{GL}}

\newcommand{\Tr}{\mathrm{Tr}}
\newcommand{\tr}{\mathrm{tr}}
\newcommand{\Pf}{\mathrm{Pf}}
\newcommand{\Sp}{\mathrm{Sp}}
\renewcommand{\tr}{\mathrm{tr}}

\colorlet{darkblue}{blue!70!black}
\colorlet{darkred}{red!70!black}
\colorlet{darkgreen}{green!70!black}
\colorlet{darkmagenta}{magenta!70!black}
\colorlet{lightyellow}{yellow!30!white}
\colorlet{lightyellowa}{yellow!50!white}

\newcommand{\old}[1]{}

\newenvironment{tric}
    {\begin{tikzpicture}[scale= 0.867,thick,draw=darkblue,double distance=1.1,
        baseline={([yshift=-.8ex]current bounding box.center)}] }
    {\end{tikzpicture}}

\newenvironment{trich}
    {\begin{tikzpicture}[scale= 0.65,thick,draw=darkblue,double distance=1.1,
        baseline={([yshift=-.8ex]current bounding box.center)}] }
    {\end{tikzpicture}}

\newenvironment{trica}
    {\begin{tikzpicture}[scale= 0.55,thick,draw=darkblue,double distance=1.1,
        baseline={([yshift=-.8ex]current bounding box.center)}] }
    {\end{tikzpicture}}

\newenvironment{trics}
    {\begin{tikzpicture}[scale= 0.45,thick,draw=darkblue,double distance=1.1,
        baseline={([yshift=-.8ex]current bounding box.center)}] }
    {\end{tikzpicture}}

\newtheorem{proposition}[thm]{Proposition}

\newtheorem{definition}[thm]{Definition}
\newtheorem{notation}[thm]{Notation}
\newtheorem{example}[thm]{Example}

\newtheorem{problem}{Problem}
\newtheorem{remark}[thm]{Remark}

\DeclareMathOperator{\wt}{\text{wt}}

\DeclareMathOperator{\Fund}{\bf{Fund}}
\DeclareMathOperator{\Rep}{\bf{Rep}}
\DeclareMathOperator{\Hom}{\rm{Hom}}

\DeclareMathOperator{\CDDC}{\textbf{Web}(Sp(4))}

\DeclareMathOperator{\triv}{\mathbf{1}}

\DeclareMathOperator{\ObjOne}{A_1}
\DeclareMathOperator{\ObjTwo}{A_2}
\DeclareMathOperator{\ObjK}{A_k}
\DeclareMathOperator{\ObjKP}{A_{k+1}}
\DeclareMathOperator{\ObjN}{A_{n}}

\begin {document}
\title {Webs and multiwebs for the symplectic group}
\author{Richard Kenyon\footnote{Department of Mathematics, Yale University, New Haven; richard.kenyon at yale.edu} \and
Haihan Wu\footnote{Department of Mathematics, Johns Hopkins University, Baltimore; hwu125 at jh.edu}}
\date{}
\maketitle

\begin{abstract}
We define $2n$-multiwebs on planar graphs and discuss their relation with $\mathrm{Sp}(2n)$-webs.
On a planar graph with a symplectic local system we define a matrix whose Pfaffian is the sum of traces of 
$2n$-multiwebs. As application we generalize Kasteleyn's theorem from dimer covers
to $2n$-multiweb covers of planar graphs with $U(n)$ gauge group.

For $\mathrm{Sp}(4)$ we relate Kuperberg's ``tetravalent vertex'' to the determinant,
and classify reduced $4$-webs on 
some simple surfaces: the annulus, torus, and pair of pants.

We likewise define, for $\mathrm{Sp}(2n)$ and $q=1$, a $2n$-valent vertex 
corresponding to the determinant, and classify reduced $2n$-webs on an annulus.

\end{abstract}

\section {Introduction}

Let $\G=(V,E)$ be a planar graph with $N$ vertices. 
An \emph{$\Sp(2n)$-connection $\Phi$ on $\G$}, also known as an
\emph{$\Sp(2n)$-local system}, is the choice, 
for each directed edge $e=uv$, of 
an element $\phi_{uv}\in \Sp(2n)$, so that $\phi_{vu}=(\phi_{uv})^{-1}$.

A \emph{$2n$-multiweb} in $\G$ (see \cite{DKS}) is a function $m:E\to\{0,1,\dots,2n\}$,  summing to $2n$ at every vertex:
$\sum_{u\sim v}m_{uv} = 2n$ for all $v\in V$. We call $m_e$ the \emph{multiplicity} of edge $e$ in $m$.
For example a $2$-multiweb is (ignoring edges of multiplicity zero) a disjoint collection of simple loops and doubled edges. 
We let $\Omega_{2n}=\Omega_{2n}(\G)$ be the set of all $2n$-multiwebs in $\G$.

Multiwebs are variants of \emph{webs}, which are certain types of graphs used in the study of the representation theory of linear groups,
in particular $\SL(n)$ and $\Sp(2n)$, see \cite{Kuperberg, lam_15}. To each multiweb is associated a corresponding web.
  
Each $2n$-multiweb $m\in\Omega_{2n}$ has a \emph{trace}, depending on the connection $\Phi$, defined below.
Our main result, Theorem \ref{main}, is the construction of a modified weighted adjacency matrix $\tilde H=\tilde H_\Phi$ of $\G$ whose Pfaffian is the sum of traces of all 
$2n$-multiwebs of $\G$. An analogous result for the Lie group $\SL(n)$ for bipartite graphs $\G$ was recently obtained in \cite{DKS}.

As application, to a graph $\G$ embedded in a planar domain $\Sigma$ we associate a 
regular function $\Pf(\tilde H)$ on the $\Sp(2n)$-character variety of $\Sigma$ which,
encodes information about flat $\Sp(2n)$ local systems on $\G$. One can use these functions 
to study the geometry of the character variety, 
but also (as we do here) the combinatorics
and probability theory of random multiwebs in $\G$.

The cases $n=1$ and $n=2$, that is, for the groups $\Sp(2)$ and $\Sp(4)$, are of special combinatorial interest.

\subsection{Kasteleyn's theorem}

For $\Sp(2)$ we have some concrete combinatorial applications to enumeration of perfect matchings.
A $2$-multiweb is
topologically very simple: it is just a disjoint collection of simple loops and doubled edges. 
Theorem \ref{main} in this setting can be used to study the dimer model and double dimer model.
Kasteleyn's celebrated theorem \cite{Kast} counts dimer covers (perfect matchings) of a planar graph via the Pfaffian of a modified (antisymmetric) adjacency matrix.
We give an alternate formulation and proof here (Theorem \ref{Sp2Kthm}) using the above $2$-multiweb trace enumeration, with an appropriate
\emph{Kasteleyn connection} on $\G$.

By using our $2n$-multiweb enumeration and a higher-rank version of the Kasteleyn connection we 
generalize Kasteleyn's theorem to an enumeration of $2n$-dimer covers interacting through the group $U(n)$ 
(see Section \ref{dimersection} and Theorem \ref{Kastthm2n}).

\def\tetravalent
{\begin{tric}
\draw (-1,1)--(1,-1);
\draw  (-1,-1)--(1,1);
\filldraw [black] (0,0) circle (3pt);
\end{tric}}

\def\Skeineadouble
{\begin{tric}
\draw[scale=0.5] (-2,-1.732)--(-1,0)--(-2,1.732) (2,1.732)--(1,0)--(2,-1.732); \draw[scale=0.5,thin, decorate, decoration={snake, segment length=0.5mm, amplitude=0.45mm},darkgreen] (-1,0)--(1,0);
\end{tric}
}

\def\Skeinec
{\begin{tric}
\draw [scale=0.45] (-2,-2)..controls(-1,-1)and(-1,1)..(-2,2) (2,2)..controls(1,1)and(1,-1)..(2,-2);
\end{tric}
}

\def\Skeinebdouble
{\begin{tric}
\draw [scale=0.5]  (-1.732,2)--(0,1)--(1.732,2) (1.732,-2)--(0,-1)--(-1.732,-2); 
\draw [scale=0.5,thin, decorate, decoration={snake, segment length=0.5mm, amplitude=0.45mm},darkgreen](0,-1)--(0,1);
\end{tric}
}

\def\Skeined
{\begin{tric}
\draw  [scale=0.45] (-2,2)..controls(-1,1)and(1,1)..(2,2) (2,-2)..controls(1,-1)and(-1,-1)..(-2,-2) ;
\end{tric}
}

\def\TetraIsDet
{\begin{tric}
\draw [scale=0.7] 
      (0,0)..controls(1.2,-0.6)and(1.5,-1.4)..(1.5,-2)
      (0,0)..controls(0.3,-0.6)and(0.5,-1.4)..(0.5,-2)
      (0,0)..controls(-0.3,-0.6)and(-0.5,-1.4)..(-0.5,-2)
      (0,0)..controls(-1.2,-0.6)and(-1.5,-1.4)..(-1.5,-2);
\filldraw [black] (0,-0.02) circle (3pt);
\end{tric}
}

\def\TetraIsDetSwap
{\begin{tric}
\draw [scale=0.7] 
      (0,0)..controls(1.2,-0.6)and(1.5,-1.4)..(1.5,-2)
      (0,0)..controls(0.3,-0.6)and(0.5,-1.4)..(0.5,-2)
      (0,0)..controls(-0.3,-0.6)and(-0.5,-1.4)..(-0.5,-2)
      (0,0)..controls(-1.2,-0.6)and(-1.5,-1.4)..(-1.5,-2);
\draw [scale=0.7] (-1.5,-2)node[below,scale=0.7] {$e_j$}
                  (-0.5,-2)node[below,scale=0.7] {$e_i$};
\filldraw [black] (0,-0.02) circle (3pt);
\end{tric}
}

\def\TetraIsDetSwapA
{\begin{tric}
\draw [scale=0.7] 
      (0,0)..controls(1.2,-0.6)and(1.5,-1.4)..(1.5,-2)
      (0,0)..controls(0.3,-0.6)and(0.5,-1.4)..(0.5,-2)
      (0,0)..controls(-0.3,-0.6)and(-0.5,-1.4)..(-0.5,-2)
      (0,0)..controls(-1.2,-0.6)and(-1.5,-1.4)..(-1.5,-2);
\draw [scale=0.7] (-1.5,-2)node[below,scale=0.7] {$e_i$}
                  (-0.5,-2)node[below,scale=0.7] {$e_j$};
\filldraw [black] (0,-0.02) circle (3pt);
\end{tric}
}

\subsection{The $\Sp(4)$ web category}

Theorem \ref{main}, our main result, inspires a change of viewpoint on the $\Sp(4)$ web category,
since it highlights the importance of the tetravalent vertex.
Kuperberg \cite{Kuperberg} defined \emph{the $\Sp(4)$ web category} by generators and relations (Definition \ref{C2Spider} below), 
where one of the relations is $\frac{\pi}{2}$-rotational invariant:
$$\Skeineadouble\ \   - \  \frac{1}{2} \ \Skeinec \ \    =  \  \Skeinebdouble \ - \  
\frac{1}{2} \ \Skeined. $$
This allowed him to define a tetravalent vertex by  $$\tetravalent :=  \  \Skeinebdouble \ - \  
\frac{1}{2} \ \Skeined,$$
and he deduced skein relations with tetravalent vertices instead of trivalent vertices (Proposition \ref{TetravalentSkein} below).

In order to study the connection between $4$-webs (that is, webs with only tetravalent vertices)
and $\Sp(4)$-webs, we observe in the following theorem that the tetravalent vertex corresponds to the determinant of a $4$ by $4$ matrix, i.e. the exterior product of $4$ vectors in the $4$-dimensional defining representation of $\Sp(4)$.

\def\TetraIsDet
{\begin{tric}
\draw [scale=0.7] 
      (0,0)..controls(1.2,-0.6)and(1.5,-1.4)..(1.5,-2)
      (0,0)..controls(0.3,-0.6)and(0.5,-1.4)..(0.5,-2)
      (0,0)..controls(-0.3,-0.6)and(-0.5,-1.4)..(-0.5,-2)
      (0,0)..controls(-1.2,-0.6)and(-1.5,-1.4)..(-1.5,-2);
\filldraw [black] (0,-0.02) circle (3pt);
\end{tric}
}

\def\TetraIsDeta
{\begin{tric}
\draw[scale=0.7] (1.5,-2)--(1,-1)--(0.5,-2)
                 (-1.5,-2)--(-1,-1)--(-0.5,-2);
\draw[scale=0.7,thin, decorate, decoration={snake, segment length=0.5mm, amplitude=0.45mm},darkgreen](1,-1)..controls(1,-0.6)and(0.4,0)..(0,0)..controls(-0.4,0)and(-1,-0.6)..(-1,-1); 
\end{tric}
}

\def\TetraIsDetb
{\begin{tric}
\draw[scale=0.7]
     (1.5,-2)..controls(1.5,-1)and(1.2,-0.5)..(1,-0.5)
     (0.5,-2)..controls(0.5,-1)and(0.8,-0.5)..(1,-0.5)
     (-0.5,-2)..controls(-0.5,-1)and(-0.8,-0.5)..(-1,-0.5)
     (-1.5,-2)..controls(-1.5,-1)and(-1.2,-0.5)..(-1,-0.5) ;
\end{tric}
}

\def\TetraIsCodet
{\begin{tric}
\draw [scale=0.7] 
      (0,0)..controls(1.2,0.6)and(1.5,1.4)..(1.5,2)
      (0,0)..controls(0.3,0.6)and(0.5,1.4)..(0.5,2)
      (0,0)..controls(-0.3,0.6)and(-0.5,1.4)..(-0.5,2)
      (0,0)..controls(-1.2,0.6)and(-1.5,1.4)..(-1.5,2);
\filldraw [black] (0,0.02) circle (3pt);
\end{tric}
}

\begin{thm} \label{Sp4det}
Denote $\textbf{det}: = \TetraIsDet\ \in {\Hom}(\ObjOne^{\otimes 4},\mathbb{R})$ , 
We have : 
\begin{align*}
  \Phi(\textbf{det}): \ \ \ \ \ \ \ \  \ \ \ \ \ \  V^{\otimes 4} \longrightarrow &\mathbb{R} \\
 v_1 \otimes v_2 \otimes v_3 \otimes v_4 
 \mapsto & |v_1 \wedge v_2 \wedge v_3 \wedge v_4| 
\end{align*}
Denote $\textbf{codet}: = \TetraIsCodet \in{\Hom}(\mathbb{R}, \ObjOne^{\otimes 4})$, 
We have : 
\[
  \Phi(\textbf{codet}):  \mathbb{R}   \longrightarrow V^{\otimes 4} \]
\[ \ \ \ \ \ \ \ \ \ \ \ \ \ \ \ \ \ \ \ \ \ \ \ \ \ \ \ \ \ \ \ \ \ \ \ \ \ \ \ \ \ \ \ \ \ \ \ \ \ \ \ \ \ \ \ \ \ \   1  \mapsto \sum _{\sigma \in \mathfrak{S}_4} (-1)^{\sigma} e_{\sigma(1)} \otimes e_{\sigma(2)}\otimes e_{\sigma(3)} \otimes e_{\sigma(4)} 
\]
\end{thm}

For the proof of Theorem \ref{Sp4det} see Section \ref{thm1proof}.
Although we will not use it here, we give a quantum version (for the quantum group $U_q(\mathfrak{sp}_4)$) of Theorem \ref{Sp4det} in the Appendix, Section \ref{quantSp4}.

If our graph $\G$ is drawn on a multiply connected domain
in $\Cx$ with flat connection, these $4$-web skein relations, which preserve the trace, allow us to replace each 
$4$-web with a linear combination of 
\emph{reduced webs}: a $4$-web on a surface is said to be \emph{reduced} if it has no contractible faces of degree $0,1,2$ or $3$ (see \cite{Kuperberg}).
The weighted enumeration of $4$-multiwebs becomes a weighted
enumeration of reduced $4$-webs.
We carry out this enumeration procedure for the simplest case of an annulus. See Section \ref{4websurfacesection}.
For the next simplest surfaces, a torus and a pair of pants, we classify reduced webs and give some basic results (Sections 
\ref{torussection} and \ref{pantssection}).

\subsection{The $\Sp(2n)$ case}
The definition of the $\Sp(4)$ web category was generalized to the $\Sp(2n)$ case for any rank $n$ in \cite{bodish2021type}. As above we define 
a $2n$-valent vertex (Definition \ref{2nVertex}) in the $\Sp(2n)$ web category with the help of the braiding structure, so that the vertex corresponds to the determinant of a $2n$ by $2n$ matrix.

\def\NTetraIsDet
{\begin{tric}
\draw [scale=0.7] 
      (0,0)..controls(1.7,-0.6)and(2,-1.4)..(2,-2)
      (0,0)..controls(0.8,-0.6)and(1,-1.4)..(1,-2)
      (0,0)..controls(-0.8,-0.6)and(-1,-1.4)..(-1,-2)
      (0,0)..controls(-1.7,-0.6)and(-2,-1.4)..(-2,-2);
\filldraw[black,scale=0.7] (-0.4,-1.8) circle (1pt) 
(0,-1.8) circle (1pt) (0.4,-1.8) circle (1pt) ;
\filldraw[black] (0,-0.05) circle (3pt) ;
\end{tric}
}

\def\NTetraIsDetA
{\begin{tric}
\draw [scale=0.7,decoration={markings,mark=at position 0.4 with {\arrow{stealth}}},postaction={decorate}] 
      (2,-2)..controls(2,-1.4)and(1.7,-0.6)..(0,0);
\draw [scale=0.7,decoration={markings,mark=at position 0.47 with {\arrow{stealth}}},postaction={decorate}]       
      (1,-2)..controls(1,-1.4)and(0.8,-0.6)..(0,0);
\draw [scale=0.7,decoration={markings,mark=at position 0.47 with {\arrow{stealth}}},postaction={decorate}] 
       (-1,-2)..controls(-1,-1.4)and(-0.8,-0.6)..(0,0);
\draw [scale=0.7,decoration={markings,mark=at position 0.4 with {\arrow{stealth}}},postaction={decorate}]       
     (-2,-2)..controls(-2,-1.4)and(-1.7,-0.6)..(0,0);
      
\filldraw[black,scale=0.7] (-0.4,-1.8) circle (1pt) 
(0,-1.8) circle (1pt) (0.4,-1.8) circle (1pt) ;
\filldraw[black] (0,-0.05) circle (3pt) ;

\end{tric}
}

\begin{thm}\label{Sp2ndet}
Denote $\textbf{det}: = \NTetraIsDet \in \Hom(\ObjOne^{\otimes 2n}, \mathbb{R} )$. Then
   \begin{align*}
       \Phi (\textbf{det}):  \qquad \qquad \qquad 
       { V_{\varpi_1}}^{\otimes 2n} &\rightarrow \mathbb{R} \\
        v_1\otimes v_2 \otimes \cdot \cdot \cdot \otimes v_{2n} &\mapsto  |v_1\wedge v_2 \wedge \cdot \cdot \cdot \wedge v_{2n}| ,
   \end{align*} 
where for $w \in \bigwedge^{2n} V_{\varpi_1}$, we denote by $|w|$ the element $r \in \R$ such that $w=r (e_1 \wedge e_2 \wedge \cdot \cdot \cdot  \wedge e_{2n})$. 
\end{thm}

For the proof see Section \ref{thm2proof}.
This construction allows us to relate $2n$-multiwebs to $\Sp(2n)$ webs.
Using this we give an explicit classification of reduced $\Sp(2n)$ webs on an annulus in Theorem \ref{Sp2nAnnulus}.

Although we won't use it here, a version of the degree-$n$ vertex and Theorem \ref{Sp2ndet} for the quantum group $U_q(\mathfrak{sl}_n)$ is given in the Appendix, Section \ref{quantSLn}. See \cite{CD} for an alternate formulation.
We don't have, at this moment, the analogous result in the quantum symplectic case.
\bigskip

\noindent{\bf Acknowledgments.} This research was supported by NSF grant DMS-1940932 and CCF-2317280 and the Simons Foundation grant 327929 and 994328. 
We thank Elijah Bodish, Daniel Douglas, Joel Hass, Vijay Higgins, Greg Kuperberg, Nicholas Ovenhouse and Sri Tata for discussions and insights.

\section{A history of web categories}

The discovery of the Jones polynomial \cite{JonesPolynomial} triggered significant mathematical developments in 
many areas including knot theory and quantum algebra. The Jones polynomial is a link invariant, written as a Laurent polynomial in one variable $q$.
It was discovered by Reshetikhin and Turaev \cite{RTinv} that the Jones polynomial can be defined by using the braiding structure in the ribbon category of $U_q(\mathfrak{sl}_2)$. This ribbon category is universally constructed for any simple Lie algebra $\mathfrak{g}$, giving a generalization of the Jones polynomial to a family of quantum link invariants, 
one for each simple Lie algebra. When $q$ is at a root of unity, 
these ribbon categories also give invariants of a three-manifold by coloring the framed link along which Dehn surgery
is performed.

 The ribbon category that computes the Jones polynomial can be presented as the Temperley-Lieb category \cite{TemLie}, which is equivalent to the subcategory of $\Rep(U_q(\mathfrak{sl}_2))$ generated by the $q$-analogue of the defining representation. Hence we can use diagrams and graphical calculations in the Temperley-Lieb category to study the representation theory of quantum group $U_q(\mathfrak{sl}_2)$. 
 
 The web category of $\mathfrak{g}_2$ was first introduced by Kuperberg \cite{Kupe-first-G2} to compute the quantum  $\mathfrak{g}_2$ invariants for links and knots. The definition of web categories was later generalized to include all the rank two simple Lie algebras $\mathfrak{g}$, that is $\mathfrak{g} = \mathfrak{sl}_3$ , $\mathfrak{sp}_4$, or $\mathfrak{g}_2$ \cite{Kuperberg}. It was shown there that the web category of $\mathfrak{g}$ is equivalent to the category of fundamental representations of the quantum group $U_q(\mathfrak{g})$, generalizing the relation between the Temperley-Lieb category and the representations of $U_q(\mathfrak{sl}_2)$.

Bearing in mind the goal of giving graphically presented generators and relations descriptions of the categories of fundamental representations of quantum groups, the definition of web categories was later extended to types A \cite{CKM} and C \cite{bodish2021type}.
Web categories are widely used to tackle problems in quantum topology, including 
the categorification of quantum link invariants generalizing Khovanov homology \cite{categorjones}, 
and giving formulas for irreducible representations of quantum groups generalizing the Jones-Wenzl projectors \cite{Jon2,Wenzl}.

The web category of $\GL_n$ was related to the dimer model in \cite{DKS}. 
We extend here these recent results to $\Sp(2n)$ when q=1. 
For recent work on dimers and $\Rep(U_q(\mathfrak{sl}_n))$
see \cite{DKOPT}.


\def\BigIk{
\begin{tric}
\draw [scale=0.6](0,0)node[left,scale=0.7,black]{$n$};
\draw 
[scale=0.6, decoration={markings,mark=at position 0.6 with {\arrow{stealth}}},postaction={decorate}]
(0,-1)--(0,0) ;

\draw 
[scale=0.6, decoration={markings,mark=at position 0.35 with {\arrow{stealth}}},postaction={decorate}]
(-1,-2)..controls(-1,-1)..(0,-1)
       node[midway,left,scale=0.7,black]{$n-1$};

\draw 
[scale=0.6,decoration={markings,mark=at position 0.6 with {\arrow{stealth}}},postaction={decorate}]
       (1,-7)..controls(1,-2)..(0,-1) 
       (1,-7)node[below,scale=0.7,black]{$1$};
       
\draw 
[scale=0.6,decoration={markings,mark=at position 0.35 with {\arrow{stealth}}},postaction={decorate}]
(-2,-3)..controls(-2,-2)..(-1,-2)
        node[midway,left,scale=0.7,black]{$n-2$};

\draw 
[scale=0.6,decoration={markings,mark=at position 0.6 with {\arrow{stealth}}},postaction={decorate}]
(0,-7)..controls(0,-3)..(-1,-2) (0,-7)node[below,scale=0.7,black]{$1$};

\filldraw[scale=0.6,black] (-2.25,-3.25) circle (1pt) 
                 (-2.5,-3.5) circle (1pt) (-2.75,-3.75) circle (1pt)
      (-0.9,-6.5) circle (1pt) (-1.25,-6.5) circle (1pt) (-1.6,-6.5) circle (1pt);

\draw 
[scale=0.6,decoration={markings,mark=at position 0.45 with {\arrow{stealth}}},postaction={decorate}]
(-3.5,-5)..controls(-3.5,-4.5)and(-3.5,-4)..(-3,-4)node[midway,left,scale=0.7,black]{$3$};

\draw 
[scale=0.6,decoration={markings,mark=at position 0.35 with {\arrow{stealth}}},postaction={decorate}]
(-4.5,-6)..controls(-4.5,-5)..(-3.5,-5)node[midway,left,scale=0.7,black]{$2$};

\draw 
[scale=0.6,decoration={markings,mark=at position 0.35 with {\arrow{stealth}}},postaction={decorate}]
(-5.5,-7)..controls(-5.5,-6)..(-4.5,-6)    (-5.5,-7)node[below,scale=0.7,black]{$1$};

\draw 
[scale=0.6,decoration={markings,mark=at position 0.4 with {\arrow{stealth}}},postaction={decorate}]
(-2.5,-7)..controls(-2.5,-5.5)..(-3.5,-5) (-2.5,-7)node[below,scale=0.7,black]{$1$};

\draw 
[scale=0.6,decoration={markings,mark=at position 0.35 with {\arrow{stealth}}},postaction={decorate}]
(-3.5,-7)..controls(-3.5,-6)..(-4.5,-6)(-3.5,-7)node[below,scale=0.7,black]{$1$};
\end{tric}
}

\def\QuantumDet
{\begin{tric}
\draw [scale=0.7,decoration={markings,mark=at position 0.4 with {\arrow{stealth}}},postaction={decorate}] 
      (2,-2)..controls(2,-1.4)and(1.7,-0.6)..(0,0);
\draw [scale=0.7,decoration={markings,mark=at position 0.47 with {\arrow{stealth}}},postaction={decorate}]       
      (0.8,-2)..controls(0.8,-1.4)and(0.6,-0.6)..(0,0);
\draw [scale=0.7,decoration={markings,mark=at position 0.5 with {\arrow{stealth}}},postaction={decorate}] 
      (0.2,-2)..controls(0.2,-1.4)and(0.1,-0.6)..(0,0);
\draw [scale=0.7,decoration={markings,mark=at position 0.45 with {\arrow{stealth}}},postaction={decorate}]       
     (-1.2,-2)..controls(-1.2,-1.4)and(-0.8,-0.6)..(0,0);
\draw [scale=0.7,decoration={markings,mark=at position 0.4 with {\arrow{stealth}}},postaction={decorate}]      
      (-2,-2)..controls(-2,-1.4)and(-1.7,-0.6)..(0,0);

 \draw[scale=0.7]
 (0.2,-2) node[below,scale=0.7]{$e_j$};

  \draw[scale=0.7]
 (0.8,-2) node[below,scale=0.7]{$e_i$};

\filldraw[black,scale=0.7] (-0.4,-1.8) circle (1pt) 
(-0.6,-1.8) circle (1pt) (-0.8,-1.8) circle (1pt) ;

\filldraw[black,scale=0.7] (1.6,-1.8) circle (1pt) 
(1.2,-1.8) circle (1pt) (1.4,-1.8) circle (1pt) ;

\filldraw[black] (0,-0.05) circle (3pt) ;

\end{tric}
}

\def\QuantumDetA
{\begin{tric}
\draw [scale=0.7,decoration={markings,mark=at position 0.4 with {\arrow{stealth}}},postaction={decorate}] 
      (2,-2)..controls(2,-1.4)and(1.7,-0.6)..(0,0);
\draw [scale=0.7,decoration={markings,mark=at position 0.47 with {\arrow{stealth}}},postaction={decorate}]       
      (0.8,-2)..controls(0.8,-1.4)and(0.6,-0.6)..(0,0);
\draw [scale=0.7,decoration={markings,mark=at position 0.5 with {\arrow{stealth}}},postaction={decorate}] 
      (0.2,-2)..controls(0.2,-1.4)and(0.1,-0.6)..(0,0);
\draw [scale=0.7,decoration={markings,mark=at position 0.45 with {\arrow{stealth}}},postaction={decorate}]       
     (-1.2,-2)..controls(-1.2,-1.4)and(-0.8,-0.6)..(0,0);
\draw [scale=0.7,decoration={markings,mark=at position 0.4 with {\arrow{stealth}}},postaction={decorate}]      
      (-2,-2)..controls(-2,-1.4)and(-1.7,-0.6)..(0,0);

 \draw[scale=0.7]
 (0.2,-2) node[below,scale=0.7]{$e_i$};

  \draw[scale=0.7]
 (0.8,-2) node[below,scale=0.7]{$e_j$};

\filldraw[black,scale=0.7] (-0.4,-1.8) circle (1pt) 
(-0.6,-1.8) circle (1pt) (-0.8,-1.8) circle (1pt) ;

\filldraw[black,scale=0.7] (1.6,-1.8) circle (1pt) 
(1.2,-1.8) circle (1pt) (1.4,-1.8) circle (1pt) ;

\filldraw[black] (0,-0.05) circle (3pt) ;

\end{tric}
}

\def\QuantumDetB
{\begin{tric}
\draw [scale=0.7,decoration={markings,mark=at position 0.4 with {\arrow{stealth}}},postaction={decorate}] 
      (2,-2)..controls(2,-1.4)and(1.7,-0.6)..(0,0);
\draw [scale=0.7,decoration={markings,mark=at position 0.47 with {\arrow{stealth}}},postaction={decorate}]       
      (0.8,-2)..controls(0.8,-1.4)and(0.6,-0.6)..(0,0);
\draw [scale=0.7,decoration={markings,mark=at position 0.5 with {\arrow{stealth}}},postaction={decorate}] 
      (0.2,-2)..controls(0.2,-1.4)and(0.1,-0.6)..(0,0);
\draw [scale=0.7,decoration={markings,mark=at position 0.45 with {\arrow{stealth}}},postaction={decorate}]       
     (-1.2,-2)..controls(-1.2,-1.4)and(-0.8,-0.6)..(0,0);
\draw [scale=0.7,decoration={markings,mark=at position 0.4 with {\arrow{stealth}}},postaction={decorate}]      
      (-2,-2)..controls(-2,-1.4)and(-1.7,-0.6)..(0,0);

 \draw[scale=0.7]
 (0.2,-2) node[below,scale=0.7]{$e_i$};

  \draw[scale=0.7]
 (0.8,-2) node[below,scale=0.7]{$e_i$};

\filldraw[black,scale=0.7] (-0.4,-1.8) circle (1pt) 
(-0.6,-1.8) circle (1pt) (-0.8,-1.8) circle (1pt) ;

\filldraw[black,scale=0.7] (1.6,-1.8) circle (1pt) 
(1.2,-1.8) circle (1pt) (1.4,-1.8) circle (1pt) ;

\filldraw[black] (0,-0.05) circle (3pt) ;
\end{tric}
}

\section{Multiwebs, cilia, traces}

\subsection{Symplectic group}
The Lie group $\Sp(2n)=\Sp(2n,\R)$, the \emph{symplectic group}, is the set of matrices in $\SL(2n,\R)$ preserving the standard symplectic form
$$\Sp(2n) = \{M\in \SL(2n,\R)~|~M^tJM=J\},$$ 
where $J$ is the matrix
$$J=\begin{pmatrix}0_n&I_n\\-I_n&0_n
\end{pmatrix}
$$
and $I_n, 0_n$ are the $n\times n$ identity matrix and zero matrix, respectively.
The operator $J$ defines the standard symplectic form $\omega$ on $\R^{2n}$ via $\omega(u,v) = u^tJv$.

\subsection{Cilia and orientations}
Let $\G=(V,E)$ be a planar graph with an $\Sp(2n)$ connection $\Phi=\{\phi_{uv}\}_{uv\in E}$.
The definition of trace of a multiweb on $\G$ 
requires some extra combinatorial data associated to $\G$. We need an \emph{orientation} of the edges of $\G$
and a choice of \emph{cilia}. 
By cilia we mean the following. At each vertex of $\G$ choose a total order of the incident edges, which is compatible with the counterclockwise circular order.
To do this it suffices to choose a starting edge. Notationally we put a mark, called cilium, in the wedge between the 
first edge and last edge, at each vertex. See Figure \ref{traceexample}.

The edge orientations and cilia choices can be arbitrary; different choices affect the definition of trace of any given multiweb by a sign.
However to get coherence of signs when summing traces of different multiwebs, we need to impose restrictions on both the orientations and the choice of cilia. One way is as follows. 
Given a generic planar embedding of $\G$, orient the edges upwards according to the $y$-coordinate of their endpoints.
At each vertex, put the cilium horizontally to its left. We call this the \emph{standard choice}
of cilia and orientations.
See an example in Figure \ref{traceexample}.


\subsection{Trace} \label{TraceDef}
Let $m$ be a multiweb in $\G$. The trace $\Tr(m)$ is defined as follows.
First assume $m$ is \emph{simple}, that is, $m_e=0$ or $m_e=1$ for every edge.
At a vertex $v$, let $vv_1,\dots,vv_{2n}$ be the adjacent edges in the support of $m$, in counterclockwise 
order starting from the cilium.
Let $W_i\cong \R^{2n}$ be a vector space associated to the half-edge of $vv_i$ near $v$. 
Let $\{e^i_1,\dots,e^i_{2n}\}$ be a standard basis of $W_i$ (we will drop the superscripts in the future). We
define a vector $\gamma_v\in W_1\otimes \dots\otimes W_{2n}$,
the \emph{codeterminant}, via the formula
$$\gamma_v = 
\sum_{\sigma \in S_{2n}}(-1)^\sigma e_{\sigma(1)}\otimes e_{\sigma(2)}\otimes \dots\otimes e_{\sigma(2n)}.$$

The trace of $m$ is now the contraction of the tensor product of the $\gamma_v$ along edges using the connection and symplectic form:
In the tensor product $\bigotimes_v\gamma_v$, the tensor components for the two halves of edge $v\to v'$
are contracted using $J\phi_{vv'}$. That is, for a simple tensor $\dots \otimes a_v\otimes\dots\otimes a_{v'}\otimes \dots$
where $a_v,a_{v'}$ correspond to the two halves of edge $vv'$, directed from $v$ to $v'$
the component of the contraction is $(a_{v'})^tJ\phi_{vv'}a_v$. 
Note that if we reverse the orientation of the edge $vv'$, the contribution from this edge changes sign: 
$$a_v^tJ\phi_{v'v}a_{v'} = -(a_{v'})^tJ\phi_{vv'}a_v$$
see (\ref{as}) below.

If $m$ is not simple, for each edge $e=vv'$ of multiplicity $m_e>1$ and connection matrix $\phi_{vv'}$, 
replace $e$ by $m_e$ parallel edges, with the same endpoints $v,v'$, same orientation, and with same connection matrix
$\phi_{vv'}$. This splitting results in a simple multiweb $m^*$
on a new graph $\G^*$. We define 
\begin{equation} \label{MultiPara}
    \Tr(m) := \frac{\Tr(m^*)}{\prod_e m_e!}.
\end{equation}

\def\SkeinmultKsp
{\begin{tric}
\draw (0.7,0)--(1.5,0.7) node[right]{$1$}
     (-0.7,0)--(-1.5,0.7) node[left]{$\mathbcal{1}$}
     (0.7,0)--(1.5,0.2) node[right]{$\mathbcal{1}$}
     (-0.7,0)--(-1.5,0.2) node[left]{$\mathbcal{1}$}
     (0.7,0)--(1.5,-0.7) node[right]{$\mathbcal{1}$}
     (-0.7,0)--(-1.5,-0.7) node[left]{$\mathbcal{1}$}
      (0.7,0)--(-0.7,0) node[midway,above]{$m_e$}; 
      
\filldraw [black] (1.4,0) circle (0.5pt);
\filldraw [black] (1.4,-0.2) circle (0.5pt);
\filldraw [black] (1.4,-0.4) circle (0.5pt);

\filldraw [black] (-1.4,0) circle (0.5pt);
\filldraw [black] (-1.4,-0.2) circle (0.5pt);
\filldraw [black] (-1.4,-0.4) circle (0.5pt);

\filldraw [black] (0.71,0) circle (3pt);
\filldraw [black] (-0.71,0) circle (3pt);
\end{tric}
}

\def\SkeinKtracesp
{\begin{tric}
\draw (0.7,0)..controls(0.6,-1)and(-0.6,-1)..(-0.7,0) 
     (0.7,0)..controls(0.6,1)and(-0.6,1)..(-0.7,0)
      (0.7,0)..controls(0.6,0.6)and(-0.6,0.6)..(-0.7,0) ;

\filldraw [black] (0,0) circle (0.5pt);
\filldraw [black] (0,-0.2) circle (0.5pt);
\filldraw [black] (0,-0.4) circle (0.5pt);
     
\draw (0.7,0)--(1.5,0.7) 
     (-0.7,0)--(-1.5,0.7) 
     (0.7,0)--(1.5,0.2) 
     (-0.7,0)--(-1.5,0.2) 
     (0.7,0)--(1.5,-0.7) 
     (-0.7,0)--(-1.5,-0.7) ; 
      
\filldraw [black] (1.4,0) circle (0.5pt);
\filldraw [black] (1.4,-0.2) circle (0.5pt);
\filldraw [black] (1.4,-0.4) circle (0.5pt);

\filldraw [black] (-1.4,0) circle (0.5pt);
\filldraw [black] (-1.4,-0.2) circle (0.5pt);
\filldraw [black] (-1.4,-0.4) circle (0.5pt);

\filldraw [black] (0.71,0) circle (3pt);
\filldraw [black] (-0.71,0) circle (3pt);
\end{tric}
}

 $$\text{In other words, } \text{Tr}\left(\SkeinmultKsp\right):=\frac{1}{m_e!} \ \text{Tr} \left(\SkeinKtracesp\right).$$


Note that changing the cilium at a single vertex by
one notch, crossing edge $e$, changes the sign of the trace of $m$ by $(-1)^{m_e}$. 
Changing the orientation of an edge of multiplicity $m_e$ changes the sign of the trace by 
$(-1)^{m_e}$.

\subsection{Trace for bipartite graphs}

\begin{prop} If $\G$ is bipartite, and the edges are oriented from black to white,
the above definition of trace agrees with the $\SL(2n)$ trace
(up to a global sign independent of $m$).
\end{prop}

Note that in \cite{sikora_01, DKS} the convention for the $\SL(2n)$ trace is to use a clockwise orientation
of half-edges at white vertices and counterclockwise orientation at black vertices. 
For this proof we use the counterclockwise orientation at both types of vertices; this is the
trace convention as defined in Kuperberg \cite{Kuperberg}. (Changing cyclic orientation from clockwise to counterclockwise 
multiplies the codeterminant
at that vertex, and thus the trace of any multiweb, by $(-1)^n$. So the net effect is to multiply the sum of traces by
$(-1)^{n|W|}$ where $|W|$ is the number of white vertices.)

\begin{proof}Let $m$ be a simple multiweb. At each white vertex, apply the operator $(-J)^{\otimes 2n}$
to the codeterminant. On the one hand this has no effect since the codeterminant is invariant under
$\SL(2n)$ change of basis, and $-J\in\SL(2n)$. 
On the other hand this has the effect of removing the $J$'s from the edges, that is,
the edge contribution $(a_{v'})^tJ\phi_{vv'}a_v$ becomes $(a_{v'})^t\phi_{vv'}a_v$,
which is the edge contribution for the $\SL(2n)$ trace. 
\end{proof}

\subsection{Trace for the identity connection}\label{idtracesection}

The trace of a multiweb for the identity connection can be defined in terms of edge colorings, as follows.
Let $\C=\{1,2,\dots,2n\}$ be the set of colors. Let $\sigma:\C\to\C$ be the involution switching $i$ and $i+n$ for each $1\le i\le n$. We refer to $\sigma$ as \emph{complementation}.

\begin{figure}[htbp]
\begin{center}\includegraphics[width=2in]{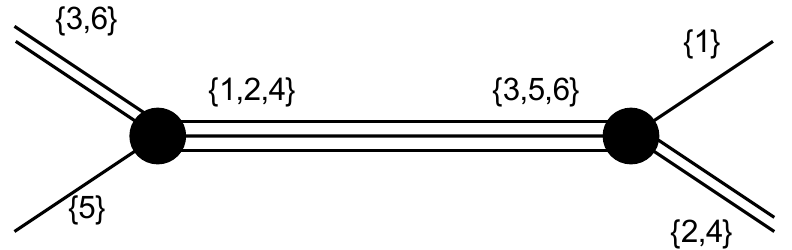}\end{center}
\caption{\label{spcoloring}An $\Sp(6)$-coloring of edges in a $6$-multiweb. Edge multiplicity is indicated by thickness.}
\end{figure}

An \emph{$\Sp(2n)$-coloring} of $m$ is defined as follows.
To an edge $e$
of multiplicity $m_e$ we associate two complementary subsets $S_e,T_e$ of $\C$, each of size $m_e$, one associated with each half-edge of $e$. At each vertex the color subsets on the adjacent half-edges are required to partition $\C$. See Figure \ref{spcoloring}.

We claim that the trace of $m$ is the signed sum of such $\Sp(2n)$-colorings. 
Let us first suppose $m$ is simple: $m_e=0$ or $1$ for every edge. 
Given a coloring $c$, at each vertex $v$, starting from the cilia and going counterclockwise
the order of the colors encountered (ignoring edges with $m_e=0$) is a permutation of $\{1,2,\dots,2n\}$ and thus
has a signature, $c_v$. For every edge
$uv$ with $m_e=1$, using the assigned orientation we have a sign $c_e$ which is $+1$ if the colors are increasing
(that is, $i,i+n$) and $-1$ if decreasing (that is, $i+n,i$).  
The sign of the coloring $c$, denoted $(-1)^c$, is defined to be the
product of these vertex signs and edge signs:
$$(-1)^c =\prod_{v}c_v\prod_{e}c_e.$$
Finally 
\be\label{trcolor}\Tr(m) := \sum_{c}(-1)^c\ee
where the sum is over edge colorings.

If $m$ is not simple, for each edge $e$ of multiplicity $m_e>1$, replace $e$ with $m_e$ parallel edges,
with the same orientation, to get a simple web $m^*$ on a new graph $\G^*$. Then 
\be\label{multitrace}\Tr(m) := \frac{\Tr(m^*)}{\prod_em_e!}.\ee

\begin{thm}\label{traceID}
For the identity connection, the definition of trace in Equations \eqref{trcolor} and \eqref{multitrace} agrees with the tensor network definition in Section \ref{TraceDef}.
\end{thm}

\begin{proof}
We identify the standard basis vectors of $\R^{2n}$ with the $2n$ colors, so that color $i$ corresponds to $e_i$.
A codeterminant then corresponds to all possible choices of ways to color the half-edges out of a vertex,
along with a sign given by the signature $c_v$. For the identity connection, 
a coloring at one vertex pairs nontrivially with a coloring at an adjacent
vertex only if the colors on the common edge are complementary, and in that case the pairing along the edge is
$c_e$: The edge sign corresponds to the operation of $J$ on basis vectors: $e_{i+n}^tJe_i=1$ corresponds to color
$i,i+n$ having sign $+1$, and $e_{i}^tJe_{i+n}=-1$ corresponds to color
$i+n,i$ having sign $-1$.  
\end{proof}

\begin{figure}[htbp]
\begin{center}\includegraphics[width=1.5in]{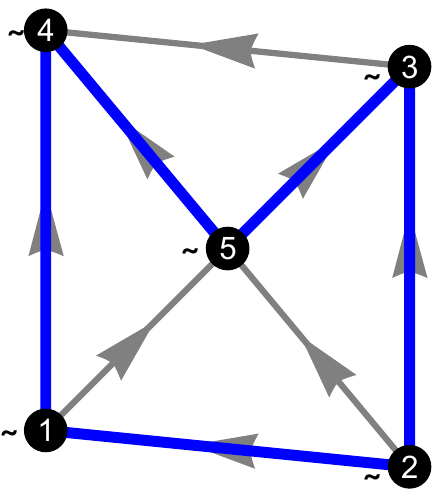}\hskip1cm\includegraphics[width=1.5in]{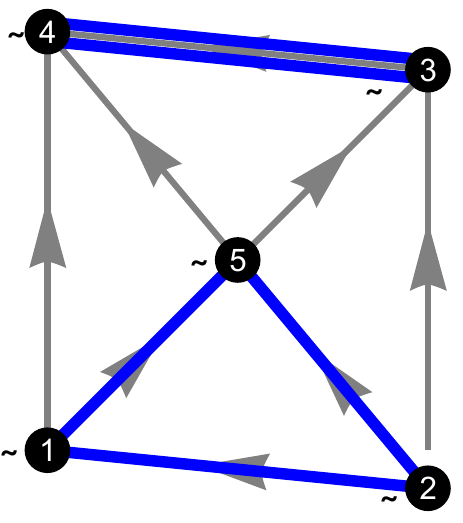}\end{center}
\caption{\label{traceexample}For $\Sp(2)$, this graph has $8$ multiwebs (two are shown, in blue). For the orientation and the cilia shown,
the traces are $+2$ for the first and $-2$ for the second.}
\end{figure}

See an example in Figure \ref{traceexample}.

There is generally no coherence in the signs of different colorings, that is, terms in (\ref{trcolor}) do not all have the same sign,
so the absolute value of the trace is not equal to the number of colorings. 
See however Section \ref{dimersection}
where sign coherence can be obtained by tensoring with a ``Kasteleyn connection''.

For a general $Sp(2n)$ connection we have the following trace formula.

\begin{prop}
Given $\mathcal{G}$ with an $Sp(2n)$ connection $\Phi=\{\phi_{uv}\}_{uv\in E}$, and standard cilia and edge orientations, for a simple multiweb $m$ we have
\be \label{TraceAsSum} \Tr(m) := \sum_{c} \prod_{v}c_v \prod_{uv} (e_{c,vu})^tJ\phi_{uv}e_{c,uv} ,   \ee
where the sum is over all colorings of half edges of $m$ such that all colors in $\mathcal{C}$ are used around each vertex, the first product is over vertices $v$ and the second is over oriented edges $uv$, $e_{c,vu}$ corresponds to the color of the half edge of $uv$ adjacent to $v$ in coloring $c$, and $e_{c,uv}$ corresponds to the color of the half edge of $uv$ adjacent to $u$ in coloring $c$.
   
If $m$ is not simple, for each edge $e$ of multiplicity $m_e>1$, replace $e$ with $m_e$ parallel edges,
with the same orientation, to get a simple web $m^*$ on a new graph $\G^*$. Then 
\be\label{MTraceAsSum}\Tr(m) := \frac{\Tr(m^*)}{\prod_em_e!}.\ee

\end{prop}

\begin{proof}
This works as in the case of the identity connection (Theorem \ref{traceID}), except that color sets $S_e,T_e$ are not necessarily complementary: when the identity connection is replaced with a general $Sp(2n)$ connection $\Phi$, the edge contribution to $\Tr(m)$ (for a simple $2n$-web) is changed from $(e_{c,vu})^tJe_{c,uv} $ to $(e_{c,vu})^tJ\phi_{uv}e_{c,uv} $.
\end{proof}

\subsection{$2$-multiweb traces}

\begin{prop}\label{2mwtrace}For a planar graph $\G$ with $N$ vertices, with an $\Sp(2)$-connection $\Phi$, and 
for a choice of standard orientation and cilia, the trace of a $2$-multiweb $m$ is
\be\label{Sp2trace}\Tr(m) = (-1)^{c_1+N}\prod_\gamma(-\tr(\phi_\gamma)),\ee
where $c_1$ is the number of doubled edges, the product is over nontrivial loops $\gamma$ of $m$, $\phi_\gamma$ is the 
monodromy of $\Phi$ along $\gamma$ (with either orientation) and $\tr$ is the matrix trace. 
\end{prop}

\begin{proof}Let us first consider the case where $\Phi$ is the identity connection. 
It suffices to assume $m$ is simple, so there are no doubled edges, by (\ref{multitrace}).
Let $\gamma$ be a loop, oriented counterclockwise (cclw), of length $L$. Color each edge of $\gamma$ with colors $1,2$ in 
cclw order (the first half edge has color $1$, the second half color $2$).
There are two sources of signs: from the cilia and from the edge signs.
The contribution to the sign from the edges of $\gamma$ is $-1$ per descending edge, that is,
edge where the second vertex has lower $y$ coordinate than the first.
The contribution to the sign from the cilia of $\gamma$ is $-1$ per cilium along $\gamma$
which is in the region enclosed by $\gamma$, that is, $-1$ per pair of adjacent ascending edges, 
$-1$ per local minima traversed right to left, and $-1$ per local maxima traversed left to right
(see Figure \ref{loopsigns}). 
\begin{figure}[htbp]
\begin{center}\includegraphics[width=1.3in]{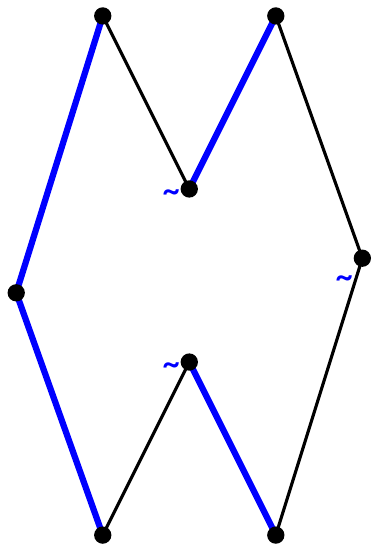}\end{center}
\caption{\label{loopsigns} This loop has $s=3$ interior cilia and $d=4$ descending edges (when oriented cclw) marked in blue, for a total sign of $(-1)^7=-1$.}
\end{figure}

By Lemma \ref{isotopy} below, $\gamma$
contributes a net sign of $(-1)^{L+1}$. If we reverse the colors, so that each edge is colored $2,1$ in cclw order,
the reversal induces a sign change at each edge and each vertex, so no net sign change. 
Thus including both colorings of $\gamma$,
the contribution to the trace is $2(-1)^{L+1}$.

Taking the product over all components the sign is $(-2)^{c_2}(-1)^N$.

Now assume $\Phi$ is a general $\Sp(2)$ connection.
Let $\gamma$ be a loop $v_1,\dots,v_L$, traversed cclw. 
Let $A_i=\phi_{v_iv_{i+1}}$ be the connection matrix on edge $v_iv_{i+1}$ oriented from $v_i$ to $v_{i+1}$. 
We reverse the standard orientation on edges that are downward-pointing along $\gamma$, 
so that edges
are oriented consistently along $\gamma$. This reversal introduces a sign $(-1)^d$,
where $d$ is the number of downward-oriented edges along $\gamma$.

Assume for the moment
that all cilia along $\gamma$ are external to $\gamma$. Then the codeterminant at $v_i$ is
$w_i:=e_1^{i+1}\otimes e_2^{i}-e_2^{i+1}\otimes e_1^i$
(with cyclic indices).

The contribution from $\gamma$ to the trace is then by definition the contraction
$$(JA_L)w_{L}(JA_{L-1})w_{L-1}\dots(JA_1)w_1.$$
Reparenthesizing and using cyclicity, this is the contraction $$A_L(w_{L}J)A_{L-1}(w_{L-1}J)\dots JA_1(w_1J).$$
At $v_{i+1}$ we thus apply the corresponding $J$ along edge $v_iv_{i+1}$ to the codeterminant
at $v_{i+1}$, yielding 
$$w_iJ=e_1^{i+1}\otimes(Je_2^i)-e_2^{i+1}\otimes(Je_1^i) = e_1^{i+1}\otimes e_1^i + e_2^{i+1}\otimes e_2^i.$$ 
Contracting the $A$'s with these tensors corresponds to matrix multiplication; the trace becomes
$$\tr(A_LA_{L-1}\dots A_1)=\tr(\phi_\gamma).$$

Now consider the case where not all cilia are external.
Each cilium on $\gamma$ enclosed by $\gamma$ changes the sign of the trace, compared
to the ``all external cilia" case.
Taking into account the additional sign contribution from the downward-oriented edges,
as in the case of the identity connection the contribution to the trace is $(-1)^{L+1}\tr(\phi_\gamma).$ Taking the product over all loops yields the formula.
\end{proof}

\begin{lemma}\label{isotopy}
Let $\gamma$ be a closed polygonal path $x_0,x_1,\dots,x_n=x_0$ in $\R^2$, possibly with repeated vertices, but
with no horizontal steps. 
Put a cilium at every vertex $x_i$, to the west of it.
Let $s$ be the number of cilia to the left of $\gamma$ (that is, for which $\gamma$ passes the cilium on its left).
Then the number of downward-pointing edges $d$ of $\gamma$ satisfies $d\equiv s+n+1\mod 2$.  
\end{lemma}

See Figure \ref{loopsigns} for an example.

\begin{proof}
This is true if $\gamma$ has a single downward-pointing edge. Starting from this $\gamma$, we deform $\gamma$ by moving vertices $x_i$ continuously in $\R^2$. 
Both $d$ and $s$ change by exactly $1$
each time an edge of $\gamma$ changes from upward-pointing to downwards or vice versa.
\end{proof}

\section{The $H$ matrix}

Suppose that $\G$ is planar with standard orientation and cilia, and with an $\Sp(2n)$ connection $\Phi=\{\phi_{uv}\}_{uv\in E}$. 
We define a matrix $H=H(\Phi)$ which does not depend on the orientation or cilia, as follows:
$H\in M_{V\times V}(M_{2n})$ is the matrix with entries 
$$H_{uv}=\begin{cases}J\phi_{uv}&u\sim v\\{\bf 0}&\text{else}\end{cases}$$
where ${\bf 0}$ is the $0$ matrix in $\Sp(2n)$.
Note that we have 
\be\label{as}H_{uv} = J\phi_{uv} = J\phi_{vu}^{-1} =\phi_{vu}^tJ =  -(J\phi_{vu})^t=-(H_{vu})^t. \ee

Let $N=|V|$ be the number of vertices of $\G$. 
Let $\tilde H$ be the $2nN\times 2nN$ matrix obtained from $H$ by replacing each entry with its $2n\times 2n$ array
of reals. Note that $\tilde H$ is antisymmetric, by (\ref{as}).

The Pfaffian $\Pf \tilde H$ of an antisymmetric matrix $\tilde H$ is defined by
\be\label{PfH}\Pf \tilde H = \sum_{\sigma}(-1)^\sigma \tilde H_{\sigma(1)\sigma(2)}\dots \tilde H_{\sigma(2M-1)\sigma(2M)}\ee
where the sum is over pairings $\sigma=\{\sigma(1),\sigma(2)\},\{\sigma(3),\sigma(4)\},\dots$ 
of the indices, with $\sigma(2i-1)<\sigma(2i)$ for each $i$.


\begin{thm}\label{main} We have $\Pf \tilde H= \pm \sum_{m\in\Omega_{2n}}\Tr(m).$
\end{thm}

\begin{proof}
It is convenient to assume vertices of $\G$ are in generic position and to index them according to increasing $y$-coordinate.
This ordering does not affect $\Pf\tilde H$.

Let $\mathcal{C}:=\{1,\dots,2n\}$. Let $\R^{2n}$ have standard basis $e_1,\dots,e_{2n}$. 
Let $M=nN$, so that $\tilde H$ is a $2M\times 2M$ matrix.

We let $\tilde\G$ be the graph with vertices $\tilde V = V\times \mathcal{C}$ and edges $(u,i)\to(v,j)$ for each edge $uv$ of $\G$,
and $i,j\in \mathcal{C}$. The edge $(u,i)\to(v,j)$ is given edge weight $e_j^tJ\phi_{uv}e_i$.

Consider the expression $\Pf\tilde H$ of (\ref{PfH}). Note that by our choice of vertex indexing in terms of increasing $y$ coordinate, $\tilde H_{\sigma(2i-1),\sigma(2i)}$ is now ordered according to the standard edge orientation.
Each nonzero summand in (\ref{PfH}) corresponds to a dimer cover of $\tilde\G$, also called $\sigma$. 
When we project the edges of such a dimer cover $\sigma$ to $\G$, it forms a $2n$-multiweb $\pi(\sigma)$ in $\G$.

So we can group the sum (\ref{PfH}) as a sum over multiwebs:
\be\label{Pfm}\Pf \tilde H=\sum_{m\in\Omega_{2n}}\,\sum_{\sigma: \pi(\sigma)=m}\tilde H_\sigma\ee
where 
$$\tilde H_\sigma:=(-1)^\sigma \tilde H_{\sigma(1)\sigma(2)}\dots \tilde H_{\sigma(2M-1)\sigma(2M)}$$ and
where $\pi$ is the projection from a dimer cover $\sigma$ of $\tilde\G$ to a $2n$-multiweb of $\G$. 
To complete the proof it suffices to show that the inner sum in (\ref{Pfm}) is, up to a global sign independent of $m$, equal to $\Tr(m)$.

We identify the $2n$ vertices of $\tilde\G$ lying over a vertex $v\in\G$ with the $2n$ colors. 
Each dimer cover $\sigma$ of $\tilde\G$ therefore corresponds to a 
coloring of the half-edges of its projection $\pi(\sigma)$: a dimer $(u,i)\to(v,j)$ projects to edge $uv$ with colors $i$ and $j$
at its two endpoints.

At each vertex of $\G$, all colors appear in $\pi(\sigma)$, so each dimer cover $\sigma\in\pi^{-1}(m)$
corresponds to a term in the expansion of $\Tr(m)$ in \eqref{TraceAsSum} and \eqref{MTraceAsSum}, and conversely. Moreover the product of the entries in $\tilde H$ given by $\sigma$
corresponds to the product of factors in the trace: a dimer $(u,i)\to(v,j)$ has weight $(J\phi_{uv})_{ij}$,
which is by definition the contraction along edge $uv$ of the term in the codeterminants at $u$ and $v$ with entry
$e_i$ and $e_j$ at $u,v$ respectively.
So there is a weight-preserving bijection between terms in $\Tr(m)$ and dimer covers in $\pi^{-1}(m)$. 
It remains to compare their signs. More specifically, we need to compare the signs in \eqref{PfH} of two
general dimer covers $\sigma,\sigma'$, covering possibly different webs $m,m'$, and 
the signs in \eqref{TraceAsSum} and \eqref{MTraceAsSum} of their corresponding colorings.

Let us take two dimer covers $\sigma,\sigma'$ of $\tilde\G$ and compare their signs. The union $\sigma\cup\sigma'$
is a union of even-length loops and doubled edges of $\tilde\G$; we can move from $\sigma$ to $\sigma'$ by sliding dimers
one notch along each loop. So it suffices to consider the case that $\sigma$ and $\sigma'$ differ in a single loop $\gamma$.

We can assume at this point that $m$ and $m'$ are simple, that is, have no multiple edges, by replacing
each edge with multiplicity by parallel edges in $\G$.

Let $\gamma$ be an even length loop in $\tilde\G$, with vertices $i_1,\dots,i_{2\ell}$ in cyclic order.
Along $\gamma$ let $\sigma$ be the pairing $\sigma=\{i_1,i_2\}\{i_3,i_4\}\dots\{i_{2\ell-1},i_{2\ell}\}$ 
and $\sigma'$ be the pairing
$\sigma'=\{i_2,i_3\}\{i_4,i_5\},\dots,\{i_{2\ell},i_1\}$; 
assume $\sigma$ and $\sigma'$ agree off of $\gamma$.
Using the definition (\ref{PfH}), the signature of $\sigma$ is a constant (depending on indices not involved in $\gamma$)
times the signature of the permutation in line form $i_1i_2\dots i_{2\ell}$, times $-1$ per
pair of $\sigma$ which is out of order. Likewise the signature of $\sigma'$ is the same constant times
the signature of $i_2i_3\dots i_{2\ell}i_1$, times $-1$ per pair of $\sigma'$ which is out of order.
The ratio of signatures is then $-1$ (for the cyclic permutation $(i_1i_2\dots i_{2\ell})$ of length $2\ell$) 
times $-1$ per pair of both $\sigma$ and $\sigma'$ which is out of order,
that is, $(-1)^{d+1}$ where $d$ is the number of downward edges in $\gamma$, when $\gamma$
is traversed cyclically in either direction.

Let us now compute the ratio of the signs of the colorings in \eqref{TraceAsSum}. 
Suppose first that $\gamma$ in $\tilde\G$ projects to a loop $\pi(\gamma)$ in $\G$. Suppose  $\pi(\gamma)$ has vertices $v_1, v_2, ..., v_t$ of $\G$ in order ($v_i$ and $v_j$ are not necessarily distinct when $i\ne j$). The sign contribution at a vertex $v_i$ is computed as follows.
As we traverse $\pi(\gamma)$, at vertex $v_i$ let $k_{v_i}$ be the number of edges in $\pi(\sigma)$
to the left of $\pi(\gamma)$, that is, in the clockwise sector between the edge in $\pi(\gamma)$ leading to $v_i$ and the edge in $\pi(\gamma)$ out from $v_i$. 
Suppose the cilium at $v_i$ is also in that sector. (See Figure \ref{signexample}).

\begin{figure}[htbp]
\begin{center}\includegraphics[width=2in]{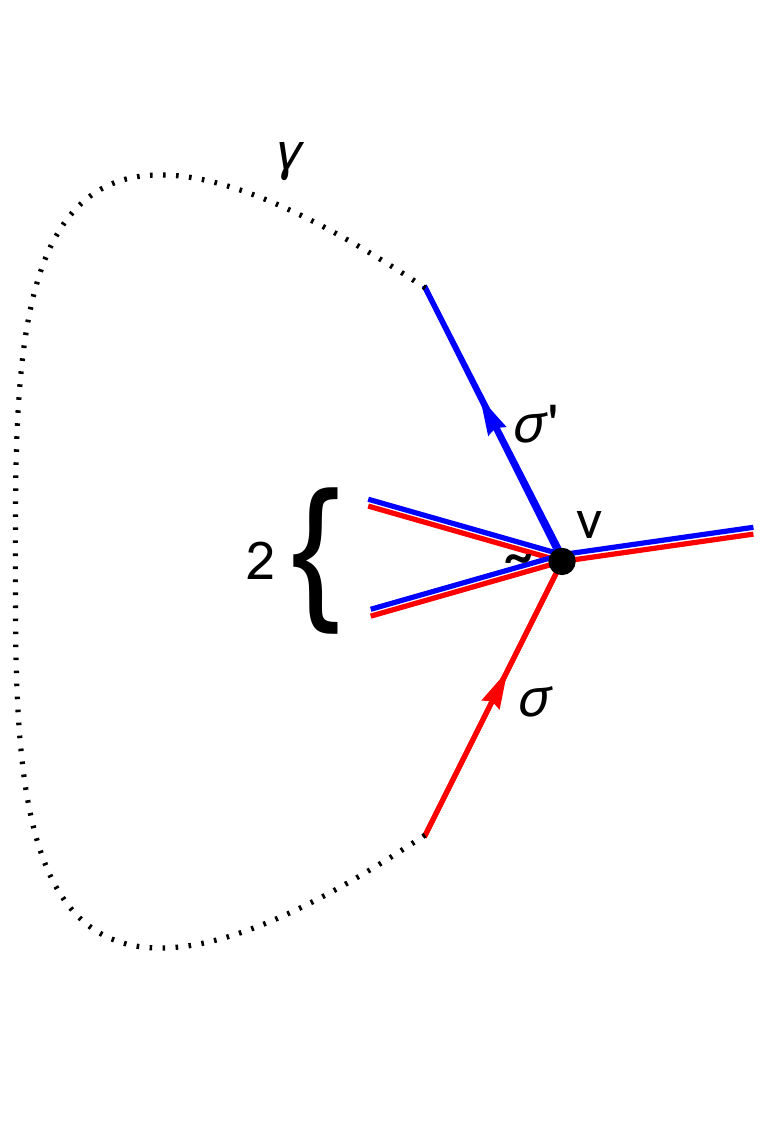}\end{center}
\caption{\label{signexample}At this vertex of $\gamma$ we have $k_v=2$.}
\end{figure}

Then when moving from 
$\sigma$ to $\sigma'$ (sliding dimers by one notch along $\gamma$),
the sign change at $v_i$ is 
thus $(-1)^{k_{v_i}+1}$. If the cilium is to the right of $\pi(\gamma)$ (not in the sector), the sign change is $(-1)^{k_{v_i}}$.

We claim that $\sum_{i=1}^{t} k_{v_i}$ is always even. Consider a closed path in the plane starting just left of the first edge of $\pi(\gamma)$ 
and following $\pi(\gamma)$ all the way around, keeping just to its left. It crosses an even number of edges of $\pi(\sigma)$, counted with multiplicity, 
since $\pi(\sigma)$ has even degree at each vertex. But it also crosses exactly $k_{v_i}$ edges of $\pi(\sigma)$ while near vertex $v_i$, and so  
$\sum_{i=1}^{t} k_{v_i}$ is even.

Taking the product of signs around $v_1, v_2, ... , v_t$, $\sum_{i=1}^{t} k_{v_i}$ is even, so the net sign change is
$(-1)^s$ where $s$ is the number of cilia on the left of $\pi(\gamma)$. 
By Lemma \ref{isotopy} this is
$(-1)^{d+1}$ where $d$ is the number of downward edges in $\gamma$. 
So the ratio of the signs of the colorings in \eqref{TraceAsSum} is equal to the ratio of the signs of terms in \eqref{PfH}.

We have shown that sign changes due to changing colorings (in the traces) and sign changes due to changing signatures
(in the Pfaffian) agree for $\sigma$ and $\sigma'$. This completes the proof.

\end{proof}

A small generalization of Theorem \ref{main} allows for the addition of edge weights.
Let $w:E\to\R_{>0}$ be a positive function on edges of $\G$, the \emph{edge weight} (with $w_{uv}=w_{vu}$).
For a multiweb $m\in\Omega_{2n}$ define its weight $w(m)$ to be the product of its edge weights: $w(m) = \prod_{E}w_e^{m_e}$.
Define the weighted matrix $H=H(\Phi,w)$ by
$$H_{uv}=\begin{cases}J\phi_{uv}w_{uv}&u\sim v\\{\bf 0}&\text{else}\end{cases}.$$
Then we have
\begin{thm}\label{mainwt} We have $\Pf \tilde H= \pm \sum_{m\in\Omega_{2n}}\Tr(m)w(m).$
\end{thm}

The proof is nearly identical to that of Theorem \ref{main}, with the addition of the lift of the edge weights to edges of $\tilde H$.

\section{The dimer model}\label{dimersection}

Let $\G$ be a planar graph and $w:E\to\R_{>0}$ a positive edge weight function.

A \emph{dimer cover of $\G$}, also known as \emph{perfect matching of $\G$}, is a set of edges covering all vertices exactly once.
Equivalently, it is a $1$-multiweb; the set of dimer covers is $\Omega_1$. 
The weight of a dimer cover $m\in\Omega_1$ is the product of its edge weights:
$w(m)=\prod_{e: m_e=1}w_e$.
Let $Z_d=Z_d(\G)$, the \emph{dimer partition function}, be the weighted sum of dimer covers, $Z_d=\sum_{m\in\Omega_1}w(m)$.

We define $\Phi_K$,
the \emph{Kasteleyn connection}, be the $\Sp(2n)$-connection (well-defined up to gauge equivalence)
in which each face has counterclockwise monodromy 
$J^{\ell-2}$, where $\ell$ is the length of the face. For example if $\G$ is a triangulation, then 
$\Phi_K$ has monodromy $J$ around every face. Generally for a loop $\gamma$, the monodromy will then
be $J^A$ where $A$ is the area enclosed, in terms of the number of triangles (quadrilateral faces count as area $2$,
pentagonal faces as area $3$, and so on).

The area $A(\gamma)$ of a loop $\gamma$ in $\G$ can be related to its length $L(\gamma)$ via the Euler characteristic:
\be\label{euler}A(\gamma)=L(\gamma)+2V_{enc}(\gamma)-2,\ee 
where $V_{enc}(\gamma)$ is the number of vertices enclosed by the loop $\gamma$.

\begin{thm}\label{Sp2Kthm}
Let $\G$ be a planar graph with an even number $N$ of vertices and $\Phi_{K}$ be the $\Sp(2)$ Kasteleyn connection. 
Let $K=H(\Phi_K)$ be the associated $H$ matrix with edge weights $w$.
Then $$|\Pf\tilde K|=(Z_d)^2.$$
\end{thm}

\begin{proof} By (\ref{Sp2trace}) the trace of a $2$-web $m$ is zero if any loop in $m$ has odd area (and thus odd length),
because $\tr(J^{A(\gamma)})=0$ if $A(\gamma)$ is odd.
If all loops in $m$ have even area, by (\ref{Sp2trace}) we have
\be\label{2webtraceK}\Tr(m) = (-1)^{c_1+N}\prod_\gamma -\tr(J^{A(\gamma)})=(-1)^{c_1+N}(-2)^{n_0}2^{n_2}\ee
where the product is taken over all nontrivial loops $\gamma$ (that is, not counting doubled edges) of $m$, $c_1$ is the number of doubled edges, $n_0$ is the number of loops of area $0\bmod 4$ and $n_2$ is the number of loops of area $2\bmod 4$.
If all loops are even then $V_{enc}(\gamma)$ in (\ref{euler}) is always even and $A(\gamma)=L(\gamma)-2\bmod 4$ by (\ref{euler}).

Now, consider a doubled edge to be a loop which has area $0$. Then taking the product over all loops in $m$, including doubled edges, and recalling that $N$ is even, the sign in (\ref{2webtraceK}) is

$$(-1)^{c_1+n_0} = \prod_{\gamma}(-1)^{A(\gamma)/2+1}=\prod_{\gamma}(-1)^{L(\gamma)/2}= (-1)^{N/2}.$$

The trace is thus
$\Tr(m) = (-1)^{N/2}2^{n_0+n_2}.$ Note that the sign is independent of $m$. Summing over all $m\in\Omega_2$ and applying Theorem \ref{main}
yields 
$$|\Pf\tilde K|=\sum_{m\in\Omega_2} 2^{n_0+n_2}.$$
Note however that the superposition of two dimers covers is a $2$-multweb with even-length loops, and the map from pairs of dimer covers to
such ``even $2$-multiwebs" has preimage exactly $2^\ell$ where $\ell=n_0+n_2$ is the number of nontrivial loops in the $2$-multiweb.
Thus 
$$(Z_d)^2 = \sum_{m\in\Omega_2} 2^{n_0+n_2} = |\Pf\tilde K|.$$
\end{proof}

We define $Z_{dd}=(Z_d)^2$; this is the partition function of the \emph{double dimer model}, see \cite{KenyonWilson}.

\subsection{Rotation matrices}

By perturbing the Kasteleyn
connection one can use it to study probabilistic properties of the double dimer model. One simple change one can make to 
the Kasteleyn connection is to tensor with another connection which commutes with it.

Rotation matrices in $\SL(2)$ are in $\Sp(2)$ and also commute with $J$. We can thus tensor the Kasteleyn connection 
$\Phi_K$ with an $S^1$-connection $\Theta$,
that is, an $\Sp(2)$-connection with connection matrices which are rotations 
$$R_\theta=\begin{pmatrix}\cos\theta&\sin\theta\\-\sin\theta&\cos\theta\end{pmatrix}.$$
The resulting Pfaffian of $\tilde K=\tilde K(\Phi_K\otimes\Theta)$ computes traces of $2$-multiwebs on $\G$ where each loop $\gamma$ now has weight $-\tr(J^{A}R_\theta)$, where $R_\theta$ is the monodromy of $\Theta$ around $\gamma$ and $A$ is the area enclosed by
$\gamma$ (still in terms of the number of triangles: a face of length $k$ has area $k-2$).

This gives trace 
\be\label{Rsigns} (2\cos\theta,-2\sin\theta,-2\cos\theta,2\sin\theta)
\ee respectively
for loops of area respectively $(0,1,2,3)\bmod 4$, where $R_\theta$ is the monodromy of $\Theta$ around $\gamma$.

\subsection{Double dimers on the annulus}

From a probability perspective, if we want to study the double dimer model using the $2$-multiweb model,
we want only configurations in which all loops have even length; those with odd-length loops should have weight $0$.
This is only possible if in (\ref{Rsigns}) we have $\theta=0$ or $\theta=\pi$: the monodromy of $R_\theta$ around odd-length
loops should be a multiple of $\pi$. 
The remaining sign in (\ref{Rsigns}) can then be absorbed into the total number of vertices as in Theorem \ref{Sp2Kthm}.

Suppose for example $\G$ is embedded in a planar annulus, with inner face $\gamma_0$ of length $\equiv 2\bmod 4$. 
Equip $\G$ with the $\Sp(2)$ Kasteleyn connection;
then Theorem \ref{Sp2Kthm} computes the partition function $Z_{dd}=(Z_d)^2$ of the double-dimer model on $\G$.
Now tensor the Kasteleyn connection with a flat $R_\pi=-I$ connection on the annulus, so that loops $\gamma$ running around the annulus
now have monodromy $-J^A$, where $A$ is the area between $\gamma$ and $\gamma_0$. 
A multiweb $m$ has nonzero trace only when all loops, including those winding around the annulus, are even. Let $m_0$ and $m_2$ be the number of loops winding around the annulus
of length $0,2\bmod 4$ respectively.
Let $n_0,n_2$ be the total number of loops of length $0,2\bmod 4$. 
Then as in the proof of Theorem \ref{Sp2Kthm} the trace is 
$$\Tr(m)=(-1)^{N/2}2^{n_0+n_2}(-1)^{m_0+m_2}.$$
Let $Z_{dd}(-1):=\Pf(\tilde K)$ be the associated partition function. The ratio 
$Z_{dd}(-1)/Z_{dd}$ can be written
$Z_{dd}(-1)/Z_{dd} = \E[(-1)^{n}]$ where $n$ is the number of loops (necessarily of even length) separating
the two boundaries.

For a double dimer configuration on a planar graph we define a \emph{spin} variable $\tau(f)\in\Z/2\Z$ on each face $f$,
by defining it to be $0$
on a fixed ``basepoint'' face $f_0$ and changing by $1\bmod 2$ when crossing any nontrivial loop. For the above annulus with external faces $f_1,f_2$ the spin correlation $\E[\tau(f_1)\tau(f_2)]$ is computed to be
$$\E[\tau(f_1)\tau(f_2)]=Z_{dd}(-1)/Z_{dd}.$$

A similar calculation allows us to compute multiple spin correlations of the type
$\E[\tau(f_1)\dots\tau(f_{2k})]$ for any $2k$ given faces on the planar graph $\G$ (assuming for simplicity that each face $f_i$ has length $2\bmod 4$): 
it suffices to compute the partition function for the Kasteleyn connection tensored 
with the $\pm1$ connection with monodromy $-1$ around each $f_i$
and $+1$ around all other faces.

\subsection{$\Sp(2n)$ and the Kasteleyn connection}\label{Kast2nsection}

Theorem \ref{Sp2Kthm} generalizes easily to $\Sp(2n)$.
\begin{thm}\label{Kastthm2n}
Let $\G$ be a planar graph and $\Phi_K$ be the $\Sp(2n)$-Kasteleyn connection. 
Let $K=H(\Phi_K)$ be the associated $H$ matrix with edge weights $w$.
Then $$|\Pf\tilde K|=(Z_d)^{2n}.$$
\end{thm}

\begin{proof}
The matrix $\tilde K$ is just a direct sum of $n$ copies of the $\Sp(2)$ $\tilde K$ matrix,
 so this follows from Theorem \ref{Sp2Kthm}.
\end{proof}

Given a $2n$-multiweb $m$, we can compute its trace for the Kasteleyn connection
as a sum of colorings of its half-edges: Proposition \ref{2mwtrace} applies since
the Kasteleyn connection decomposes into a direct sum of $\Sp(2)$ connections. A coloring of $m$ decomposes $m$ into a 
union of $n$ differently-colored $2$-multiwebs, $m_1,\dots,m_n$, one for each pair of complementary colors. Because we are using the Kasteleyn connection, each $m_i$ has loops of even area (and hence even length) only. We call such loops \emph{even}. The $\Sp(2)$-trace of $m_i$ is given by $(-1)^{N/2}2^{l_i}$ where $l_i$ is the number of loops of $m_i$. 
We have proved:

\begin{thm}
For a $2n$-web $m$ on a planar graph with the $\Sp(2n)$ Kasteleyn connection, the trace of $m$ is 
$$\Tr(m) = (-1)^{nN/2}\sum_c 2^{l(c)}w(m)$$
where the sum is over all decompositions $c$ of $m$ into colored even $2$-multiwebs, and $l$ is the total number of loops in the decomposition $c$.
\end{thm}

\begin{figure}[htbp]
\begin{center}\includegraphics[width=1.2in]{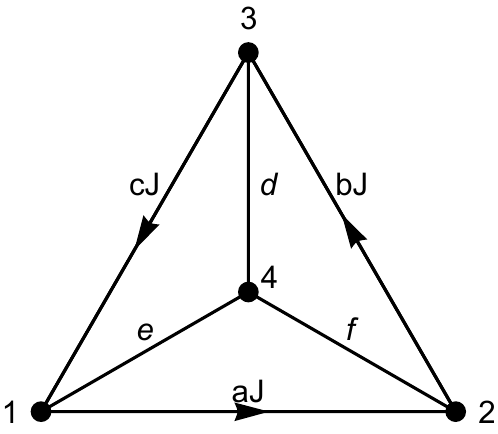}\hskip.3cm\includegraphics[width=1.2in]{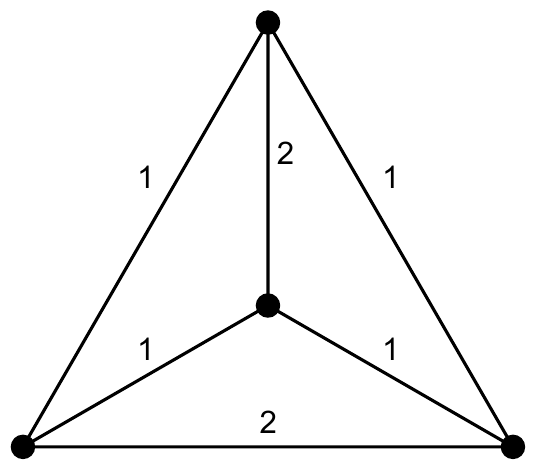}\hskip.3cm\includegraphics[width=1.2in]{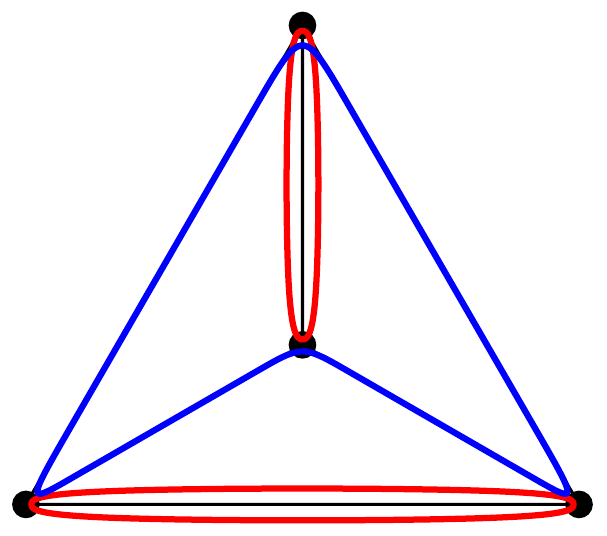}\hskip.3cm\includegraphics[width=1.2in]{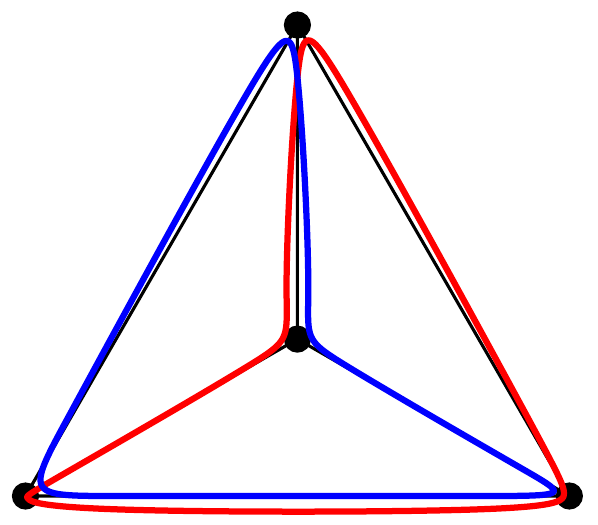}\end{center}
\caption{\label{2by3}For the graph $\G$ on the left with edge weights $a,...,f$ as shown, and Kasteleyn connection, we have 
$K$ given by (\ref{K}). The Pfaffian of $\tilde K$ is
$(ad+be+cf)^4$. The $\Sp(4)$ multiweb $m$ in the second panel can be decomposed into colored loops in four ways (2 of which are shown; the other two
are obtained by reversing colors). Counting $2^{\text{\# loops}}$ (not counting doubled edges), these configurations have weight $2$ and $4$
respectively, for a total trace of $\Tr(m) = 2(2+4)=12$. The coefficient of $a^2bcd^2ef$ in $\Pf\tilde K$ is indeed $12$.}
\end{figure}

As an example see Figure \ref{2by3}, with Kasteleyn matrix 
\be\label{K}K=\begin{pmatrix}0&-aI&cI&eJ\\aI&0&-bI&fJ\\-cI&bI&0&dJ\\eJ&fJ&dJ&0\end{pmatrix}.\ee

\subsection{$U(n)$ connections}

As in the $\Sp(2)$ case, we can perturb the Kasteleyn
connection by tensoring with another connection which commutes with it.

A matrix $A\in\Sp(2n,\R)$ which is both symplectic and commutes with $J$ is orthogonal:
$J=A^tJA=A^tAJ$ implies $A^tA=I$. Conversely the intersection of $\Sp(2n,\R)$ and $O(2n,\R)$
consists of the symplectic matrices which commute with $J$. This group is isomorphic to $U(n)$, the unitary group. 
The embedding $\rho$ of $U(n)$ in the symplectic group is as follows. Given $M\in U(n)$, we define
$$\rho(M) = \begin{pmatrix}\Re(M)&\Im(M)\\-\Im(M)&\Re(M)\end{pmatrix}.$$
By abuse of notation we refer to this subgroup as $U(n)$.

Consequently it is natural to consider a planar graph $\G$ with a $\Sp(2n)$-Kasteleyn connection tensored with a $U(n)$ connection.
Applying Theorem \ref{main} to the connection with perturbed traces gives information about the shape of a random $2n$-multiweb.
We just need to see how the trace of a $2n$-multiweb changes under the perturbation.
For $n>1$ this is complicated by the fact that the topology of a $2n$ web is typically itself complicated.

We leave this as a potentially interesting problem.

\begin{problem} Study the interacting $2n$-dimer model in the presence of a $U(n)$ connection.
\end{problem}

One result for $\Sp(4)$ is discussed in the next section.

\subsection{$4$-dimer model on an annulus}

Let us consider the simplest case in which
a graph $\G$ is embedded in a planar annulus, with inner face $\gamma_0$ of length $\equiv 2\bmod 4$. 
Equip $\G$ with the $\Sp(4)$ Kasteleyn connection.
Tensor this connection with a flat connection with monodromy $R\in U(2)$ around the annulus.

By Theorem \ref{Sp4ann}, a reduced $\Sp(4)$-web on $\G$ has components which are single loops or ``doubled loops". 
So the trace of a $4$-multiweb on $\G$ as a function of $R$ is a linear combination of products of 
the trace of a single loop and the trace of a doubled loop.

The general matrix $R\in U(2)$ has the form 
$$\begin{pmatrix}\cos(\theta)e^{i\alpha}&\sin(\theta)e^{i\beta}\\
-\sin(\theta)e^{-i\beta}e^{i\eps}&\cos(\theta)e^{-i\alpha}e^{i\eps}\end{pmatrix}$$
for real $\theta,\alpha,\beta,\eps$.

For this $R$, the trace of an odd single loop is $0$; a short calculation shows that the trace of an odd doubled-loop is
\be\label{odl}1-\cos(2\alpha-\eps)+2\cos(\eps)-2\cos^2(\alpha-\frac{\eps}2)\cos(2\theta).\ee
 
Likewise the trace of an even single loop is
\be\label{esl}2(\cos\alpha+\cos(\alpha-\eps))\cos\theta\ee
and the trace of an even doubled-loop is 
\be\label{edl}1+\cos(2\alpha-\eps)+2\cos(\eps)+2\cos^2(\alpha-\frac{\eps}2)\cos(2\theta).\ee

In order to have no odd loops we set the odd doubled-loop trace (\ref{odl}) to zero, by setting 
$$\cos2\theta=\frac{1-\cos(2\alpha-\eps)+2\cos(\eps)}{2\cos^2(\alpha-\frac{\eps}2)}.$$
Then the trace of a single even loop (\ref{esl}) becomes $\pm(2+2\cos\eps)$ and the trace of a doubled even loop (\ref{edl}) becomes $2+4\cos\eps$,
and in particular these are independent of $\beta,\alpha$. 
 
Therefore there is no loss in generality in setting $\alpha=0=\beta$, in which case $\theta=0$ or $\theta=\pi$ and
$R=\pm\begin{pmatrix}1&0\\0&e^{i\eps}\end{pmatrix}.$
Suppose $\theta=0$, so that the weight of a single loop is $2+2\cos\eps$.
This is $1$ plus half the weight of a doubled loop.

\begin{figure}[htbp]
\begin{center}\includegraphics[width=2.5in]{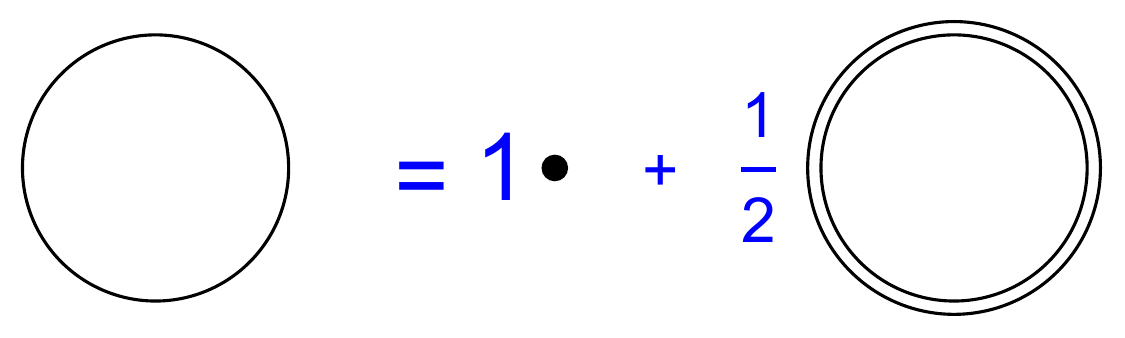}\end{center}
\caption{\label{supplementary}A ``supplementary" skein relation on an annulus.}
\end{figure}

In terms of enumeration, 
we do not have enough information to count both the number of single loops and the number of doubled loops, since their traces only depend on one parameter $\eps$.
If we add the supplementary skein relation of Figure 
\ref{supplementary}, however,
which replaces each single loop with a linear combination of the empty configuration
and $1/2$ of a doubled loop, 
then every configuration can be reduced to a nonnegative linear combination of 
doubled loops:
$$Z_4(\eps) = \sum_{k=0}^\infty C_k (2+4\cos\eps)^k$$
where $C_k$ is the sum of traces of configurations resolving to $k$ doubled loops.
By varying $\eps$ we can extract $C_k$ for each $k$.

\section{$\rm{Sp(4)}$ webs and representations}

\subsection{Multiwebs and webs}

A \emph{$2n$-web} is a $2n$-valent graph, embedded in a disk or on a surface, possibly with some boundary vertices of degree $1$ on the boundary of the disk or surface.

For a graph $\G$ embedded in an oriented surface, to each $2n$-multiweb $m$ is associated a $2n$-web $\hat m$,
with the same vertices as $m$ and such that an edge of $m$ of multiplicity $m_e$ is replaced with $m_e$ parallel edges connecting
its endpoints. The web $\hat m$ inherits a ribbon structure from that of $m$,
as well as cilia and edge orientations.

While multiwebs play a combinatorial role in this paper, webs themselves are representation-theoretic objects, defining invariants in tensor
products of representations, see \cite{Kuperberg}.
Skein relations on webs, which a priori arise from the underlying representation theory, give rise to analogous relations for multiwebs;
these are relations between traces of multiwebs.
We focus here primarily
on the case of $\Sp(4)$ webs, whose skein relations were studied in \cite{Kuperberg}.

\subsection{Definition of the $\rm{Sp(4)}$ web category}

\def\Skeina
{\begin{tric}
\draw (0.75,0) circle (0.75);
\end{tric}
}

\def\Skeinb
{\begin{tric}
\draw[thin, decorate, decoration={snake, segment length=0.49mm, amplitude=0.45mm},darkgreen] (0.75,0) circle (0.75);
\end{tric}
}

\def\Skeinebdouble
{\begin{tric}
\draw [scale=0.5]  (-1.732,2)--(0,1)--(1.732,2) (1.732,-2)--(0,-1)--(-1.732,-2); 
\draw [scale=0.5,thin, decorate, decoration={snake, segment length=0.5mm, amplitude=0.45mm},darkgreen](0,-1)--(0,1);
\end{tric}
}

\def\Skeinec
{\begin{tric}
\draw [scale=0.45] (-2,-2)..controls(-1,-1)and(-1,1)..(-2,2) (2,2)..controls(1,1)and(1,-1)..(2,-2);
\end{tric}
}

\def\DoubleCap
{\begin{tric}
\draw[thin, decorate, decoration={snake, segment length=0.5mm, amplitude=0.45mm},darkgreen] (0,0)..controls(0,1)and(1,1)..(1,0); 
\draw (0,0) node[black,below,scale=0.7]{$\ObjTwo$}
      (1,0) node[black,below,scale=0.7]{$\ObjTwo$}; 
\end{tric}
}

\def\DoubleCapNL
{\begin{tric}
\draw[thin, decorate, decoration={snake, segment length=0.5mm, amplitude=0.45mm},darkgreen] (0,0)..controls(0,1)and(1,1)..(1,0); 
\end{tric}
}

\def\SingleCap
{\begin{tric}
\draw (0,0)..controls(0,1)and(1,1)..(1,0); 
\draw (0,0) node[black,below,scale=0.7]{$\ObjOne$}
      (1,0) node[black,below,scale=0.7]{$\ObjOne$};
\end{tric}
}

\def\SingleCapNL
{\begin{tric}
\draw (0,0)..controls(0,1)and(1,1)..(1,0); 
\end{tric}
}

\def\SingleCup
{\begin{tric}
\draw (0,0)..controls(0,-1)and(1,-1)..(1,0); 
\draw (0,0) node[black,above,scale=0.7]{$\ObjOne$}
      (1,0) node[black,above,scale=0.7]{$\ObjOne$};
\end{tric}
}

\def\SingleCupNL
{\begin{tric}
\draw (0,0)..controls(0,-1)and(1,-1)..(1,0); 
\end{tric}
}

\def\DoubleCup
{\begin{tric}
\draw[thin, decorate, decoration={snake, segment length=0.5mm, amplitude=0.45mm},darkgreen] (0,0)..controls(0,-1)and(1,-1)..(1,0);
\draw (0,0) node[black,above,scale=0.7]{$\ObjTwo$}
      (1,0) node[black,above,scale=0.7]{$\ObjTwo$};
\end{tric}
}

\def\DoubleCupNL
{\begin{tric}
\draw[thin, decorate, decoration={snake, segment length=0.5mm, amplitude=0.45mm},darkgreen] (0,0)..controls(0,-1)and(1,-1)..(1,0);
\end{tric}
}

\def\Vertb
{\begin{tric}
\draw[scale=0.8,thin, decorate, decoration={snake, segment length=0.5mm, amplitude=0.45mm},darkgreen] (0,0)--(90:1) ;
\draw[scale=0.8] (0,0)--(225:1) (0,0)--(315:1);
\draw (90:1)node[black,anchor=south,scale=0.7]{$\ObjTwo$}
      (225:1)node[black,anchor=north,scale=0.7]{$\ObjOne$}
      (315:1)node[black,anchor=north,scale=0.7]{$\ObjOne$};
\end{tric}
}

\def\ZigZagVertb
{\begin{tric}
\draw[scale=0.55,thin, decorate, decoration={snake, segment length=0.5mm, amplitude=0.45mm},darkgreen] (2,-2)--(2,0)..controls(2,1.5)and(1.5,2).. (1,2)..controls(0.5,2)and(0,1.5)..(0,1)--(0,0);
\draw[scale=0.55] (0,0)..controls(0.3,-0.3)and(0.5,-0.7)..(0.5,-1)
                  ..controls(0.5,-1.5)and(-0.3,-2)..(-1,-2)
                  ..controls(-1.7,-2)and(-2.5,-1.5)..(-2.5,0)--(-2.5,2)
                  
                (0,0)--(-0.5,-0.5)..controls(-0.7,-0.7)and(-0.8,-0.8)..(-1,-0.8)..
                                  controls(-1.5,-0.8)and(-1.8,0)..(-1.8,1)--(-1.8,2);
\end{tric}
}

\def\ZigZagVerta
{\begin{tric}
\draw[scale=0.8,thin, decorate, decoration={snake, segment length=0.5mm, amplitude=0.45mm},darkgreen] (0,0)--(0,-2) ;
\draw[scale=0.55] (0,0)--(0.3,0.3)..controls(0.5,0.5)and(0.7,1)..(0.7,2) 
(0,0)--(-0.3,0.3)..controls(-0.5,0.5)and(-0.7,1)..(-0.7,2) ;
\end{tric}
}

\def\ZigZagVert
{\begin{tric}
\draw[scale=0.55,thin, decorate, decoration={snake, segment length=0.5mm, amplitude=0.45mm},darkgreen]
(-2,-2)--(-2,0)..controls(-2,1.5)and(-1.5,2).. (-1,2)..controls(-0.5,2)and(0,1.5)..(0,1)--(0,0);

\draw[scale=0.55] (0,0)..controls(-0.3,-0.3)and(-0.5,-0.7)..(-0.5,-1)
                  ..controls(-0.5,-1.5)and(0.3,-2)..(1,-2)
                  ..controls(1.7,-2)and(2.5,-1.5)..(2.5,0)--(2.5,2)
                  
                (0,0)--(0.5,-0.5)..controls(0.7,-0.7)and(0.8,-0.8)..(1,-0.8)..
                                  controls(1.5,-0.8)and(1.8,0)..(1.8,1)--(1.8,2);
\end{tric}
}

\def\SingleZigZag
{\begin{tric}
\draw [scale=0.7](0,0)..controls(0,1)and(1,1)..(1,0);
\draw [scale=0.7](1,0)..controls(1,-1)and(2,-1)..(2,0);
\draw [scale=0.7](0,0)--(0,-2) (2,0)--(2,2);
\end{tric}
}

\def\SingleZigZaga
{\begin{tric}
\draw [scale=0.7](0,-2)--(0,2);
\end{tric}}

\def\SingleZigZagb
{\begin{tric}
\draw [scale=0.7](0,0)..controls(0,1)and(-1,1)..(-1,0);
\draw [scale=0.7](-1,0)..controls(-1,-1)and(-2,-1)..(-2,0);
\draw [scale=0.7](0,0)--(0,-2) (-2,0)--(-2,2);
\end{tric}
}

\def\DoubleZigZag
{\begin{tric}
\draw [scale=0.7,thin, decorate, decoration={snake, segment length=0.5mm, amplitude=0.45mm},darkgreen](0,-2)--(0,0)..controls(0,1)and(1,1)..(1,0)..controls(1,-1)and(2,-1)..(2,0)--(2,2);
\end{tric}
}

\def\DoubleZigZaga
{\begin{tric}
\draw [scale=0.7,thin, decorate, decoration={snake, segment length=0.5mm, amplitude=0.45mm},darkgreen](0,-2)--(0,2);
\end{tric}}

\def\DoubleZigZagb
{\begin{tric}
\draw [scale=0.7,thin, decorate, decoration={snake, segment length=0.5mm, amplitude=0.45mm},darkgreen](0,-2)--(0,0)..controls(0,1)and(-1,1)..(-1,0)..controls(-1,-1)and(-2,-1)..(-2,0)--(-2,2);
\end{tric}
}

\def\Skeini{
\begin{tric}
\draw [thin, decorate, decoration={snake, segment length=0.5mm, amplitude=0.45mm},darkgreen](0,0.7)--(0,1.5) (0,-1.5)--(0,-0.7);
\draw (0,0.7)..controls(-0.5,0.7)and(-0.5,-0.7)..(0,-0.7)  
      (0,0.7)..controls(0.5,0.7)and(0.5,-0.7)..(0,-0.7);
\end{tric}
}

\def\Skeinia{
\begin{tric}
\draw[thin, decorate, decoration={snake, segment length=0.5mm, amplitude=0.45mm},darkgreen] (0,1.5)--(0,-1.5);
\end{tric}
}

\def\Skeinj{
\begin{tric}
\draw[thin, decorate, decoration={snake, segment length=0.5mm, amplitude=0.45mm},darkgreen](0,-1.5)--(0,0);
\draw (0,0)..controls(0.7,0.5)and(0.7,1.5)..(0,1.5)..controls(-0.7,1.5)and(-0.7,0.5)..(0,0);
\end{tric}
}

\def\Skeinldouble
{\begin{tric}
\draw [scale=0.65] (-1,0)--(1,0)--(0,1.732)--cycle;
\draw[scale=0.65,thin, decorate, decoration={snake, segment length=0.5mm, amplitude=0.45mm},darkgreen](-2,-0.577)--(-1,0)(2,-0.577)--(1,0)  (0,2.982)--(0,1.732);
\end{tric}
}

The connection between graphs and the representations of a Lie group is usefully described using the language of category theory.

A category is a collection of objects and morphisms, where a morphism relates two objects called the source and the target of the morphism. In a graphical category, a morphism is given by a formal linear combination of graphs with the same boundary points, and an object is given by a tuple of boundary points. We read a graph vertically as a morphism, which means that the tuple of boundary points beneath the graph is the source, and the tuple of boundary points above the graph is the target. Composition of morphisms corresponds to vertical stacking of graphs.

If a category $\mathbf{C}$ is equipped with a tensor: 
$\bigotimes: \mathbf{C} \times \mathbf{C} \rightarrow \mathbf{C}$, such that $\bigotimes$ is associative and for which 
there is an identity object, then we call  $\mathbf{C}$ a monoidal category or a tensor category. A graphical category can be equipped with a tensor simply by putting graphs side by side. To be more precise, the tensor product of objects is horizontal concatenation of tuples of boundary points of graphs. The tensor product of morphisms is horizontal concatenation of graphs. The identity object is given by the empty diagram, which can be seen as a tuple of no boundary points.

\def\compoexamplea
{\begin{tric}
\draw [scale=0.6] (0,0)--(0,3);
\draw[scale=0.6] (1.5,1.5)--(1.5,3);
\draw [scale=0.6](1,0)..controls(1,0.75)..(1.5,1.5) (2,0)..controls(2,0.75)..(1.5,1.5);
\end{tric}
}

\def\compoexampleb
{\begin{tric}
\draw[scale=0.6](2.5,0)--(2.5,-1.5);
\draw[scale=0.6](0.5,-1)--(0.5,-2);
\draw[scale=0.6](1,0)..controls(1,-0.5)..(0.5,-1) (0,0)..controls(0,-0.5)..(0.5,-1);
\draw[scale=0.6](1,-3)..controls(1,-2.5)..(0.5,-2) (0,-3)..controls(0,-2.5)..(0.5,-2);
\draw[scale=0.6](2.5,-1.5)..controls(2,-2.25)..(2,-3)
                (2.5,-1.5)..controls(3,-2.25)..(3,-3);
\end{tric}
}

\def\compoexamplec
{\begin{tric}
\draw [scale=0.6] (0,0)--(0,3);
\draw[scale=0.6] (1.5,1.5)--(1.5,3);
\draw [scale=0.6](1,0)..controls(1,0.75)..(1.5,1.5) (2,0)..controls(2,0.75)..(1.5,1.5);
\draw[scale=0.6](5.5,3)--(5.5,1.5);
\draw[scale=0.6](3.5,2)--(3.5,1);
\draw[scale=0.6](4,3)..controls(4,2.5)..(3.5,2) (3,3)..controls(3,2.5)..(3.5,2);
\draw[scale=0.6](4,0)..controls(4,0.5)..(3.5,1) (3,0)..controls(3,0.5)..(3.5,1);
\draw[scale=0.6](5.5,1.5)..controls(5,0.75)..(5,0)
                (5.5,1.5)..controls(6,0.75)..(6,0);
\end{tric}
}

\def\compoexampled
{\begin{tric}
\draw [scale=0.7] (0,2)--(0,4);
\draw[ scale=0.7] (1.5,3)--(1.5,4);
\draw [scale=0.7](1,2)..controls(1,2.5)..(1.5,3) (2,2)..controls(2,2.5)..(1.5,3);
\draw[scale=0.7](2,0)--(2,2);
\draw[scale=0.7](0.5,1)--(0.5,0);
\draw[scale=0.7](1,2)..controls(1,1.5)..(0.5,1) (0,2)..controls(0,1.5)..(0.5,1);
\end{tric}
}

\def\compoexampled
{\begin{tric}
\draw [scale=0.6] (0,0)--(0,3);
\draw[scale=0.6] (1.75,1.5)--(1.75,3);
\draw [scale=0.6](1,0)..controls(1,0.75)..(1.75,1.5) (2.5,0)..controls(2.5,0.75)..(1.75,1.5);

\draw[scale=0.6](2.5,0)--(2.5,-1.5);
\draw[scale=0.6](0.5,-1)--(0.5,-2);
\draw[scale=0.6](1,0)..controls(1,-0.5)..(0.5,-1) (0,0)..controls(0,-0.5)..(0.5,-1);
\draw[scale=0.6](1,-3)..controls(1,-2.5)..(0.5,-2) (0,-3)..controls(0,-2.5)..(0.5,-2);
\draw[scale=0.6](2.5,-1.5)..controls(2,-2.25)..(2,-3)
                (2.5,-1.5)..controls(3,-2.25)..(3,-3);
\end{tric}
}

For example, consider $f=\ \compoexamplea $ \ and \
 $g= \ \compoexampleb \ $. Then $f$ is a morphism from a tuple of $3$ points to a tuple of $2$ points, and $g$ is a morphism from a tuple of $4$ points to a tuple of $3$ points.  We have 
 
 $$f\otimes g =\ \compoexamplec   \quad 
 \text{and} \quad f\circ g =\ \compoexampled  \ .$$\\

When defining a graphical category, we allow different types of edges in a graph, and therefore allow different types of boundary points of the graph.

\begin{defn} [{\cite[Section 4]{Kuperberg}}]  \label{C2Spider} 
 The $\rm{Sp(4)}$ web category, denoted by $\CDDC$, is a monoidal category, whose objects are generated by objects \{$\ObjOne$, $\ObjTwo$\}, and whose morphisms are generated by the following graphs:  
 $$
 \SingleCap\!\!\in {\Hom}(\ObjOne\otimes\ObjOne,\mathbb{R}),
 \SingleCup\!\!\in {\Hom}(\mathbb{R},\ObjOne\otimes\ObjOne),
\DoubleCap\!\! \in {\Hom} (\ObjTwo\otimes\ObjTwo,\mathbb{R}),$$
$$
\DoubleCup\!\!\in {\Hom}(\mathbb{R},\ObjTwo\otimes\ObjTwo), 
 \Vertb \in {\Hom}(\ObjOne\otimes\ObjOne,\ObjTwo), 
 $$
 modulo the tensor-ideal generated by the following skein relations.

\begin{align}
&\SingleZigZag \quad = \quad \SingleZigZaga \quad = \quad \SingleZigZagb   
 \quad  \qquad  \qquad 
\DoubleZigZag\quad = \quad \DoubleZigZaga \quad = \quad \DoubleZigZagb \notag \\
&\quad  \quad  \qquad \qquad 
\ZigZagVerta  \quad := \quad  \ZigZagVert \quad = \quad \ZigZagVertb \notag
\end{align}

\begin{align}
& \Skeina =-4 \ \ \ \ \ \ \ \ \ \ \ \ \ \ \ \ \ \ \ \ \ \  \ \ \ \ \ \ \ 
\Skeinb = 5 \ \ \  \notag \\
&\Skeinj=0\ \ \ \ \ \ \ \ \ \ \ \ \ \ \ \ \ \ \ \ \ \ \ 
\Skeini= \ -2 \ \ \ \Skeinia  \ \ \ \ \ \ \ \ \ \ \ \ \ \ \ \ \ \ \ \ \ \ \  \Skeinldouble=0  \notag \\
&\qquad \qquad \Skeineadouble\ \   - \  \frac{1}{2} \ \Skeinec \ \    =  \  \Skeinebdouble \ - \  
\frac{1}{2} \ \Skeined  \label{C2IH}
\end{align}
\end{defn}

\begin{remark}
    $\SingleCapNL$ and $\DoubleCapNL$ are known as evaluation maps.
    $\SingleCupNL$ and $\DoubleCupNL$ are known as coevaluation maps. 
\end{remark}

\old{
A morphism in the $\Sp(4)$ web category is called an $\Sp(4)$-web. Note that this is different from our notion of $4$-webs above. 
In particular an $\Sp(4)$ web is in general a linear combination of graphs, each of which may have inputs and outputs. However, these notions of $\Sp(4)$-web and $4$-webs are related by Definition \ref{}.

  For $P_k\in \{ \ObjOne, \ObjTwo \}, 1 \le k \le n,$ the invariant space is defined as 
    $$\rm{Inv}_{\CDDC} (P_1 \otimes P_2 \otimes \cdot \cdot \cdot \otimes P_n)
   :=\rm{Hom}_{\CDDC} (P_1 \otimes P_2 \otimes \cdot \cdot \cdot \otimes P_n,\mathbb{R}). $$
    Correspondingly, for $V_k\in \{ V_{(1,0)}, V_{(0,1)}  \}, 1 \le k \le n,$  $$\rm{Inv}_{\Fund(\rm{Sp(4)})}(V_1 \otimes V_2 \otimes \cdot \cdot \cdot \otimes V_n)
   :=\rm{Hom}_{\Fund(\rm{Sp(4)})}(V_1 \otimes V_2 \otimes \cdot \cdot \cdot \otimes V_n,\mathbb{R}). $$
   Note that $\rm{Hom}(A, B) \cong \rm{Inv}(A^* \otimes B)$.
   }

\begin{definition} [{\cite[Section 4]{Kuperberg}}]  \label{Tetravalent}
  By Equation \eqref{C2IH},  we can define a tetravalent vertex:
\begin{align}
   \tetravalent := \Skeineadouble\ \   - \  \frac{1}{2} \Skeinec \ \    =  \  \Skeinebdouble \ - \  \frac{1}{2}\Skeined  \label{Tetraa}
\end{align}
\end{definition}

\begin{remark}In the $Sp(4)$ web category, the diagrams are read vertically and the tetravalent vertex is well-defined without reference to cilia, because there is already a natural order of inputs and outputs. When we study $4$-multiwebs or $4$-webs on a planar graph, however, 
cilia are introduced to indicate how diagrams are read locally around each vertex. See Section \ref{rotationsection} below.
\end{remark}

\def\ReidemAa
{\begin{tric}
\draw (0.1,-1)..controls (-0.1,0.3) and (-0.2,0.5) .. (-0.4,0.5) 
       (-0.4,0.5)..controls(-0.5,0.5)and(-0.7,0.3)..(-0.7,0);
\draw (0.1,1)..controls  (-0.1,-0.3)and (-0.2,-0.5) .. (-0.4,-0.5) 
       (-0.4,-0.5)..controls(-0.5,-0.5)and(-0.7,-0.3)..(-0.7,0);
\filldraw [black] (-0.09,0) circle (2pt);
\end{tric}
}

\def\ReidemAb
{\begin{tric}
\draw (0,-1)--(0,1);
\end{tric}
}

\def\ReidemBa
{\begin{tric}
\draw (0,-1)..controls(1,-0.3) and (1,0.3) .. (0,1)
      (1,-1)..controls(0,-0.3) and (0,0.3) .. (1,1);
\filldraw [black] (0.5,0.56) circle (2pt);
\filldraw [black] (0.5,-0.56) circle (2pt);
\end{tric}
}

\def\ReidemBaa
{\begin{tric}
\draw (0,-1)--(1,1)   (1,-1)--(0,1);
\filldraw [black] (0.5,0) circle (2pt);
\end{tric}
}

\def\ReidemBb
{\begin{tric}
\draw (0,-1)-- (0,1)
      (1,-1)-- (1,1);
\end{tric}
}

\def\ReidemCa
{\begin{tric}
\draw (70:1)--(230:1) (110:1)--(310:1) (190:1)--(350:1);
\filldraw [black] (0,0.35) circle (2pt);
\filldraw [black] (-0.31,-0.175) circle (2pt);
\filldraw [black] (0.31,-0.175) circle (2pt);
\end{tric}
}

\def\ReidemCc
{\begin{tric}
\draw(70:1)..controls(80:0.2)and(100:0.2)..(110:1) 
     (230:1)--(350:1) 
     (310:1)--(190:1);
\filldraw [black] (0,-0.535) circle (2pt);
\end{tric}
}

\def\ReidemCd
{\begin{tric}
\draw(190:1)..controls(200:0.2)and(220:0.2)..(230:1) 
      (70:1)--(310:1)   (110:1)--(350:1);
\filldraw [black] (0.46,0.28) circle (2pt);
\end{tric}
}

\def\ReidemCe
{\begin{tric}
\draw (310:1)..controls(320:0.2)and(340:0.2)..(350:1)
      (70:1)--(190:1)    (110:1)--(230:1);
\filldraw [black] (-0.46,0.28) circle (2pt);
\end{tric}
}

\def\ReidemCf
{\begin{tric}
\draw (70:1)..controls(80:0.2)and(100:0.2)..(110:1) 
(190:1)..controls(200:0.2)and(220:0.2)..(230:1) 
(310:1)..controls(320:0.2)and(340:0.2)..(350:1);
\end{tric}
}

\def\ReidemCb
{\begin{tric}
\draw[scale=0.75] (-1,0.866)--(0.5,-1.732) (1,0.866)--(-0.5,-1.732) (-1.5,0)--(1.5,0);
\end{tric}
}

\def\ZigZagTetra
{\begin{tric}
\draw [scale=0.7]  (0,0)--(1,1) (0,0)--(-0.5,0.5) (0,0)--(0.5,-0.5) (0,0)--(-1,-1)(-0.5,0.5)..controls(-0.7,0.7)and(-1,1)..(-1.5,1)
..controls(-2.5,1)and(-2.5,-0.5)..(-2.5,-1)
(0.5,-0.5)..controls(0.7,-0.7)and(1,-1)..(1.5,-1)
..controls(2.5,-1)and(2.5,0.5)..(2.5,1);
\filldraw [black] (0,0) circle (2pt);
\end{tric}
}

\def\ZigZagTetraa
{\begin{tric}
\draw   [scale=0.7] (0,0)--(1,1) (0,0)--(-1,1) (0,0)--(1,-1) (0,0)--(-1,-1);
\filldraw [black] (0,0) circle (2pt);
\end{tric}
}

\begin{proposition} [{\cite[Section 4]{Kuperberg}}]  \label{TetravalentSkein}
We have the following relations for the tetravalent vertex.
\begin{align}
 \ZigZagTetra&=\ZigZagTetraa,\label{pivotTetra} \\ 
\ReidemAa&= 2 \quad \ReidemAb  \quad , \label{TetraSkeinA} \\ 
\ReidemBa &= -2 \quad \ReidemBaa \quad -2 \quad \ReidemBb \quad , \label{TetraSkeinB} \\ 
\ReidemCa &=  \ReidemCc  +  \ReidemCd  +  \ReidemCe  +  2 \  \ReidemCf \quad . \label{TetraSkeinC} 
\end{align}
\end{proposition}

\begin{proof}
Apply Equation \eqref{Tetraa} to rewrite the webs with tetravalent vertices as linear combinations of webs with double edges, then apply skein relations in Definition \ref{C2Spider}. 
\end{proof}

\def\HomtoInvExp
{\begin{tric}
\draw  (0,0)--(0,2) (0.5,0)--(0.5,2) (1.5,0)--(1.5,2);
\fill[white] (0.75,1) ellipse (1.2 and 0.3);
\draw[thick,darkred](0.75,1) ellipse (1.2 and 0.3);

\filldraw [black] (0.8,1.7) circle (0.6pt) (1,1.7) circle (0.6pt) (1.2,1.7) circle (0.6pt) (0.8,0.3) circle (0.6pt) (1,0.3) circle (0.6pt) (1.2,0.3) circle (0.6pt);

\draw [very thick, decorate, decoration = {calligraphic brace}] 
(0,2.3) -- (1.5,2.3);
\draw [very thick, decorate, decoration = {calligraphic brace}] 
(1.5,-0.3) -- (0,-0.3);

\draw (0.75,1)node[scale=0.7,black]{$\mathcal{W}$}
      (0.75,2.4)node[scale=0.7,black,above]{$m$} 
      (0.75,-0.4)node[scale=0.7,black,below]{$n$}; 
      
\draw (-1.7,0)--(-1.7,2) (-2.2,0)--(-2.2,2) (-0.7,0)--(-0.7,2);   
\filldraw [black] (-1,1) circle (0.6pt) (-1.2,1) circle (0.6pt) (-1.4,1) circle (0.6pt); 
\draw [very thick, decorate, decoration = {calligraphic brace}] 
(-0.7,-0.3) -- (-2.2,-0.3) ;
\draw  (-1.45,-0.4)node[scale=0.7,black,below]{$m$} ; 
\end{tric}
}

\def\HomtoInvA
{\begin{tric}
\draw  (0,0)--(0,1.5)..controls(0,2)and(-0.7,2.3)..(-0.7,0)
(0.5,0)--(0.5,1.5)..controls(0.5,2.5)and(-1.2,3)..(-1.2,0)
(1.5,0)--(1.5,1.5)..controls(1.5,2)and(1,3)..(0,3)..controls(-1.5,3)and(-2.2,2)..(-2.2,0);
\fill[white] (0.75,1) ellipse (1.2 and 0.3);
\draw[thick,darkred](0.75,1) ellipse (1.2 and 0.3);
\filldraw [black] (0.7,1.7) circle (0.6pt) (0.9,1.7) circle (0.6pt) (1.1,1.7) circle (0.6pt)
(0.8,0.3) circle (0.6pt) (1,0.3) circle (0.6pt) (1.2,0.3) circle (0.6pt)
(-1.8,0.7) circle (0.6pt) (-1.6,0.7) circle (0.6pt) (-1.4,0.7) circle (0.6pt);
\draw [very thick, decorate, decoration = {calligraphic brace}] 
(1.5,-0.3) -- (-2.2,-0.3);
\draw (0.75,1)node[scale=0.7,black]{$\mathcal{W}$}
      (-0.3,-0.4)node[scale=0.7,black,below]{$m+n$}; 
\end{tric}
}

\def\HomtoInvC
{\begin{tric}
\draw  (0.2,0)--(0.2,1) (0.5,0)--(0.5,1) (1.5,0)--(1.5,1)
(0,1)--(0,1.5)..controls(0,1.8)and(-0.65,2.3)..(-0.65,0)
(0.6,1)--(0.6,1.5)..controls(0.6,2.5)and(-1.2,3)..(-1.2,0)
(0.9,1)--(0.9,1.5)..controls(0.9,2.1)and(0.5,2.5)..(-0.1,2.5)..controls(-1.2,2.5)and(-1.5,1.5)..(-1.5,0)
(1.2,1)--(1.2,1.5)..controls(1.2,2.5)and(3,3)..(3,0)
(1.75,1)--(1.75,1.5)..controls(1.75,1.8)and(2.4,2.3)..(2.4,0) ;

\fill[white] (0.84,1) ellipse (1.2 and 0.3);
\draw[thick,darkred](0.84,1) ellipse (1.2 and 0.3);

\filldraw [black] (0.15,1.5) circle (0.6pt) (0.275,1.5) circle (0.6pt) (0.4,1.5) circle (0.6pt)
(-1.05,0.3) circle (0.6pt) (-0.925,0.3) circle (0.6pt) (-0.8,0.3) circle (0.6pt)
(0.8,0.3) circle (0.6pt) (1,0.3) circle (0.6pt) (1.2,0.3) circle (0.6pt)
(1.35,1.5) circle (0.6pt) (1.475,1.5) circle (0.6pt) (1.6,1.5) circle (0.6pt)
(2.55,0.3) circle (0.6pt) (2.675,0.3) circle (0.6pt) (2.8,0.3) circle (0.6pt)
;

\draw [very thick, decorate, decoration = {calligraphic brace}] 
(1.5,-0.22) -- (0.2,-0.22);
\draw [very thick, decorate, decoration = {calligraphic brace}] (-0.65,-0.2) -- (-1.5,-0.2) ;
\draw [very thick, decorate, decoration = {calligraphic brace}] 
(3,-0.2) -- (2.4,-0.2);

\draw (0.75,1)node[scale=0.7,black]{$\mathcal{W}$}
      (0.85,-0.35)node[scale=0.7,black,below]{$n$}
       (-1,-0.3)node[scale=0.7,black,below]{$m-k$}
        (2.7,-0.3)node[scale=0.7,black,below]{$k$}; 
\end{tric}
}

A morphism in the $\Sp(4)$ web category is a linear combination of trivalent graphs. These graphs are called $\Sp(4)$-webs. Note that this is different from our notion of $4$-webs above. However, these notions of $\Sp(4)$-web and $4$-webs are related by Definition \ref{Tetravalent}.

To be more precise, taking any $\Sp(4)$-web $\mathcal{I} \in {\Hom}(\ObjOne^{\otimes n} ,\mathbb{R})$, using Definition \ref{Tetravalent} we can write $\mathcal{I}$ as a linear combination of tetravalent graphs, each of which is a $4$-web.

\begin{definition}
A $4$-web on a surface is said to be \emph{reduced} if it has no contractible faces of degree $0,1,2$ or $3$. 
\end{definition}
Note that from Proposition \ref{TetravalentSkein}, a web with a faces of degree $<4$ can be written as a linear combination of webs with fewer faces. This gives us a confluent reduction procedure for $4$-webs.

\subsection{Rotation of webs}\label{rotationsection}

For any web $\mathcal{W} \in {\Hom}(\ObjOne^{\otimes n} ,\ObjOne^{\otimes m})$, 
we can associate to  $\mathcal{W}$ a web $\mathcal{W}_0
\in {\Hom}(\ObjOne^{\otimes (m+n)}, \mathbb{R})$  by the following.
First we have 
$${(\text{id}_{\ObjOne})}^{\otimes m} \otimes  \mathcal{W} = \HomtoInvExp \in {\Hom}(\ObjOne^{\otimes (m+n)} ,\ObjOne^{\otimes 2m}).$$ Composing  ${(\text{id}_{\ObjOne})}^{\otimes m} \otimes  \mathcal{W}$ with evaluation maps, 
we have $$\mathcal{W}_0 =  \HomtoInvA \in {\Hom}(\ObjOne^{\otimes (m+n)}, \mathbb{R}).$$

Furthermore, one can associate to $\mathcal{W}$ different webs  
$$\mathcal{W}_k = \HomtoInvC \in {\Hom}(\ObjOne^{\otimes (m+n)}, \mathbb{R}),\qquad\qquad k=0,1,..., m.$$ 
These $\mathcal{W}_k$ can be identified with each other by rotation of the diagrams. 
These rotations $\mathcal{W}_k$ of web $\mathcal{W}$ 
correspond to different choices of cilia, since they rotate the inputs.
Any $\mathcal{W}_k$ can be indicated by a web in a disk with a choice of cilium.
  \def\TetraCilia
{\begin{tric}
\draw (-1,1)--(1,-1);
\draw  (-1,-1)--(1,1);
\filldraw [black] (0,0) circle (3pt);
\draw[black, very thick, dash pattern=on 3pt off 2pt] (0,0) circle (1.42); 
\draw [scale=0.7](0,0) node[ultra thick, right, black, rotate=90]{$\sim$}; 
\end{tric}}
\def\TetraCiliaa
{\begin{tric}
\draw [scale=0.7] 
      (0,0)..controls(1.2,-0.6)and(1.5,-1.4)..(1.5,-2)
      (0,0)..controls(0.3,-0.6)and(0.5,-1.4)..(0.5,-2)
      (0,0)..controls(-0.3,-0.6)and(-0.5,-1.4)..(-0.5,-2)
      (0,0)..controls(-1.2,-0.6)and(-1.5,-1.4)..(-1.5,-2);
\filldraw [black] (0,-0.02) circle (3pt);
\end{tric}
}
For example, $$ \TetraCilia= \TetraCiliaa .$$

\subsection{The representation category of $\Sp(4)$}

Now we recall some basic representation theory of $\Sp(4)$:
 
The defining $\Sp(4)$ representation $V$ is $4$ dimensional, with basis vectors:
$$v_{(1,0)}=e_1 ,\qquad v_{(-1,1)}=e_2 , \qquad v_{(1,-1)}=e_4,  \quad \text{and} \quad v_{(-1,0)}=-e_3, $$ 
where the subscripts of the vectors are weights in the representation. The defining representation $V$ is the first fundamental representation of $\Sp(4)$, also denoted $V_{(1,0)}$.

We have $V \wedge V \cong V_{(0,1)} \oplus \mathbb{R}$, where $ V_{(0,1)}$ is the second fundamental representation of $\Sp(4)$ and has dimension $5$. We can write down the basis vectors of $ V_{(0,1)}$ as vectors in  $V \wedge V$: 
$$v_{(0,1)}=e_1 \wedge e_2 , \qquad  v_{(2,-1)}=e_1 \wedge e_4 , \qquad v_{(0,0)}=e_2 \wedge e_4 - e_1 \wedge e_3, $$
$$ v_{(-2,1)}= -e_2 \wedge e_3 , \quad \text{and} \quad v_{(0,-1)}= - e_4 \wedge e_3. $$

\old{

\begin{defn}
    The $\Sp(4)$ representation category, denoted by $\Rep(\Sp(4))$, is a category whose objects are finite dimensional representations of $\Sp(4)$, and whose morphisms are $\Sp(4)$-equivariant maps between $\Sp(4)$ representations. 
\end{defn}

A subcategory of a category $\mathbf{C}$ is a category $\mathbf{S}$ whose objects are objects in $\mathbf{C}$ and whose morphisms are morphisms in $\mathbf{C}$ with the same identities and composition of morphisms. We say that $\mathbf{S}$ is a full subcategory of $\mathbf{C}$ if for each pair of objects $X$ and $Y$ of $\mathbf{S}$, $ \mathrm{Hom}_{\mathbf{S}}(X,Y)= \mathrm{Hom}_{\mathbf{C}}(X,Y)$.

\begin{defn}
    The category of $\Sp(4)$ fundamental representations, denoted by $\Fund(\rm{Sp(4)})$, is the full monoidal subcategory of the 
    representation category $\Rep(\rm{Sp(4)})$ generated by the fundamental representations $V_{(1,0)}$ and $ V_{(0,1)}$. 
\end{defn}

}

\begin{defn}
    The $\Sp(4)$ fundamental representation category, denoted by $\Fund(\rm{Sp(4)})$, is a monoidal category, whose objects are generated by the $\Sp(4)$ fundamental representations $\{ V_{(1,0)}, V_{(0,1)}\}$, and whose morphisms are $\Sp(4)$ homomorphisms between $\Sp(4)$ representations. 
\end{defn}

The $\Sp(4)$ web category $\CDDC$ is essentially the same as the $\Sp(4)$ fundamental representation category $\Fund(\Sp(4))$, by recalling an equivalence between the two categories in the following theorem by Kuperberg.


\begin{thm}[{\cite[Theorem 5.1 and 6.9]{Kuperberg}}]   \label{WebAndRep}
There is an equivalence of monoidal categories by the following functor 
\[
\Phi: \CDDC \rightarrow \Fund(\rm{Sp(4)})
\]
such that $\Phi(\ObjOne) = V_{(1,0)}$, $\Phi(\ObjTwo) = V_{(0,1)}$. 
\end{thm}

Furthermore, one can write down the image of an $\Sp(4)$-web under the functor $\Phi$ as a homomorphism between $\Sp(4)$ representations, see \cite{Sp4tilt}.
Since $\bigwedge^4 V$ is $1$-dimensional with basis $e_1 \wedge e_2 \wedge e_3 \wedge e_4$, 
for $w \in \bigwedge^4 V$, we denote by $|w|$ the element $r \in \R$ such that $w=r (e_1 \wedge e_2 \wedge e_3 \wedge e_4)$.

\begin{proposition} [{\cite[Section 3]{Sp4tilt}}]
The functor $\Phi$ sends the generating morphisms in $\CDDC$ to morphisms in $\Fund(\rm{Sp(4)})$ by the following.

\begin{align*}
       &\Phi \bigg( \SingleCap \bigg): \quad  V \otimes V \longrightarrow \mathbb{R}  
    \qquad \qquad \Phi \bigg( \DoubleCap \bigg): \quad  V_{(0,1)} \otimes V_{(0,1)} \longrightarrow \mathbb{R}  \\
       & \qquad \qquad \qquad \qquad 
       u \otimes v \mapsto v\cdot Ju  
       \qquad \qquad \qquad \qquad \qquad \qquad \quad \ u \otimes v \mapsto | u \wedge v | 
\end{align*}

\begin{align*}
    & \Phi \bigg( \SingleCup \bigg):  \quad  \mathbb{R} \longrightarrow   V \otimes V \\
 &\qquad \qquad \qquad \qquad 
 1 \mapsto e_1 \otimes e_3 - e_3 \otimes e_1 + e_2 \otimes e_4  - e_4 \otimes e_2 \\
      & \Phi \bigg( \DoubleCup \bigg):  \quad  \mathbb{R} \longrightarrow V_{(0,1)} \otimes V_{(0,1)} \\
   & 1 \mapsto  (e_1 \wedge e_2)\otimes (e_3 \wedge e_4) 
   + (e_1 \wedge e_4) \otimes (e_2 \wedge e_3) \\
   & + \frac{1}{2} (e_2 \wedge e_4 - e_1 \wedge e_3) \otimes 
   (e_2 \wedge e_4 - e_1 \wedge e_3)
   + (e_2 \wedge e_3) \otimes  (e_1 \wedge e_4) 
   + (e_3 \wedge e_4) \otimes (e_1 \wedge e_2) 
   \end{align*}

\begin{align*}
 \Phi\Bigg(\Vertb \Bigg): & V \otimes V \longrightarrow  V_{(0,1)}\\
 &u \otimes v \mapsto v \wedge u, 
\text{ when } Span\{u,v\} \ne 
Span\{e_1,e_3\} \text{ or } Span\{e_2,e_4\}, \\ 
&e_2 \otimes e_4 \mapsto - \frac{1}{2} (e_2 \wedge e_4 - e_1 \wedge e_3), \qquad
e_4 \otimes e_2 \mapsto  \frac{1}{2} (e_2 \wedge e_4 - e_1 \wedge e_3),\\
&e_1 \otimes e_3 \mapsto \frac{1}{2} (e_2 \wedge e_4 - e_1 \wedge e_3), \qquad
e_3 \otimes e_1 \mapsto -\frac{1}{2} (e_2 \wedge e_4 - e_1 \wedge e_3).
\end{align*}
\end{proposition}

We are now ready to prove Theorem \ref{Sp4det}. 
\label{thm1proof}
\begin{proof}[Proof of Theorem \ref{Sp4det}]
By Definition \ref{Tetravalent}, we know that $$\TetraIsDet=\TetraIsDeta -\frac{1}{2} \ \TetraIsDetb  {\ }_{.} $$ We can show that $\Phi(\textbf{det})$ takes a tensor product of $4$ vectors in $V$ to their wedge product in $\wedge^4 V \cong \mathbb{R}$ by explicit vector calculations from Theorem \ref{WebAndRep}. Similar arguments work for $\Phi(\textbf{codet})$. 
\end{proof}

\def\TetraBraid
{\begin{tric}
\draw[scale=0.8](-1.5,-2)..controls(-1.2,-1)and(-1,-0.5)..(-0.5,-0.5)..controls(0,-0.5)and(0.2,-1)..(0.5,-2) ;
\draw[double=darkblue,ultra thick,white,double distance=0.8pt,line width=4pt,scale=0.8]  (1.5,-2)..controls(1.2,-1)and(1,-0.5)..(0.5,-0.5)..controls(0,-0.5)and(-0.2,-1)..(-0.5,-2);
\end{tric}}

\def\TetraBraida
{\begin{tric}
\draw[scale=0.7]
     (1.5,-2)..controls(1.5,-1)and(1.2,-0.5)..(1,-0.5)
     (0.5,-2)..controls(0.5,-1)and(0.8,-0.5)..(1,-0.5)
     (-0.5,-2)..controls(-0.5,-1)and(-0.8,-0.5)..(-1,-0.5)
     (-1.5,-2)..controls(-1.5,-1)and(-1.2,-0.5)..(-1,-0.5) ;
\end{tric}
}

\def\TetraBraidb
{\begin{tric}
\draw[scale=0.7]
     (1.5,-2)..controls(1.2,-1)and(0.5,-0.5)..(0,-0.5)..controls(-0.5,-0.5)and(-1.2,-1)..(-1.5,-2)
     (0.5,-2)..controls(0.3,-1)and(-0.3,-1)..(-0.5,-2) ;
\end{tric}
}

\def\TensorSwap{
\begin{tric}
\draw[scale=0.9] (0,0)--(1,1) (0,0)--(-1,-1)
       (-0.2,0.2)--(-1,1) (0.2,-0.2)--(1,-1);
\end{tric}
}

\def\TensorSwapa{
\begin{tric}
\draw[scale=0.9] (0,0)--(-1,1) (0,0)--(1,-1)
       (0.2,0.2)--(1,1) (-0.2,-0.2)--(-1,-1);
\end{tric}
}

\def\TensorSwapb
{
\begin{tric}
\draw [scale=0.7]  (0,0)--(-0.5,0.5) (0,0)--(0.5,-0.5) (-0.5,0.5)..controls(-0.7,0.7)and(-1,1)..(-1.5,1)
..controls(-2.5,1)and(-2.5,-0.5)..(-2.5,-1)
(0.5,-0.5)..controls(0.7,-0.7)and(1,-1)..(1.5,-1)
..controls(2.5,-1)and(2.5,0.5)..(2.5,1);
\draw[double=darkblue,ultra thick,white,double distance=0.8pt,line width=4pt,scale=0.7]   (1,1)--(-1,-1);
\end{tric}
}

We have furthermore the following relation between the tetravalent vertex and tensor swapping. 
\begin{definition} \label{sp4braid}
     Define the crossing  $ \beta_{\ObjOne,\ObjOne} \in {\Hom}_{\CDDC}(\ObjOne\otimes\ObjOne,\ObjOne\otimes\ObjOne)$ as 
     $$\beta_{\ObjOne,\ObjOne}= \TensorSwap:=\Skeinec \   + \ \Skeined + \ \tetravalent $$
\end{definition}
Notice that by \eqref{pivotTetra},
\[
\TensorSwapa= \TensorSwapb = \Skeined \   + \ \Skeinec  + \ \ZigZagTetra = 
\TensorSwap,
\]
so there is no distinction between an over crossing and an under crossing.

\begin{proposition}  We have
    \begin{align*}
    \Phi(\beta_{\ObjOne,\ObjOne}): V \otimes V &\rightarrow  V \otimes V \\
        u \otimes v &\mapsto v \otimes u.
    \end{align*}
\end{proposition}

\begin{proof}
By Definition \ref{sp4braid}, 
$$\TetraIsDet \  = \ \TetraBraid \ - \ \TetraBraida \ - \  \TetraBraidb \ . $$
By direct calculations,
 $$s\wedge t \wedge u \wedge v = (u\cdot J s)(v \cdot J t)- (t \cdot J s)(v \cdot J u) - (v \cdot J s)(u\cdot J t). $$
Comparing parts we are left with 
$$\Phi\left( \TetraBraid \right): s \otimes t \otimes u \otimes v \mapsto (u\cdot J s)(v \cdot J t) $$
which is equivalent to the statement. 
\end{proof}


\section{$4$-webs on simple surfaces}
\label{4websurfacesection}

We study the space of closed $4$-webs on a simple surface. 
Sikora and Westbury \cite{SurfaceSkein} proved that reduced $\Sp(4)$-webs form a basis for the space of $\Sp(4)$ webs on any surface. We classify all closed reduced $4$-webs on an annulus, a torus, and a pair of pants, and therefore give a basis description for the space of closed $\Sp(4)$-webs on the corresponding simple surface.

\subsection{Annulus}
We consider here the case where the graph $\mathcal{G}$ is embedded on an annulus.

\def\TriMonogon
{\begin{trich}
 \draw [thick,black,fill=yellow!30] (0,0.7) circle (1.8cm);
   \draw [thick,black,fill=white] (0,0.7) circle (0.4cm);
   
\draw (0,0)..controls(2,0)and(1,1)..(0,2)
      (0,0)..controls(-2,0)and(-1,1)..(0,2);
\draw (0,0)..controls(1.5,0.3)and(0.5,1.4)..(0,1.4)
      (0,0)..controls(-1.5,0.3)and(-0.5,1.4)..(0,1.4);
\draw (0,2)--(0.6,2.2) (0,2)--(-0.6,2.2);
\filldraw [black] (0,0) circle (3pt)
                  (0,2) circle (3pt);
\end{trich}
}

\def\TriMonogona
{\begin{trich}
\draw [thick,black,fill=yellow!30] (0,0.7) circle (1.8cm);
   \draw [thick,black,fill=white] (0,0.7) circle (0.4cm);
   
\draw (0,0)..controls(2,0)and(1.5,1.7)..(0,1.7)
      (0,0)..controls(-2,0)and(-1.5,1.7)..(0,1.7);
\draw (0,0)..controls(1.5,0.3)and(1,1.4)..(0,1.4)
      (0,0)..controls(-1.5,0.3)and(-1,1.4)..(0,1.4);
\draw (0,1.9)..controls(0.3,1.9)and(0.4,2)..(0.6,2.2) 
      (0,1.9)..controls(-0.3,1.9)and(-0.4,2)..(-0.6,2.2);
\filldraw [black] (0,0) circle (3pt);
\end{trich}
}

\def\TriMonogonb
{\begin{trich}
 \draw [thick,black,fill=yellow!30] (0,0.7) circle (1.8cm);
   \draw [thick,black,fill=white] (0,0.7) circle (0.4cm);
   
\draw (0.3,0)..controls(2,0)and(1.2,1.2)..(0,1.7)
      (-0.3,0)..controls(-2,0)and(-1.2,1.2)..(0,1.7);
\draw (0.3,0)..controls(1.5,0.3)and(0.5,1.4)..(0,1.4)
      (-0.3,0)..controls(-1.5,0.3)and(-0.5,1.4)..(0,1.4);
\draw (0,2)--(0.6,2.2) (0,2)--(-0.6,2.2);
\draw[thin, decorate, decoration={snake, segment length=0.5mm, amplitude=0.45mm},darkgreen] (0,1.7)--(0,2) ;
\draw[thin, decorate, decoration={snake, segment length=0.5mm, amplitude=0.45mm},darkgreen] (-0.3,0)--(0.3,0); 
\end{trich}
}

\def\TriMonogonc
{\begin{trich}
 \draw [thick,black,fill=yellow!30] (0,0.7) circle (1.8cm);
   \draw [thick,black,fill=white] (0,0.7) circle (0.4cm);
   \draw (0.3,-0.1)..controls(2,0)and(1.2,1.2)..(0,1.7)
      (-0.3,-0.1)..controls(-2,0)and(-1.2,1.2)..(0,1.7);
\draw (0.3,0.1)..controls(1.5,0.3)and(0.5,1.4)..(0,1.4)
      (-0.3,0.1)..controls(-1.5,0.3)and(-0.5,1.4)..(0,1.4);
\draw (0.3,-0.1)..controls(0.1,-0.1)and(0.1,0.1)..(0.3,0.1) 
      (-0.3,-0.1)..controls(-0.1,-0.1)and(-0.1,0.1)..(-0.3,0.1);  
\draw (0,2)--(0.6,2.2) (0,2)--(-0.6,2.2);
\draw[thin, decorate, decoration={snake, segment length=0.5mm, amplitude=0.45mm},darkgreen] (0,1.7)--(0,2) ; 
\end{trich}
}

\def\TriMonogond
{\begin{trich}
 \draw [thick,black,fill=yellow!30] (0,0.7) circle (1.8cm);
   \draw [thick,black,fill=white] (0,0.7) circle (0.4cm);
   
\draw (0.3,0)..controls(2,0)and(1.5,1.7)..(0,1.7)
      (-0.3,0)..controls(-2,0)and(-1.5,1.7)..(0,1.7);
\draw (0.3,0)..controls(1.5,0.3)and(0.5,1.4)..(0,1.4)
      (-0.3,0)..controls(-1.5,0.3)and(-0.5,1.4)..(0,1.4);
\draw (0.6,2.2)..controls(0.3,1.9)and(-0.3,1.9)..(-0.6,2.2);
\draw[thin, decorate, decoration={snake, segment length=0.5mm, amplitude=0.45mm},darkgreen]
(-0.3,0)--(0.3,0); 
\end{trich}
}

\def\TriMonogone
{\begin{trich}
 \draw [thick,black,fill=yellow!30] (0,0.7) circle (1.8cm);
   \draw [thick,black,fill=white] (0,0.7) circle (0.4cm);
   
\draw (0.3,-0.1)..controls(2,0)and(1.5,1.7)..(0,1.7)
      (-0.3,-0.1)..controls(-2,0)and(-1.5,1.7)..(0,1.7);
\draw (0.3,0.1)..controls(1.5,0.3)and(0.5,1.4)..(0,1.4)
      (-0.3,0.1)..controls(-1.5,0.3)and(-0.5,1.4)..(0,1.4);
\draw (0.3,-0.1)..controls(0.1,-0.1)and(0.1,0.1)..(0.3,0.1) 
      (-0.3,-0.1)..controls(-0.1,-0.1)and(-0.1,0.1)..(-0.3,0.1);  
\draw  (0.6,2.2)..controls(0.3,1.9)and(-0.3,1.9)..(-0.6,2.2);
\end{trich}
}

\def\TetraIsDeta
{\begin{tric}
\draw[scale=0.7] (1.5,-2)--(1,-1)--(0.5,-2)
                 (-1.5,-2)--(-1,-1)--(-0.5,-2);
\draw[scale=0.7,thin, decorate, decoration={snake, segment length=0.5mm, amplitude=0.45mm},darkgreen](1,-1)..controls(1,-0.6)and(0.4,0)..(0,0)..controls(-0.4,0)and(-1,-0.6)..(-1,-1); 
\end{tric}
}

\begin{lemma}
 For a $4$-web which has a contractible triangle face with two overlapping vertices, the following skein relation holds.  
\begin{equation}
        \TriMonogon = -\frac{1}{2} \TriMonogona \label{TetraSkeinD}
    \end{equation}
\end{lemma}

\begin{proof}
By Definition \ref{Tetravalent},
$$ \TriMonogon = \TriMonogonb -\frac{1}{2} \TriMonogonc 
-\frac{1}{2} \TriMonogond  +\frac{1}{4} \TriMonogone.$$

By Definition \ref{C2Spider}, the first two terms are equal to $0$. And by Definition \ref{Tetravalent} again, the sum of the last two terms is equal to $ -\cfrac{1}{2} \TriMonogona $.
\end{proof}

\def\ANNSpOneN{
\begin{trica}
   \draw [thick,black,fill=yellow!30] (0,0) circle (3cm);
   \draw [thick,black,fill=white] (0,0) circle (1cm);
   \filldraw[black] (60:2) circle (3pt) (120:2) circle (3pt)
   (0:2) circle (3pt) (180:2) circle (3pt) ;
   \draw (2,0) arc
    [
        start angle=0,
        end angle=225,
        x radius=2cm,
        y radius =2cm
    ] ;
     \draw (2,0) arc
    [
        start angle=0,
        end angle=-45,
        x radius=2cm,
        y radius =2cm
    ] ;
      \filldraw[black] (245:2) circle (1pt) (255:2) circle (1pt)
   (265:2) circle (1pt) (275:2) circle (1pt) (285:2) circle (1pt) (295:2) circle (1pt);
   \draw (0,2)node[above,black,scale=0.7]{$\textbf{3}$}
         (150:2)node[left,black,scale=0.7]{$\textbf{1}$}
         (30:2)node[right,black,scale=0.7]{$\textbf{1}$}
         (210:2)node[left,black,scale=0.7]{$\textbf{3}$}
         (-30:2)node[right,black,scale=0.7]{$\textbf{3}$};
\end{trica}
}

\def\ANNSpOneNexpanded{
\begin{trica}
   \draw [thick,black,fill=yellow!30] (0,0) circle (3cm);
   \draw [thick,black,fill=white] (0,0) circle (1cm);
   \filldraw[black] (60:2) circle (3pt) (120:2) circle (3pt)
   (0:2) circle (3pt) (180:2) circle (3pt) ;
   \draw (2,0) arc
    [
        start angle=0,
        end angle=225,
        x radius=2cm,
        y radius =2cm
    ] ;
     \draw (2,0) arc
    [
        start angle=0,
        end angle=-45,
        x radius=2cm,
        y radius =2cm
    ] ;
      \filldraw[black] (245:2) circle (1pt) (255:2) circle (1pt)
   (265:2) circle (1pt) (275:2) circle (1pt) (285:2) circle (1pt) (295:2) circle (1pt);
\draw(60:2)..controls(80:3)and(100:3)..(120:2); 
\draw(60:2)..controls(80:1.2)and(100:1.2)..(120:2); 
\draw(2,0)..controls(2.3,-0.7)and(2.4,-1)..(2,-1.5)
(-2,0)..controls(-2.3,-0.7)and(-2.4,-1)..(-2,-1.5);
\draw (2,0)..controls(1.5,-0.5)and(1.2,-0.8)..(1,-1.2)
      (-2,0)..controls(-1.5,-0.5)and(-1.2,-0.8)..(-1,-1.2);
\end{trica}
}

\def\ANNSpTwoN{
\begin{trica}
   \draw [thick,black,fill=yellow!30] (0,0) circle (3cm);
   \draw [thick,black,fill=white] (0,0) circle (1cm);
   \filldraw[black] (60:2) circle (3pt) (120:2) circle (3pt)
   (0:2) circle (3pt) (180:2) circle (3pt) ;
   \draw (2,0) arc
    [
        start angle=0,
        end angle=225,
        x radius=2cm,
        y radius =2cm
    ] ;
     \draw (2,0) arc
    [
        start angle=0,
        end angle=-45,
        x radius=2cm,
        y radius =2cm
    ] ;
      \filldraw[black] (245:2) circle (1pt) (255:2) circle (1pt)
   (265:2) circle (1pt) (275:2) circle (1pt) (285:2) circle (1pt) (295:2) circle (1pt);
   \draw (0,2)node[above,black,scale=0.7]{$\textbf{2}$}
         (150:2)node[left,black,scale=0.7]{$\textbf{2}$}
         (30:2)node[right,black,scale=0.7]{$\textbf{2}$}
         (210:2)node[left,black,scale=0.7]{$\textbf{2}$}
         (-30:2)node[right,black,scale=0.7]{$\textbf{2}$};
\end{trica}
}

\def\ANNSpTwoNexpanded{
\begin{trica}
   \draw [thick,black,fill=yellow!30] (0,0) circle (3cm);
   \draw [thick,black,fill=white] (0,0) circle (1cm);
   \filldraw[black] (60:2) circle (3pt) (120:2) circle (3pt)
   (0:2) circle (3pt) (180:2) circle (3pt) ;
   \draw (2,0) arc
    [
        start angle=0,
        end angle=225,
        x radius=2cm,
        y radius =2cm
    ] ;
     \draw (2,0) arc
    [
        start angle=0,
        end angle=-45,
        x radius=2cm,
        y radius =2cm
    ] ;
      \filldraw[black] (245:2) circle (1pt) (255:2) circle (1pt)
   (265:2) circle (1pt) (275:2) circle (1pt) (285:2) circle (1pt) (295:2) circle (1pt);
\draw(60:2)..controls(80:1.2)and(100:1.2)..(120:2); 
\draw (2,0)..controls(1.5,-0.5)and(1.2,-0.8)..(1,-1.2)
      (-2,0)..controls(-1.5,-0.5)and(-1.2,-0.8)..(-1,-1.2);
\draw (180:2)..controls(160:1.2)and(140:1.2)..(120:2)
      (0:2)..controls(20:1.2)and(40:1.2)..(60:2); 
\end{trica}
}

\def\ANNSp{
\begin{trics}
   \draw [thick,black,fill=yellow!30] (0,0) circle (3cm);
   \draw [thick,black,fill=white] (0,0) circle (1cm);
   \draw [thick](0,0) circle (2cm);
   
\end{trics}
}

\def\ANNSpa{
\begin{trics}
   \draw [thick,black,fill=yellow!30] (0,0) circle (3cm);
   \draw [thick,black,fill=white] (0,0) circle (1cm);
   
\draw[thick] (0,0.4cm)  circle (1.8cm)
             (0,0)  circle (2.2cm);

\draw[fill=black] (0,2.2cm)  circle (0.07cm); 
\end{trics}
}

\begin{definition}
An $\Sp(4)$ $1$-loop with $2n$ vertices on an annulus is defined as:
$$\mathscr{L}_{2n}^{(1)}:=\ANNSpOneN=\frac{1}{6^n} \ \ANNSpOneNexpanded$$
An $\Sp(4)$ 2-loop with $n$ vertices on an annulus is defined as :
$$\mathscr{L}_{n}^{(2)}:=\ANNSpTwoN=\frac{1}{2^n} \ \ANNSpTwoNexpanded$$
\end{definition}

\begin{proposition}
For a flat connection on the annulus, we have $\mathscr{L}_{2n}^{(1)}=\mathscr{L}_{2n-2}^{(1)}$, 
so for any $n\in \mathbb{Z}_{\ge 0}$:  
$$\mathscr{L}_{2n}^{(1)}=\mathscr{L}_{0}^{(1)}= \ANNSp $$
and $\mathscr{L}_{n}^{(2)}=-\mathscr{L}_{n-1}^{(2)}+2 $, so $\mathscr{L}_{n}^{(2)}=\mathscr{L}_{n-2}^{(2)} $. 
Thus for any $n\in \mathbb{Z}_{\ge 0}$, 
$$\mathscr{L}_{2n+1}^{(2)}=\mathscr{L}_{1}^{(2)}= \ANNSpa $$ and $\mathscr{L}_{2n}^{(2)}=\mathscr{L}_{2}^{(2)}=-\mathscr{L}_{1}^{(2)}+2 .$
\end{proposition}

\begin{proof}
    Apply relations in Definition \ref{C2Spider} and Definition \ref{Tetravalent}. 
\end{proof}

\begin{thm}\label{Sp4ann}
By use of skein relations \eqref{TetraSkeinA}, \eqref{TetraSkeinB}, \eqref{TetraSkeinC}, and \eqref{TetraSkeinD}, any $4$-web 
on an annulus with a flat $\Sp(4)$-local system can be reduced to a unique linear combination of disjoint unions of noncontractible 1-loops and 2-loops.
\end{thm}

\begin{proof}
For a connected tetravalent graph $\mathcal{G}$ on a sphere which is not a cycle, we have (by Euler characteristic)
\begin{equation*}
    8=3n_1 + 2n_2 + 1n_3 + 0n_4 - 1n_5 + \dots + (4-k)n_k + \dots 
\end{equation*} 
where $n_k$ is the number of faces with $k$ edges.

When a tetravalent graph $\G$ is embedded on an annulus, at most two faces contain boundary components. So there is at least one contractible monogon face or one contractible bigon face or two contractible triangle faces.

Denote the number of vertices of $\mathcal{G}$ by $V$. We finish the proof by induction on $V$. 
\begin{itemize}
\item When there is a contractible monogon face, we can reduce $V$ by using relation \eqref{TetraSkeinA}.

\item When there is a contractible bigon face such that the two vertices of the bigon are different, we can reduce $V$ by using relation \eqref{TetraSkeinB}.

\item When there is a contractible bigon face such that the two vertices of the bigon are the same (and no contractible monogon), we know that $\mathcal{G}$ is a 2-cycle.

\item When there is a contractible triangle face such that all the three vertices of the triangle are different, we can reduce $V$ by using relation \eqref{TetraSkeinC}.

\item When there is a contractible triangle face such that two of the three vertices of the bigon are the same, we can reduce $V$ by using relation \eqref{TetraSkeinD}.

\end{itemize}

Applying the above inductive step reduces the number of vertices in $\mathcal{G}$ and possibly disconnects $\mathcal{G}$, unless $\mathcal{G}$ is itself a 2-cycle.  We can continue applying the inductive step to each component until every component is a cycle or 2-cycle.

\end{proof}

\subsection{Torus}\label{torussection}
We consider here the case where the graph $\mathcal{G}$ is embedded on a torus.

\begin{proposition}
Any reduced connected $4$-web on a torus is either a noncontractible cycle, a noncontractible 2-cycle, or a square grid 
on the torus (which can be described by a quotient of the square grid on $\mathbb{R}^2$ by a sublattice of translations in $\Z^2$). 
\end{proposition}

\begin{proof}
Consider a reduced $4$-web $\mathcal{W}$ on a torus. If there is a noncontractible simple loop $\gamma$ on the torus, 
such that $\gamma$ and $\mathcal{W}$
don't intersect, then we can cut the torus along $\gamma$, and  $\mathcal{W}$ becomes a reduced web on an annulus. So $\mathcal{W}$ is either a noncontractible cycle or a noncontractible 2-cycle in this case.

If $\mathcal{W}$ intersects all the noncontractible loops, then $\mathcal{W}$ doesn't have a non-reducible bigon face. 
So by skein relations \eqref{TetraSkeinA}, \eqref{TetraSkeinB}, \eqref{TetraSkeinC}, and \eqref{TetraSkeinD},  $\mathcal{W}$ doesn't have any face with less than $4$ edges. On the other hand, since the Euler characteristic of a torus is equal to $0$,  $\mathcal{W}$ doesn't have any face with more than $4$ edges. So  $\mathcal{W}$ is a square grid on a torus, which is necessarily 
a quotient of the standard square grid on $\mathbb{R}^2$ by a sublattice of translations in $\Z^2$. 
\end{proof}

\subsection{Pair of pants}\label{pantssection}

Let $\Sigma$ be a pair of pants, that is, a sphere with three punctures. 
There are two types of special reduced $4$-webs on a pair of pants, which we denote $W_\eta$ and $W_\xi$, which we now describe.

Let $\eta=(a,b)\in\Z^2$, with $\eta\ne(0,0)$. We construct the dual graph of a reduced web $W_{\eta}$ on $\Sigma$  as follows. 
Choose $S_1$ to be a grid point on the square grid. Let 
\[\overrightarrow {S_1 P}=(a,b), \quad 
\overrightarrow {P S_2}=(b,-a), \quad 
\overrightarrow {S_2 Q}=(-a,-b), \quad 
\overrightarrow {Q S_1}=(-b,a).\] 
From the square $PS_1QS_2$, glue sides $PS_1$ with $PS_2$ and $QS_1$ with $QS_2$ as shown, putting punctures at the corners, to form a 3-punctured sphere $\Sigma$. 
The dual graph of the image of the square grid is a $4$-web $W_\eta$ on $\Sigma$ with a monogon face at $P$ and $Q$ and a bigon face at $S_1=S_2$, and all other faces are quads.

\def\OneOneTwo
{\begin{tric}
\draw[darkgreen,scale=0.7,semithick] (-0.5,0)--(5.5,0) (-0.5,1)--(5.5,1)  (-0.5,2)--(5.5,2) (-0.5,3)--(5.5,3) (-0.5,4)--(5.5,4) (-0.5,5)--(5.5,5) ;
\draw[darkgreen,scale=0.7,semithick] (0,-0.5)--(0,5.5) (1,-0.5)--(1,5.5)  (2,-0.5)--(2,5.5) (3,-0.5)--(3,5.5) (4,-0.5)--(4,5.5) (5,-0.5)--(5,5.5) ;
\filldraw[darkgreen,scale=0.7] 
 (0,0) circle (1.2pt)  (1,0) circle (1.2pt)  (2,0) circle (1.2pt)
 (3,0) circle (1.2pt)  (4,0) circle (1.2pt)  (5,0) circle (1.2pt)
  (0,1) circle (1.2pt)  (1,1) circle (1.2pt)  (2,1) circle (1.2pt)
 (3,1) circle (1.2pt)  (4,1) circle (1.2pt)  (5,1) circle (1.2pt)
  (0,2) circle (1.2pt)  (1,2) circle (1.2pt)  (2,2) circle (1.2pt)
 (3,2) circle (1.2pt)  (4,2) circle (1.2pt)  (5,2) circle (1.2pt)
  (0,3) circle (1.2pt)  (1,3) circle (1.2pt)  (2,3) circle (1.2pt)
 (3,3) circle (1.2pt)  (4,3) circle (1.2pt)  (5,3) circle (1.2pt)
  (0,4) circle (1.2pt)  (1,4) circle (1.2pt)  (2,4) circle (1.2pt)
 (3,4) circle (1.2pt)  (4,4) circle (1.2pt)  (5,4) circle (1.2pt)
   (0,5) circle (1.2pt)  (1,5) circle (1.2pt)  (2,5) circle (1.2pt)
 (3,5) circle (1.2pt)  (4,5) circle (1.2pt)  (5,5) circle (1.2pt); 
\draw[ultra thick, darkred,-stealth reversed,scale=0.7] (2,0)--(3.5,1);
  \draw[ultra thick, darkred,-stealth reversed,scale=0.7] (2,0)--(1,1.5);
  
         \draw[ultra thick, darkred,-stealth reversed,scale=0.7] (1.1,1.35)--(0.9,1.65);
         
   \draw[ultra thick, darkred,-stealth reversed,scale=0.7] (3,5)--(4,3.5);
   
    \draw[ultra thick, darkred,-stealth reversed,scale=0.7] (3,5)--(1.5,4);
    
          \draw[ultra thick, darkred,-stealth reversed,scale=0.7] (1.65,4.1)--(1.35,3.9);
          
\draw[ultra thick, darkred,scale=0.7] (1.5,4)--(0,3) (1,1.5)--(0,3) 
               (3.9,3.65)--(5,2) (3.35,0.9)--(5,2);
\draw[scale=0.7] (3,5) node[above, black] {$S_2$}   
      (2,0) node[below, black] {$S_1$}
      (0,3) node[left, black] {$P$} 
      (5,2) node[right, black] {$Q$} ;          
\end{tric}
}

\begin{align*}
&\text{Dual graph of } W_{(-2,3)}:   \OneOneTwo \\
\end{align*}

\begin{figure}[htbp]
\begin{center}\includegraphics[width=2in]{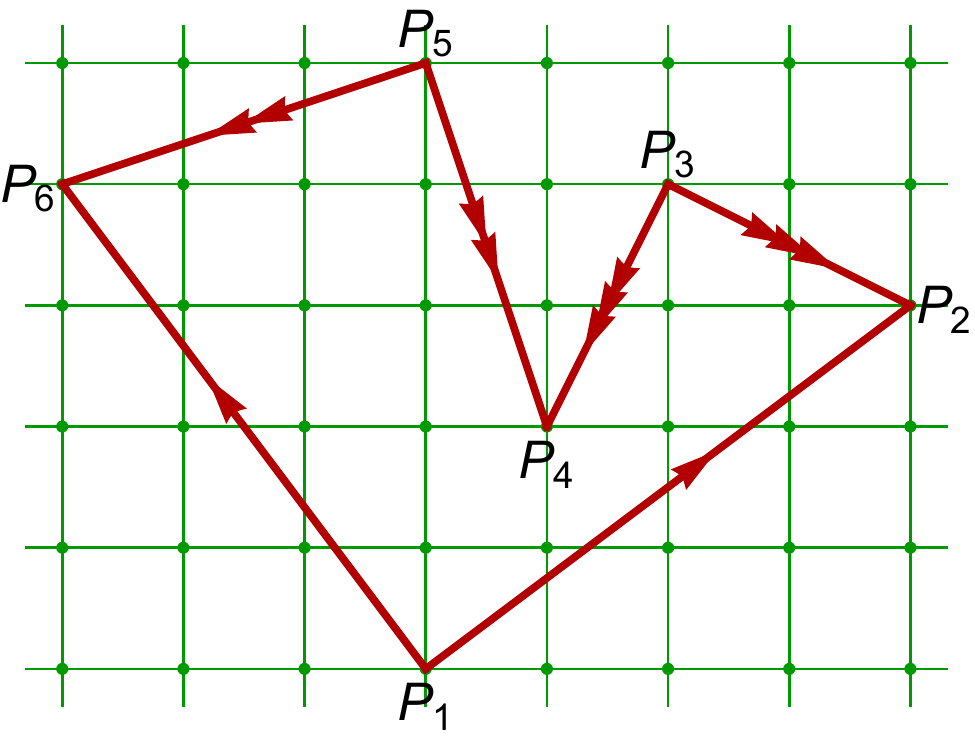}\end{center}
\caption{\label{hexagon}A hexagon $\xi$.}
\end{figure}

Let $\xi$ be a hexagon $P_1P_2\dots P_6$ with vertices in $\Z^2$, with right angles at $P_1,P_3,P_5$ and
\begin{align*}|P_1P_2|&=|P_1P_6|\\|P_3P_2|&=|P_3P_4|\\|P_5P_4|&=|P_5P_6|,\end{align*}
see Figure \ref{hexagon}.
Glue sides $P_1P_6$ and $P_1P_2$, sides $P_3P_2$ and $P_3P_4$, and sides $P_5P_4$ and $P_5P_6$
to form a closed genus $0$ surface $\Sigma_\xi$ tiled with squares, with cone angles $\pi/2$ at $P_1,P_3,P_5$ and angle $5\pi/2$ at the common image
of $P_2,P_4,P_6$. The dual graph of the image of the square grid is a $4$-web $W_\xi$ on $\Sigma_\xi$ with a monogon face surrounding
each of $P_1,P_3,P_5$, a pentagonal face around $P_2=P_4=P_6$, and other faces of degree $4$.

\begin{lemma} \label{NoIntersect}
    Consider a quadrangulation of the sphere with unit Euclidean squares and with $n$ cone points $v_1,v_2,...,v_n$, where $v_1,v_2,...,v_{n-1}$ have positive curvatures. Suppose the shortest geodesic from $v_n$ to $v_i$ is $l_i$, $1\le i \le n-1$. Then $l_i$ and $l_j$ only intersect at $v_n$ if $i \ne j$. 
\end{lemma}

\def\XGeo{
\begin{tric}
   \draw[very thick] (-0.5,-0.5)--(1,1);
   \draw[very thick,darkgreen](0.5,-0.5)--(-1,1);
   \draw[very thick] (-0.5,-0.5)..controls(-1,-1)and(-1,-1.5)..(0,-2);
   \draw[very thick,darkgreen](0.5,-0.5)..controls(1,-1)and(1,-1.5)..(0,-2);
\filldraw[black] 
          (0,-2) circle (1.5pt) node[below,scale=0.7] {$v_n$}
          (0,0) circle (1.5pt)
          (1,1) circle (1.5pt) node[above,scale=0.7] {$v_i$}
          (-1,1) circle (1.5pt)  node[above,scale=0.7] {$v_j$} ;
\draw  (0,0.1) node[above,scale=0.7] {$u$};
\draw (-0.85,-1) node[left,scale=0.7,darkblue] {$l_i$};
\draw (0.85,-1) node[right,scale=0.7,darkgreen] {$l_j$};
\draw[black] (0.2,-0.2)..controls(0.3,-0.1)and(0.3,0.1)..(0.2,0.2)
        (-0.2,-0.2)..controls(-0.3,-0.1)and(-0.3,0.1)..(-0.2,0.2);
\draw (0.2,0)node[right,scale=0.7]{$\alpha$};
\draw (-0.2,0)node[left,scale=0.7]{$\beta$};
\end{tric}
}

\begin{proof}
Suppose $l_i$ and $l_j$ intersect at $u\ne v_n$.
$$\XGeo$$
Consider the geodesic distances $D_{l_i}(v_n,u)$ and  $D_{l_j}(v_n,u)$.
If the distances are not equal, suppose $D_{l_i}(v_n,u)>D_{l_j}(v_n,u)$, then $D_{l_i}(v_n,v_i)=D_{l_i}(v_n,u)+D_{l_i}(u,v_i)>D_{l_j}(v_n,u)+D_{l_i}(u,v_i)$, which is contradicted to $l_i$ being the shortest geodesic from $v_n$ to $v_i$. 
    
Since $u\ne v_n$, $u$ is either not a cone point or is a cone point with positive curvature. So $\alpha+\beta < 2 \pi$, which means that $\alpha < \pi$ or $\beta <  \pi$.  If $D_{l_i}(v_n,u)=D_{l_j}(v_n,u)$, suppose  $\alpha < \pi$, then there exist a geodesic from $v_n$ to $v_i$ whose length is less than $D_{l_j}(v_n,u)+D_{l_i}(u,v_i)=D_{l_i}(v_n,u)+D_{l_i}(u,v_i)=D_{l_i}(v_n,v_i)$. This contradicts the facts that 
$l_i$ is a shortest geodesic from $v_n$ to $v_i$.  
\end{proof}

\begin{thm} A reduced $4$-web on a pair of pants is a union of a collection of cycles and 2-cycles and at most one of the webs of type $W_{\xi}$ or $W_{\eta}$. 
\end{thm}

\begin{proof}
Consider a connected reduced $4$-web $\mathcal{W}$ on $\Sigma$.
If $\mathcal{W}$ has a face containing more than one boundary component, it is a $1$-loop or $2$-loop, by Theorem \ref{Sp4ann}.
Otherwise, if each face of $\mathcal{W}$ contains at most one boundary component, we have (by Euler characteristic)
\begin{equation}
    8=3n_1 + 2n_2 + 1n_3 + 0n_4 - 1n_5 + \dots + (4-k)n_k + \dots \label{EulerChar}
\end{equation} 
where $n_k$ is the number of faces with $k$ edges.

By relations \eqref{TetraSkeinA}, \eqref{TetraSkeinB}, \eqref{TetraSkeinC}, and \eqref{TetraSkeinD}, any contractible face of $\mathcal{W}$ has at least $4$ edges. Then by equation \eqref{EulerChar}, $\mathcal{W}$ falls into one of the following two cases. 
\begin{itemize}
\item $\mathcal{W}$ has $3$ noncontractible monogon  faces, $1$ contractible pentagon face, and contractible square faces. 
\item $\mathcal{W}$ has $2$  noncontractible monogon faces, $1$ noncontractible bigon face, and contractible square faces. 
\end{itemize}

Take the dual graph $\overline{\mathcal{W}}$ of $\mathcal{W}$.  It is a quadrangulation $\rm{Q}$ of $S^2$.  Its cone points (for the Euclidean metric in which all faces are unit squares) are given by vertices of $\overline{\mathcal{W}}$ which are dual to the monogon faces, bigon faces, or pentagon faces of  $\mathcal{W}$.

\def\Monogon
{\begin{tric}
\draw (-1,-0.5)..controls(1,0.5)and(0.5,1)..(0,1)
      (1,-0.5)..controls(-1,0.5)and(-0.5,1)..(0,1);
\end{tric}
}

\def\DualMonogon
{\begin{tric}
\draw (-1,-0.5)..controls(1,0.5)and(0.5,1)..(0,1)
      (1,-0.5)..controls(-1,0.5)and(-0.5,1)..(0,1);
\filldraw[darkgreen] (0,0.6) circle (2pt)
                     (0,1.4) circle (2pt)
                     (0,-0.5) circle (2pt);     
\draw[darkgreen] (0,0.6)--(0,1.4) 
                 (0,-0.5)..controls(1,0)and(1,1.3)..(0,1.4) 
                 (0,-0.5)..controls(-1,0)and(-1,1.3)..(0,1.4); 
\end{tric}
}

\def\ConeMonogon
{\begin{tric}
\fill [lightyellow,semitransparent] (1,1)--(1,-1)--(-1,-1)--(-1,1)--cycle;
\draw [darkgreen](-1,1)--(1,1) (-1,-1)--(-1,1);
\draw [darkgreen,-stealth](1,-1)--(0,-1); 
\draw [darkgreen,-stealth](1,-1)--(1,0);
\draw [darkgreen] (1,-0.1)--(1,1) (0.1,-1)--(-1,-1);
\filldraw[darkgreen] (1,1) circle (2pt) (-1,1) circle (2pt) 
                     (1,-1) circle (2pt) (-1,-1) circle (2pt);
\end{tric}
}

\def\ConeMonogona
{\begin{tric}
\fill [lightyellow,semitransparent]
       (1,1.5)--(0,-0.2)--(-1,1.5)..controls(-0.8,1.7)and(-0.5,2)..
       (0,2)..controls(0.5,2)and(0.8,1.7)..(1,1.5)--cycle; 
\fill [lightyellow,semitransparent]
       (1,1.5)--(0,-0.2)--(-1,1.5)..controls(-0.8,1.3)and(-0.5,1)..
       (0,1)..controls(0.5,1)and(0.8,1.3)..(1,1.5)--cycle;  
\fill [lightyellowa,semitransparent]
       (1,1.5)--(0,-0.2)--(-1,1.5)..controls(-0.8,1.3)and(-0.5,1)..
       (0,1)..controls(0.5,1)and(0.8,1.3)..(1,1.5)--cycle;  
\draw [darkgreen,-stealth] (0,-0.2)--(0.5,0.65);       
\draw [darkgreen] (0.4,0.48)--(1,1.5);
\draw [darkgreen]
      (-1,1.5)..controls(-0.8,1.7)and(-0.5,2)..(0,2)
      (1,1.5)..controls(0.8,1.7)and(0.5,2)..(0,2)
      (-1,1.5)..controls(-0.8,1.3)and(-0.5,1)..(0,1)
      (1,1.5)..controls(0.8,1.3)and(0.5,1)..(0,1); 
\filldraw[darkgreen] (0,-0.2) circle (2pt) 
                     (1,1.5) circle (2pt)  
                     (-1,1.5) circle (2pt); 
\end{tric}
}

\def\Bigon
{\begin{tric}
\draw (-1,-1)..controls(1,-0.3) and (1,0.3) .. (-1,1)
      (1,-1)..controls(-1,-0.3) and (-1,0.3) .. (1,1);
\end{tric}
}

\def\DualBigon
{\begin{tric}
\draw (-1,-1)..controls(1,-0.3) and (1,0.3) .. (-1,1)
      (1,-1)..controls(-1,-0.3) and (-1,0.3) .. (1,1);
\filldraw[darkgreen] (1,0) circle (2pt) (-1,0) circle (2pt) 
                     (0,0) circle (2pt)
                     (0,1.4) circle (2pt) (0,-1.4) circle (2pt);
\draw[darkgreen] (1,0)--(0,0) (-1,0)--(0,0) 
                 (1,0)--(0,1.4) (-1,0)--(0,1.4)
                 (1,0)--(0,-1.4) (-1,0)--(0,-1.4); 
\end{tric}
}

\def\ConeBigon
{\begin{tric}
\fill [lightyellow,semitransparent] (0,1.5)--(1.5,1.5)--(1.5,-1.5)--(0,-1.5)--cycle;
\draw [darkgreen](0,0)--(0,1.5) (0,0)--(1.5,0) (0,1.5)--(1.5,1.5);
\draw [darkgreen](0,0)--(0,-1.5) (0,0)--(1.5,0) (0,-1.5)--(1.5,-1.5);
\draw [darkgreen,-stealth](1.5,0)--(1.5,0.75); 
\draw [darkgreen,-stealth](1.5,0)--(1.5,-0.75);
\draw [darkgreen] (1.5,0.7)--(1.5,1.5) (1.5,-0.7)--(1.5,-1.5);
\filldraw[darkgreen] (0,0) circle (2pt) (1.5,0) circle (2pt) 
                     (0,1.5) circle (2pt) (1.5,1.5) circle (2pt)
                     (0,-1.5) circle (2pt) (1.5,-1.5) circle (2pt);
\end{tric}
}

\def\ConeBigona
{\begin{tric}
\fill [lightyellow,semitransparent]
 (0,0)--(1.5,1)--(0.5,1.8)--(-1.5,1)--cycle; 
\fill [lightyellow,semitransparent]
 (0,0)--(1.5,1)--(-0.5,1.8)--(-1.5,1)--cycle; 
\fill [lightyellowa,semitransparent]
 (0,0)--(1.5,1)--(0,1.6)--(-1.5,1)--cycle; 
\draw [darkgreen,-stealth] (0,0)--(0.75,0.5);       
\draw [darkgreen] (0.6,0.4)--(1.5,1);
\draw [darkgreen] (-0.5,1.8)--(1.5,1) (-1.5,1)--(-0.5,1.8)
                  (-1.5,1)--(0,0); 
\draw[color=darkgreen, dash pattern=on 3pt off 2pt] (-1.5,1)--(0,1.6); 
\draw[color=darkgreen](0,1.6)--(0.5,1.8) (0.5,1.8)--(1.5,1);
\filldraw[darkgreen] (0,0) circle (2pt) 
                     (1.5,1) circle (2pt)  
                     (-1.5,1) circle (2pt)
                     (-0.5,1.8) circle (2pt)
                     (0.5,1.8) circle (2pt); 
\end{tric}
}

\def\Pentagon
{\begin{tric}
\draw  (-30:1.5)--(-150:1.5) (42:1.5)--(-78:1.5) 
       (114:1.5)--(-6:1.5)
       (186:1.5)--(66:1.5) (-102:1.5)--(-222:1.5);
\end{tric}
}

\def\DualPentagon
{\begin{tric}
\draw  (-30:1.5)--(-150:1.5) (42:1.5)--(-78:1.5) 
       (114:1.5)--(-6:1.5)
       (186:1.5)--(66:1.5) (-102:1.5)--(-222:1.5);
\filldraw[darkgreen] (0,0) circle (2pt) 
(-54:1.5) circle (2pt) (-18:1.1) circle (2pt)
(18:1.5) circle (2pt) (54:1.1) circle (2pt)
(90:1.5) circle (2pt) (126:1.1) circle (2pt)
(162:1.5) circle (2pt) (198:1.1) circle (2pt)
(234:1.5) circle (2pt) (270:1.1) circle (2pt);
\draw[darkgreen]  
 (0,0)--(-18:1.1)   (0,0)--(54:1.1)   (0,0)--(126:1.1)  (0,0)--(198:1.1)  (0,0)--(270:1.1) 
(-54:1.5)--(-18:1.1)--(18:1.5)--(54:1.1)--(90:1.5)--(126:1.1)--(162:1.5)--(198:1.1)--(234:1.5)--(270:1.1)--cycle ;
\end{tric}
}

\def\ConePentagon
{\begin{tric}
\fill [lightyellow,semitransparent] (0,0)--(0,-1.2)--(-0.5,-1.9)--(-0.5,-0.7)--cycle;
\fill [lightyellow,semitransparent] (0,0)--(-1.2,0)--(-1.7,-0.7)--(-0.5,-0.7)--cycle;
\fill [lightyellow,semitransparent] (0,0)--(-1.2,0)--(-0.7,0.7)--(0.5,0.7)--cycle;
\fill [lightyellow,semitransparent] (0,0)--(1.2,0)--(1.7,0.7)--(0.5,0.7)--cycle;
\fill [lightyellow,semitransparent] (0,0)--(1.2,0)--(1.2,1.2)--(0,1.2)--cycle;
\fill [lightyellowa,semitransparent] (0,0)--(1.2,0)--(1.2,0.7)--(0,0.7)--cycle;
\draw [darkgreen](0,0)--(1.2,0) (0,0)--(-1.2,0) 
                  (0,0)--(-0.5,-0.7);
\draw [darkgreen,-stealth](0,0)--(0,0.6); 
\draw [darkgreen,-stealth](0,0)--(0,-0.6);
\draw [darkgreen] (0,1.2)--(1.2,1.2)--(1.2,0);
\draw [darkgreen] (-0.5,-0.7)--(-0.5,-1.9)--(0,-1.2);
\draw [darkgreen] (-0.5,-0.7)--(-1.7,-0.7)--(-1.2,0); 
\draw[color=darkgreen, dash pattern=on 3pt off 2pt] (0,0.7)--(1.2,0.7) (0,0)--(0.5,0.7); 
\draw[color=darkgreen] (0,0.7)--(-0.7,0.7)--(-1.2,0) 
                       (1.2,0)--(1.7,0.7)--(1.2,0.7); 
\draw [darkgreen] (0,0.5)--(0,1.2) (0,-0.5)--(0,-1.2);
\filldraw[darkgreen] (0,0) circle (2pt) (1.2,0) circle (2pt) 
(0,1.2) circle (2pt) (1.2,1.2) circle (2pt) 
(0,-1.2) circle (2pt) (1.7,0.7) circle (2pt)
(0.5,0.7) circle (2pt) (-0.7,0.7) circle (2pt)
(-1.2,0) circle (2pt) (-1.7,-0.7) circle (2pt)
(-0.5,-0.7) circle (2pt) (-0.5,-1.9) circle (2pt); 
\end{tric}
}

\def\ConePentagona
{\begin{tric}
\fill [lightyellow,semitransparent] (0,0)--(0.6,-0.9)--(-0.1,-1.4)--(-0.3,-0.9)--cycle;
\fill [lightyellow,semitransparent] (0,0)--(0.6,-0.9)--(1.8,-1) --cycle;
\fill [lightyellow,semitransparent] (0,0)--(-0.3,-0.9)--(-1.4,-1.3)--cycle; 
\fill [lightyellowa,semitransparent] (0,0)--(0.6,-0.9)--(-0.1,-1.4)--(-0.3,-0.9)--cycle;
\fill [lightyellowa,semitransparent] (0,0)--(0.6,-0.9)--(1.8,-1) --cycle;
\fill [lightyellowa,semitransparent] (0,0)--(-0.3,-0.9)--(-1.4,-1.3)--cycle; 
\fill [lightyellow,semitransparent] (0,0)--(-1.4,-1.3)--(-1,-1.3)--cycle; 
\fill [lightyellow,semitransparent] (0,0)--(1.4,-1.1)--(1.8,-1)--cycle; 
\fill [lightyellow,semitransparent] (0,0)--(-1,-1.3)--(-0.3,-1.8)--(0.3,-1.5)--cycle;
\fill [lightyellow,semitransparent] (0,0)--(1.4,-1.1)--(1,-1.6)--(0.3,-1.5)--cycle;
\draw[darkgreen] (0,0)--(0.3,-1.5)
                (0,0)--(1.4,-1.1) (0,0)--(-1,-1.3);
\draw[darkgreen] (-1.4,-1.3)--(-1,-1.3)
                (1.4,-1.1)--(1.8,-1)
                (-1,-1.3)--(-0.3,-1.8)--(0.3,-1.5)
                (1.4,-1.1)--(1,-1.6)--(0.3,-1.5);
\draw[color=darkgreen, dash pattern=on 3pt off 2pt] 
(0,0)--(0.6,-0.9)--(1.8,-1)
(0,0)--(-0.3,-0.9)--(-1.4,-1.3)
(0.6,-0.9)--(-0.1,-1.4)--(-0.3,-0.9);
\filldraw[darkgreen] (0,0) circle (2pt) 
(0.6,-0.9)circle (2pt)  (-0.1,-1.4)circle (2pt)  (-0.3,-0.9)circle (2pt) 
(1.8,-1)circle (2pt)  (-1.4,-1.3)circle (2pt) 
(-1,-1.3)circle (2pt) (1.4,-1.1)circle (2pt)  (-0.3,-1.8)circle (2pt) 
(0.3,-1.5)circle (2pt)  (1,-1.6)circle (2pt) ;
\end{tric}
}

The cone point dual to the monogon face has cone angle $\cfrac{\pi}{2}$, and thus curvature there $\cfrac{3\pi}{2}$. 
\begin{align*}
    &\Monogon \xrightarrow[]{\text{Dualizing}} \DualMonogon  
    \qquad \qquad
    \ConeMonogon \xrightarrow[]{\text{Gluing}}  \ConeMonogona \\
\end{align*}
The cone point dual to the bigon face has cone angle $\pi$, and curvature $\pi$. 
\begin{align*}   
&\Bigon \xrightarrow[]{\text{Dualizing}} \DualBigon  
    \qquad \qquad
    \ConeBigon \xrightarrow[]{\text{Gluing}}  \ConeBigona
\end{align*}
The cone point dual to the pentagon face has cone angle $\cfrac{5\pi}{2}$, and curvature $-\cfrac{\pi}{2}$. 
\begin{align*}   
&\Pentagon \xrightarrow[]{\text{Dualizing}} \DualPentagon  
    \qquad 
    \ConePentagon \xrightarrow[]{\text{Gluing}}  \ConePentagona
\end{align*} \\

Let $E$ be the quadrangulation of $\mathbb{C}$ by the unit square grid.

Suppose $\mathcal{W}$ has $2$  noncontractible monogon faces, $1$ noncontractible bigon face, and contractible square faces. 
Then $ \overline{\mathcal{W}}$ has three cone points with positive curvatures, adding up to $4\pi$. 
Denote singular vertices of $\rm{Q}$ by $v_i$, $i=1,2,3$.  Suppose $v_1$ and $v_2$ have curvature $\cfrac{3\pi}{2}$, and $v_3$ has curvature $\pi$. 
Consider the shortest geodesics from $v_3$ to $v_1$ and $v_2$. By Lemma \ref{NoIntersect}, we know that the geodesics are non-intersecting except for at $v_3$. Since $\bar{\mathcal{W}}$ is a quadrangulation of the sphere, we can cut the sphere open along the geodesics. 
What remains unfolds to a quadrilateral with two opposite vertices $v_1,v_2$ of angle $\pi/2$, each with two adjacent sides of the same length. This is a square.
So ${\mathcal{W}}$ is given by $W_\xi$ in this case.

Suppose $\mathcal{W}$ has $3$  noncontractible monogon faces, $1$ contractible pentagon face, and contractible square faces. $\overline{\mathcal{W}}$ has four cone points, $3$ of which have positive curvatures. 
Denote singular vertices of $\rm{Q}$ by $v_i$, $i=1,2,3,4$. Suppose $v_1$, $v_2$, and $v_3$ have curvature $\cfrac{3\pi}{2}$, and $v_4$ has curvature $-\pi/2$.

Consider the shortest geodesics from $v_4$ to $v_1, v_2,$ and $v_3$. By Lemma \ref{NoIntersect}, we know that the geodesics are non-intersecting except for at $v_4$. Since $\bar{\mathcal{W}}$ is a quadrangulation of the sphere, we can cut the sphere open along the geodesics. Unfolding, we get a hexagon of type $\xi$ above. 
So ${\mathcal{W}}$ is given by $W_\xi$ in this case. 
\end{proof}

\section{$Sp(2n)$ webs}\label{Sp2nscn}

\subsection{Definition of $\Sp(2n)$ web category}

\def\Vertc
{\begin{tric}
\draw          (0,0)--(90:1.1)    
                   node[right,pos=0.4,black,scale=0.7]{$\mathrm{k+1}$}
                (0,0)--(210:1.1) 
                node[below,pos=0.4,black,scale=0.7]{$\mathrm{1}$}
                (0,0)--(330:1.1)
                node[below,pos=0.4,black,scale=0.7]{$\mathrm{k}$};
\draw[scale=1.2]
      (90:1)node[black,anchor=south,scale=0.7]{$\ObjKP$}
      (210:1)node[black,anchor=north,scale=0.7]{$\ObjOne$}
      (330:1)node[black,anchor=north,scale=0.7]{$\ObjK$}; 
\end{tric}
}

\def\Vertd
{\begin{tric}
\draw  (0,0)--(90:1.1)    
        node[right,pos=0.4,black,scale=0.7]{$\mathrm{k+1}$}
                (0,0)--(210:1.1) 
                node[below,pos=0.4,black,scale=0.7]{$\mathrm{k}$}
                (0,0)--(330:1.1)
                node[below,pos=0.4,black,scale=0.7]{$\mathrm{1}$};
\draw [scale=1.2]
      (90:1)node[black,anchor=south,scale=0.7]{$\ObjKP$}
      (210:1)node[black,anchor=north,scale=0.7]{$\ObjK$}
      (330:1)node[black,anchor=north,scale=0.7]{$\ObjOne$}; 
\end{tric}
}

\def\Skeinaone
{\begin{tric}
\draw (0.7,0) circle (0.7);
\draw (0,0)node[left,black,scale=0.7]{$\mathrm{1}$}; 
\end{tric}
}

\def\Skeingg
{\begin{tric}
\draw[scale=0.8] (0,-1)--(0,0) (0,0)..controls(0.7,0.5)and(0.7,1.5)..(0,1.5)..controls(-0.7,1.5)and(-0.7,0.5)..(0,0);
\draw (0,-0.35)node[left,black,scale=0.7]{$\mathrm{2}$};
\draw (-0.4,0.7)node[left,black,scale=0.7]{$\mathrm{1}$};
\end{tric}
}

\def\Skeinbigon
{\begin{tric}
\draw[scale= 0.8] (0,0.7)--(0,1.5) (0,-0.7)--(0,-1.5)
        (0,1.1)node[above,black,left,scale=0.7]{$\mathrm{k}$}
        (0,-1.1)node[below,black,left,scale=0.7]{$\mathrm{k}$}
      (0,0.7)..controls(-0.5,0.7)and(-0.5,-0.7)..(0,-0.7) node[left,midway,black,scale=0.7]{$\mathrm{1}$} (0,0.7)..controls(0.5,0.7)and(0.5,-0.7)..(0,-0.7)
      node[right,midway,black,scale=0.7]{$\mathrm{k-1}$} ;
\end{tric}
}

\def\Skeinbigona
{\begin{tric}
\draw [scale=0.8] (0,1.5)--(0,-1.5)  node[right,pos=0.8,black,scale=0.7]{$\mathrm{k}$} ;
\end{tric}
}

\def\HigherColor
{\begin{tric}
\draw [scale=0.8] (0,1.5)--(0,-1.5)  node[right,pos=0.8,black,scale=0.7]{$\mathrm{n+1}$} ;
\end{tric}
}

\def\SkeinIH
{\begin{tric}
\draw[scale=0.5] (-2,-1.732)--(-1,0)--(-2,1.732) (2,1.732)--(1,0)--(2,-1.732) 
(-1.6,-1)node[left,black,scale=0.7]{$\mathrm{k}$}
(-1.6,1)node[left,black,scale=0.7]{$\mathrm{1}$}
(1.6,1)node[right,black,scale=0.7]{$\mathrm{1}$}
(1.6,-1) node[right,black,scale=0.7]{$\mathrm{k}$}
(-1,0)--(1,0)node[below,midway,black,scale=0.7]{$\mathrm{k+1}$};
\end{tric}
}

\def\SkeinIHa
{\begin{tric}
\draw [scale=0.5]  (-1.74,2)--(0,1)--(1.74,2) 
(1.74,-3)--(0,0) node[right,pos=0.7,black,scale=0.7]{$\mathrm{1}$}
(0,0)--(-1.74,-3) node[left,pos=0.3,black,scale=0.7]{$\mathrm{1}$}
(0,0)--(0,1) node[left,midway,black,scale=0.7]{$\mathrm{2}$}
(-1.65,-2.5) node[left,black,scale=0.7]{$\mathrm{k}$}
(1.65,-2.5) node[right,black,scale=0.7]{$\mathrm{k}$}
(-1,1.5) node[above,black,scale=0.7]{$\mathrm{1}$}
(1,1.5) node[above,black,scale=0.7]{$\mathrm{1}$}
(1.16,-2)--(-1.16,-2)node[below,midway,black,scale=0.7]{$\mathrm{k-1}$};
\end{tric}
}

\def\SkeinIHb
{\begin{tric}
\draw[scale=0.5] (-2,-1.732)--(-1,0)--(-2,1.732) (2,1.732)--(1,0)--(2,-1.732) 
(-1.6,-1)node[left,black,scale=0.7]{$\mathrm{k}$}
(-1.6,1)node[left,black,scale=0.7]{$\mathrm{1}$}
(1.6,1)node[right,black,scale=0.7]{$\mathrm{1}$}
(1.6,-1) node[right,black,scale=0.7]{$\mathrm{k}$}
(-1,0)--(1,0)node[below,midway,black,scale=0.7]{$\mathrm{k-1}$};
\end{tric}
}

\def\SkeinIHc
{\begin{tric}
\draw  [scale=0.5] 
(-2,2)..controls(-1,1)and(1,1)..(2,2) node[above,black,midway,scale=0.7]{$\mathrm{1}$}
(2,-2)..controls(1,-1)and(-1,-1)..(-2,-2) node[below,black,midway,scale=0.7]{$\mathrm{k}$};
\end{tric}
}

\def\Skeinassocia
{\begin{tric}
\draw[scale=0.8] (0,0)..controls(0,0.5)and(0.2,0.7)..(0.5,1)
      (1,0)..controls(1,0.5)and(0.8,0.7)..(0.5,1)
      (0.5,1)..controls(0.5,1.5)and(0.3,1.7)..(0,2) 
              node[right,black,midway, scale=0.7]{$\mathrm{k+1}$}
      (-1,0)..controls(-1,1)and(-0.6,1.5)..(0,2)
      (0,2)--(0,3)
      (0.1,0.5)node[left,black,scale=0.7]{$\mathrm{k}$}
      (0.9,0.5)node[right,black,scale=0.7]{$\mathrm{1}$}
      (-0.9,0.5)node[left,black,scale=0.7]{$\mathrm{1}$}
      (0,2.5)node[right,black,scale=0.7]{$\mathrm{k+2}$};
\end{tric}
}

\def\Skeinassoci
{\begin{tric}
\draw[scale=0.8] (0,0)..controls(0,0.5)and(-0.2,0.7)..(-0.5,1)
      (-1,0)..controls(-1,0.5)and(-0.8,0.7)..(-0.5,1)
      (-0.5,1)..controls(-0.5,1.5)and(-0.3,1.7)..(0,2) 
              node[left,black,midway, scale=0.7]{$\mathrm{k+1}$}
      (1,0)..controls(1,1)and(0.6,1.5)..(0,2)
      (0,2)--(0,3)
      (-0.1,0.5)node[right,black,scale=0.7]{$\mathrm{k}$}
      (-0.9,0.5)node[left,black,scale=0.7]{$\mathrm{1}$}
      (0.9,0.5)node[right,black,scale=0.7]{$\mathrm{1}$}
      (0,2.5)node[right,black,scale=0.7]{$\mathrm{k+2}$};
\end{tric}
}

\def\klabelCap
{\begin{tric}
\draw (0,0)..controls(0,1)and(1,1)..(1,0)node[above,midway,black,scale=0.7]{$\mathrm{k}$}; 
\draw (0,0) node[black,below,scale=0.7]{$\ObjK$}
      (1,0) node[black,below,scale=0.7]{$\ObjK$};
\end{tric}
}

\def\klabelCup
{\begin{tric}
\draw (0,0)..controls(0,-1)and(1,-1)..(1,0)node[below,midway,black,scale=0.7]{$\mathrm{k}$};
\draw (0,0) node[black,above,scale=0.7]{$\ObjK$}
      (1,0) node[black,above,scale=0.7]{$\ObjK$};
\end{tric}
}

\def\KZigZag
{\begin{tric}
\draw [scale=0.7](0,0)..controls(0,1)and(1,1)..(1,0);
\draw [scale=0.7](1,0)..controls(1,-1)and(2,-1)..(2,0);
\draw [scale=0.7](0,0)--(0,-2)node[left,black,midway,scale=0.7]{$\mathrm{k}$}
                 (2,0)--(2,2);
\end{tric}
}

\def\KZigZaga
{\begin{tric}
\draw [scale=0.7](0,-2)--(0,2)node[left,black,pos=0.3,scale=0.7]{$\mathrm{k}$};
\end{tric}}

\def\KZigZagb
{\begin{tric}
\draw [scale=0.7](0,0)..controls(0,1)and(-1,1)..(-1,0);
\draw [scale=0.7](-1,0)..controls(-1,-1)and(-2,-1)..(-2,0);
\draw [scale=0.7](0,0)--(0,-2)node[left,black,midway,scale=0.7]{$\mathrm{k}$}
(-2,0)--(-2,2);
\end{tric}
}

\def\KZigZagVertb
{\begin{tric}
\draw[scale=0.7] (0,0)--(0,1)..controls(0,1.5)and(0.5,2)..(1,2)..controls(1.5,2)and(2,1.5)..(2,0)--(2,-2);
\draw (0,0.5)node[right,black,scale=0.7]{$\mathrm{k+1}$};
\draw[scale=0.7] (0,0)..controls(0.3,-0.3)and(0.5,-0.7)..(0.5,-1)
                  ..controls(0.5,-1.5)and(-0.3,-2)..(-1,-2)
                  ..controls(-1.7,-2)and(-2.5,-1.5)..(-2.5,0)--(-2.5,2)
                  
                (0,0)--(-0.5,-0.5)..controls(-0.7,-0.7)and(-0.8,-0.8)..(-1,-0.8)..
                                  controls(-1.5,-0.8)and(-1.8,0)..(-1.8,1)--(-1.8,2);
\draw[scale=0.7] (-0.65,-0.6) node[above,black,scale=0.7]{$\mathrm{1}$};
\draw[scale=0.7] (0.6,-0.65) node[above,black,scale=0.7]{$\mathrm{k}$};
\end{tric}
}

\def\KZigZagVerta
{\begin{tric}
\draw[scale=0.7] (0,0)--(0,-2) node[left,black,midway,scale=0.7]{$\mathrm{k+1}$} ;
\draw[scale=0.7] (0,0)--(0.3,0.3)..controls(0.5,0.5)and(0.7,1)..(0.7,2) node[right,black,midway,scale=0.7]{$\mathrm{1}$}
(0,0)--(-0.3,0.3)..controls(-0.5,0.5)and(-0.7,1)..(-0.7,2) node[left,black,midway,scale=0.7]{$\mathrm{k}$};
\end{tric}
}

\def\KZigZagVert
{\begin{tric}
\draw[scale=0.7] (0,0)--(0,1)..controls(0,1.5)and(-0.5,2)..(-1,2)..controls(-1.5,2)and(-2,1.5)..(-2,0)--(-2,-2);
\draw[scale=0.7] (0,0.7) node[left,black,scale=0.7]{$\mathrm{k+1}$};
\draw[scale=0.7] (0,0)..controls(-0.3,-0.3)and(-0.5,-0.7)..(-0.5,-1)
                  ..controls(-0.5,-1.5)and(0.3,-2)..(1,-2)
                  ..controls(1.7,-2)and(2.5,-1.5)..(2.5,0)--(2.5,2)
                  
                (0,0)--(0.5,-0.5)..controls(0.7,-0.7)and(0.8,-0.8)..(1,-0.8)..
                                  controls(1.5,-0.8)and(1.8,0)..(1.8,1)--(1.8,2);
\draw[scale=0.7] (-0.6,-0.65) node[above,black,scale=0.7]{$\mathrm{1}$};
\draw[scale=0.7] (0.7,-0.6) node[above,black,scale=0.7]{$\mathrm{k}$};
\end{tric}
}

\def\KoZigZagVertb
{\begin{tric}
\draw[scale=0.7] (0,0)--(0,1)..controls(0,1.5)and(0.5,2)..(1,2)..controls(1.5,2)and(2,1.5)..(2,0)--(2,-2);
\draw (0,0.5)node[right,black,scale=0.7]{$\mathrm{k+1}$};
\draw[scale=0.7] (0,0)..controls(0.3,-0.3)and(0.5,-0.7)..(0.5,-1)
                  ..controls(0.5,-1.5)and(-0.3,-2)..(-1,-2)
                  ..controls(-1.7,-2)and(-2.5,-1.5)..(-2.5,0)--(-2.5,2)
                  
                (0,0)--(-0.5,-0.5)..controls(-0.7,-0.7)and(-0.8,-0.8)..(-1,-0.8)..
                                  controls(-1.5,-0.8)and(-1.8,0)..(-1.8,1)--(-1.8,2);
\draw[scale=0.7] (-0.65,-0.6) node[above,black,scale=0.7]{$\mathrm{k}$};
\draw[scale=0.7] (0.6,-0.65) node[above,black,scale=0.7]{$\mathrm{1}$};
\end{tric}
}

\def\KoZigZagVerta
{\begin{tric}
\draw[scale=0.7] (0,0)--(0,-2) node[left,black,midway,scale=0.7]{$\mathrm{k+1}$} ;
\draw[scale=0.7] (0,0)--(0.3,0.3)..controls(0.5,0.5)and(0.7,1)..(0.7,2) node[right,black,midway,scale=0.7]{$\mathrm{k}$}
(0,0)--(-0.3,0.3)..controls(-0.5,0.5)and(-0.7,1)..(-0.7,2) node[left,black,midway,scale=0.7]{$\mathrm{1}$};
\end{tric}
}

\def\KoZigZagVert
{\begin{tric}
\draw[scale=0.7] (0,0)--(0,1)..controls(0,1.5)and(-0.5,2)..(-1,2)..controls(-1.5,2)and(-2,1.5)..(-2,0)--(-2,-2);
\draw[scale=0.7] (0,0.7) node[left,black,scale=0.7]{$\mathrm{k+1}$};
\draw[scale=0.7] (0,0)..controls(-0.3,-0.3)and(-0.5,-0.7)..(-0.5,-1)
                  ..controls(-0.5,-1.5)and(0.3,-2)..(1,-2)
                  ..controls(1.7,-2)and(2.5,-1.5)..(2.5,0)--(2.5,2)
                  
                (0,0)--(0.5,-0.5)..controls(0.7,-0.7)and(0.8,-0.8)..(1,-0.8)..
                                  controls(1.5,-0.8)and(1.8,0)..(1.8,1)--(1.8,2);
\draw[scale=0.7] (-0.6,-0.65) node[above,black,scale=0.7]{$\mathrm{k}$};
\draw[scale=0.7] (0.7,-0.6) node[above,black,scale=0.7]{$\mathrm{1}$};
\end{tric}
}

\begin{defn}[{\cite[Definition 1.1]{bodish2021type}}] \label{CnSpider}
The category of $\rm{Sp(2n)}$ webs, denoted by $\textbf{Web}(\rm{Sp(2n)})$, is a monoidal category, whose objects are generated by objects $\{\ObjOne,\ObjTwo,\dots,\ObjN\}$, and whose morphisms are generated by the following graphs, where the edge labeling 
(not to be confused with the edge multiplicity, which we denote in bold) agrees with the labeling of the corresponding boundary points: 
 \begin{align*}
 & \klabelCap \in {\Hom}(\ObjK\otimes\ObjK,\mathbb{R}) ,\ \
 \klabelCup \in {\Hom}(\mathbb{R},\ObjK\otimes\ObjK),  \\
    & \Vertc \in {\Hom}(\ObjOne\otimes \ObjK , \ObjKP)\ \ ,  \ \
     \Vertd \in {\Hom}(\ObjK \otimes \ObjOne, \ObjKP) \ 
 \end{align*}
modulo the tensor-ideal generated by the following skein relations.

\begin{align}
&\KZigZag \quad = \quad \KZigZaga \quad = \quad \KZigZagb ,  \notag \\
&
\KZigZagVerta  \quad := \quad  \KZigZagVert \quad = \quad \KZigZagVertb , \notag\\
&
\KoZigZagVerta  \quad := \quad  \KoZigZagVert \quad = \quad \KoZigZagVertb , \notag
\end{align}

\begin{align} 
& \Skeinaone = -2n, \qquad 
\Skeingg= 0  ,\qquad
\Skeinbigon  \ =  k \quad \Skeinbigona \ \  , \notag \\
&\Skeinassoci =\Skeinassocia \ , \qquad \HigherColor=0, \notag  
\\
& \SkeinIH =  \SkeinIHa   -\frac{n-k}{n-k+1}  \SkeinIHb
  +  \frac{n-k}{n} \SkeinIHc.   \label{SpHighIH}
\end{align}
\end{defn}

\begin{thm} [{\cite[Theorem 1.4]{bodish2021type}}]
 There is an equivalence of monoidal categories 
 \begin{align*}
    \Phi: \textbf{Web}(Sp(2n)) &\rightarrow \textbf{Fund}(Sp(2n)) \\
    \ObjK &\mapsto V_{\varpi_k}
 \end{align*}
 where $V_{\varpi_k}$ is the $k$-th fundamental representation of $\Sp(2n)$. 
\end{thm}

\subsection{$2n$-valent vertex in the $\Sp(2n)$ Web Category}

\def\OneOneTwoOneOne
{\begin{tric}
\draw [scale=0.5]  (-1.732,2)--(0,1) 
                   (0,1)--(1.732,2) 
                   (1.732,-2)--(0,-1)
                   (0,-1)--(-1.732,-2); 
\draw [scale=0.5] (0,-1)--(0,1)node[left,black,midway,scale=0.7]{$\mathrm{2}$};
\end{tric}
}

\def\HOneOneTwoOneOne
{\begin{tric}
\draw[scale=0.5] (-2,-1.732)--(-1,0)--(-2,1.732) (2,1.732)--(1,0)--(2,-1.732) 
(-1,0)--(1,0)node[below,midway,black,scale=0.7]{$\mathrm{2}$};
\end{tric}
}

\begin{definition} \label{defSp2ncrossing}
    Define the crossing  $ \beta_{\ObjOne,\ObjOne} \in {\Hom}(\ObjOne\otimes\ObjOne,\ObjOne\otimes\ObjOne)$ as 
     \begin{equation}
         \beta_{\ObjOne,\ObjOne}= \TensorSwap:=\Skeinec \   +\frac{1}{n} \ \Skeined - \OneOneTwoOneOne . \label{GeneralCross}
     \end{equation}
\end{definition}

Notice that by \eqref{SpHighIH},
\[ \TensorSwapa= \TensorSwapb= \Skeined \  +\frac{1}{n} \  \Skeinec - \HOneOneTwoOneOne =  \TensorSwap, \]
so there is no distinction between an overcrossing and an undercrossing.

\begin{proposition} [{\cite[Porism 5.4]{bodish2021type}}]  We have
    \begin{align*}
    \Phi(\beta_{\ObjOne,\ObjOne}): V_{\varpi_1} \otimes V_{\varpi_1} &\rightarrow  V_{\varpi_1} \otimes V_{\varpi_1} \\
        u \otimes v &\mapsto v \otimes u 
    \end{align*}
\end{proposition}

\def\ReideA
{\begin{tric}
\draw  [scale=0.8] (-1.5,1.5)..controls(0,-1.5)and(1.5,-1.5)..(1.5,0);
\draw[scale=0.8,double=darkblue,double distance=0.8pt,ultra thick,white,line width=3pt] (-1.5,-1.5)..controls(0,1.5)and(1.5,1.5)..(1.5,0);
\end{tric}
}

\def\ReideAa
{\begin{tric}
\draw  [scale=0.8] (0,1.5)--(0,-1.5);
\end{tric}
}

\def\ReideB
{\begin{tric}
\draw [scale=0.4] (1.5,1.5)..controls(1.5,0)and(-1.5,0) ..(-1.5,-1.5) ;
\draw[scale=0.4,double distance=0.8pt, double=darkblue,ultra thick,white,line width=3pt] (-1.5,1.5)..controls(-1.5,0)and(1.5,0) ..(1.5,-1.5) ;

\draw[scale=0.4] (-1.5,-1.5)..controls(-1.5,-3)and(1.5,-3) ..(1.5,-4.5) ;
\draw[scale=0.4, double distance=0.8pt,double=darkblue,ultra thick,white,line width=3pt] (1.5,-1.5)..controls(1.5,-3)and(-1.5,-3) ..(-1.5,-4.5);
\end{tric}
}

\def\ReideBa
{\begin{tric}
\draw  [scale=0.8] (0,1.5)--(0,-1.5) (2,1.5)--(2,-1.5);
\end{tric}
}

\def\ReidemCa
{\begin{tric}
\draw [scale=0.6] (-2,2)--(2,-2);
\draw [scale=0.6,double=darkblue,semithick,double distance=0.8pt,white,line width=3pt] (0,-2)..controls(-2,-1)and(-2,1)..(0,2);
\draw [scale=0.6,double=darkblue,double distance=0.8pt,semithick,white,line width=3pt](-2,-2)--(2,2) ;
\end{tric}
}

\def\ReidemCb
{\begin{tric}
\draw [scale=0.6] (-2,2)--(2,-2);
\draw [scale=0.6,double=darkblue,double distance=0.8pt,white,line width=3pt] (0,-2)..controls(2,-1)and(2,1)..(0,2);
\draw [scale=0.6,double=darkblue,double distance=0.8pt,white,line width=3pt](-2,-2)--(2,2) ;
\end{tric}
}

\begin{proposition} [{\cite[Proposition 5.7]{bodish2021type}}] \label{ReideforCross}
The crossing $ \beta_{\ObjOne,\ObjOne}$ satisfies the following Reidemeister moves: 
\begin{align*}
    \ReideA=-\quad \ReideAa \qquad  \quad  
    \ReideB =\ \ReideBa \qquad \quad  
    \ReidemCa = \ReidemCb
\end{align*}
\end{proposition}

\def\NTetraIsDet
{\begin{tric}
\draw [scale=0.7] 
      (0,0)..controls(1.7,-0.6)and(2,-1.4)..(2,-2)
      (0,0)..controls(0.8,-0.6)and(1,-1.4)..(1,-2)
      (0,0)..controls(-0.8,-0.6)and(-1,-1.4)..(-1,-2)
      (0,0)..controls(-1.7,-0.6)and(-2,-1.4)..(-2,-2);
\filldraw[black,scale=0.7] (-0.4,-1.8) circle (1pt) 
(0,-1.8) circle (1pt) (0.4,-1.8) circle (1pt) ;
\filldraw[black] (0,-0.05) circle (3pt) ;
\end{tric}
}

\def\HexaIsDet
{\begin{tric}
\draw [scale=0.7] 
      (0,0)..controls(1.7,-0.6)and(2,-1.4)..(2,-2)
      (0,0)..controls(0.8,-0.6)and(1.2,-1.4)..(1.2,-2)
      (0,0)..controls(-0.8,-0.6)and(-1.2,-1.4)..(-1.2,-2)
      (0,0)..controls(-1.7,-0.6)and(-2,-1.4)..(-2,-2)
      (0,0)..controls(0.25,-0.6)and(0.4,-1.4)..(0.4,-2)
      (0,0)..controls(-0.25,-0.6)and(-0.4,-1.4)..(-0.4,-2);
\filldraw[black] (0,-0.05) circle (3pt) ;
\end{tric}
}

\def\HexaBraida
{\begin{tric}
\draw[scale=0.7]
     (1.5,-2)..controls(1.5,-1)and(1.2,-0.5)..(1,-0.5)
     (0.5,-2)..controls(0.5,-1)and(0.8,-0.5)..(1,-0.5)
     (0,-2)..controls(0,-1)and(-0.3,-0.5)..(-0.5,-0.5)
     (-1,-2)..controls(-1,-1)and(-0.7,-0.5)..(-0.5,-0.5) 
     (3,-2)..controls(3,-1)and(2.7,-0.5)..(2.5,-0.5)
     (2,-2)..controls(2,-1)and(2.3,-0.5)..(2.5,-0.5);
\end{tric}
}

\def\HexaBraidb
{\begin{tric}
\draw[scale=0.7]
     (2.7,-2)..controls(2.4,-1)and(1.5,-0.5)..(0.75,-0.5)..controls(0,-0.5)and(-0.9,-1)..(-1.2,-2)
     (0.5,-2)..controls(0.3,-1)and(-0.3,-1)..(-0.5,-2) 
     (2,-2)..controls(1.8,-1)and(1.2,-1)..(1,-2);
\end{tric}
}

\def\HexaBraidc
{\begin{tric}
\draw[scale=0.7]
     (2.7,-2)..controls(2.5,-1.3)and(2,-1)..(1.5,-1)..controls(1,-1)and(0.5,-1.3)..(0.3,-2)

      (3.4,-2)..controls(3.2,-1)and(2.3,-0.5)..(1.5,-0.5)..controls(0.7,-0.5)and(-0.2,-1)..(-0.4,-2)
    
     (2,-2)..controls(1.8,-1.3)and(1.2,-1.3)..(1,-2);
\end{tric}
}

\def\HexaBraidd
{\begin{tric}
\draw[scale=0.7]
     (2.7,-2)..controls(2.4,-1)and(2,-0.5)..(1.5,-0.5)..controls(1,-0.5)and(0.6,-1)..(0.3,-2)
     
     (-0.3,-2)..controls(-0.5,-1)and(-1.1,-1)..(-1.3,-2) 
     (2,-2)..controls(1.8,-1)and(1.2,-1)..(1,-2);
\end{tric}
}

\def\HexaBraide
{\begin{tric}
\draw[scale=0.7]
     (2.7,-2)..controls(2.4,-1)and(2,-0.5)..(1.5,-0.5)..controls(1,-0.5)and(0.6,-1)..(0.3,-2)
     
     (3.3,-2)..controls(3.5,-1)and(4.1,-1)..(4.3,-2) 
     (2,-2)..controls(1.8,-1)and(1.2,-1)..(1,-2);
\end{tric}
}

\def\HexaBraidf
{\begin{tric}
\draw[scale=0.7]
       (2.7,-2)..controls(2.5,-1)and(1.2,-1)..(1,-2)     
      (3.4,-2)..controls(3.2,-1)and(2.3,-0.5)..(1.5,-0.5)..controls(0.7,-0.5)and(-0.2,-1)..(-0.4,-2) ;
      
 \draw [scale=0.7,double=darkblue,double distance=0.8pt,white,line width=3pt] 
      (2,-2)..controls(1.8,-1)and(0.5,-1)..(0.3,-2) ;     
\end{tric}
}

\def\HexaBraidg
{\begin{tric}
\draw[scale=0.7]
     (0,-2)..controls(0,-1)and(-0.3,-0.5)..(-0.5,-0.5)..controls(-0.7,-0.5)and(-1,-1)..
     (-1,-2)
     
     (3,-2)..controls(3,-1)and(2.7,-0.5)..(2.25,-0.5)..controls(1.8,-0.5)and(1.5,-1)
     ..(1.5,-2);

\draw [scale=0.7,double=darkblue,double distance=0.8pt,white,line width=3pt] 
   (2.2,-2)..controls(2.2,-1)and(1.9,-0.5)..(1.45,-0.5)..controls(1,-0.5)and(0.7,-1)..(0.7,-2);
\end{tric}
}

\def\HexaBraidh
{\begin{tric}
\draw[scale=0.7]
     (4.7,-2)..controls(4.7,-1)and(4.4,-0.5)..(4.2,-0.5)..controls(4,-0.5)and(3.7,-1)..
     (3.7,-2)
     
     (3,-2)..controls(3,-1)and(2.7,-0.5)..(2.25,-0.5)..controls(1.8,-0.5)and(1.5,-1)
     ..(1.5,-2);

\draw [scale=0.7,double=darkblue,double distance=0.8pt,white,line width=3pt] 
   (2.2,-2)..controls(2.2,-1)and(1.9,-0.5)..(1.45,-0.5)..controls(1,-0.5)and(0.7,-1)..(0.7,-2);
\end{tric}
}

\def\HexaBraidi
{\begin{tric}
\draw[scale=0.7]
     (3,-2)..controls(3,-1)and(2.7,-0.5)..(2.25,-0.5)
     (1.5,-2)..controls(1.5,-1)and(1.8,-0.5)..(2.25,-0.5)
     
       (0.5,-2)..controls(0.5,-1)and(0.2,-0.5)..(-0.25,-0.5)
     (-1,-2)..controls(-1,-1)and(-0.7,-0.5)..(-0.25,-0.5);
\draw [scale=0.7,double=darkblue,double distance=0.8pt,white,line width=3pt] 
      (2.2,-2)..controls(2,-1)and(1.6,-0.5)..(1,-0.5)
     (-0.2,-2)..controls(0,-1)and(0.4,-0.5)..(1,-0.5) ;
\end{tric}
}

\def\HexaBraidj
{\begin{tric}
\draw[scale=0.7]
     (3,-2)..controls(2.7,-1)and(2.3,-0.5)..(1.5,-0.5)..controls(0.7,-0.5)and(0.3,-1)..(0,-2)
     
     (2.3,-2)..controls(2,-0.8)and(1,-0.8)..(0.7,-2);

\draw [scale=0.7,double=darkblue,double distance=0.8pt,white,line width=3pt]     
     (1.5,-2)..controls(1.8,-1)and(2.2,-0.5)..(2.6,-0.5)..controls(3,-0.5)and(3.4,-1)..(3.7,-2);
\end{tric}
}

\def\HexaBraidk
{\begin{tric}
\draw[scale=0.7]
     (3,-2)..controls(2.7,-1)and(2.3,-0.5)..(1.5,-0.5)..controls(0.7,-0.5)and(0.3,-1)..(0,-2)
     
     (2.3,-2)..controls(2,-0.8)and(1,-0.8)..(0.7,-2);

\draw [scale=0.7,double=darkblue,double distance=0.8pt,white,line width=3pt]     
     (1.5,-2)..controls(1.2,-1)and(0.8,-0.5)..(0.4,-0.5)..controls(0,-0.5)and(-0.4,-1)..(-0.7,-2);
\end{tric}
}

\def\HexaBraidl
{\begin{tric}
\draw[scale=0.7]
      (3.2,-2)..controls(2.9,-1)and(2.5,-0.5)..(2.1,-0.5)..controls(1.7,-0.5)and(1.3,-1)..(1,-2);

\draw [scale=0.7,double=darkblue,double distance=0.8pt,white,line width=3pt]     
     (2.5,-2)..controls(2.3,-1)and(2,-0.5)..(1.25,-0.5)..controls(0.5,-0.5)and(0.2,-1)..(0,-2);

\draw [scale=0.7,double=darkblue,double distance=0.8pt,white,line width=3pt]     
     (1.5,-2)..controls(1.2,-1)and(0.8,-0.5)..(0.4,-0.5)..controls(0,-0.5)and(-0.4,-1)..(-0.7,-2);
     
\end{tric}
}

\begin{definition}\label{2nVertex}
    Define a degree $2n$ vertex  $\NTetraIsDet \in \Hom(\ObjOne^{\otimes 2n}, \mathbb{R} )$ by
    \[ \NTetraIsDet := \sum_{\alpha \in P_{2n}} (-1)^{K_\alpha+\frac{n(n+1)}{2}} M_\alpha ,\]
    where  $P_{2n}= \{((i_1,j_1),(i_2,j_2), ...  , (i_n,j_n)): i_k < j_k , \ i_k<i_{k+1}, \ 1\le i_k,j_k \le 2n \}$ and  $M_\alpha \in \Hom(\ObjOne^{\otimes 2n}, \mathbb{R} )$ such that: \begin{align*}
        \Phi(M_\alpha):\qquad \qquad \qquad 
        { V_{\varpi_1}}^{\otimes 2n} &\rightarrow \mathbb{R} \\
        v_1\otimes v_2 \otimes \cdot \cdot \cdot \otimes v_{2n} &\mapsto \prod_{k=1}^n ( v_{j_k}\cdot J v_{i_k} )
    \end{align*}
    $K_\alpha$ is the number of crossings in $M_\alpha$. 
\end{definition}

 \begin{example}\label{hexavalentex}
  When n=3, the hexavalent vertex in $\textbf{Web}(Sp(6))$ is given by the following.
  \begin{align*}
      &\HexaIsDet = \HexaBraida  + \HexaBraidb + \HexaBraidc + \HexaBraidd \\ 
      &+ \HexaBraide - \HexaBraidf - \HexaBraidg - \HexaBraidh + \HexaBraidi \\
      &+ \HexaBraidj + \HexaBraidk -  \HexaBraidl
  \end{align*}
  
\end{example}

Now we prove Theorem \ref{Sp2ndet}.
\label{thm2proof}
\begin{proof}[Proof of Theorem \ref{Sp2ndet}.]
$V_{\varpi_1}$ is the defining representation of $\Sp(2n)$, with basis $e_1,e_2,...,e_{2n}$.  
    First, we show that 
\be\label{phidet1}\Phi (\textbf{det})(e_1\otimes e_2 \otimes \cdot \cdot \cdot \otimes e_{2n})=1.\ee 
For $\alpha\in P_{2n}$, we know that 
    \[ \Phi(M_\alpha)(e_1\otimes e_2 \otimes \cdot \cdot \cdot \otimes e_{2n})=0, \] 
    unless $\alpha=\alpha_0:=((1,n+1),(2,n+2),...,(n,2n))$. We also know that
    \[K_{\alpha_0}=\sum_{k=1}^{n-1} k = \frac{n(n-1)}{2}. \]
    What's more, $e_{k+n} \cdot J e_{k}=-1,$ for $ 1\le k\le n$. So
\begin{align*}
    \Phi (\textbf{det})(e_1\otimes e_2 \otimes \cdot \cdot \cdot \otimes e_{2n})
    &=(-1)^{K_{\alpha_0}+\frac{n(n+1)}{2}}\Phi (M_{\alpha_0})(e_1\otimes e_2 \otimes \cdot \cdot \cdot \otimes e_{2n}) \\
    &=(-1)^{n^2} \prod_{k=1}^n(e_{k+n} \cdot J e_{k}) 
    =(-1)^{n(n+1)}=1 
\end{align*}
Second, we show that  
\be\label{phidet2}\Phi (\textbf{det})(v_1\otimes v_2 \otimes \cdot \cdot \cdot \otimes v_k \otimes v_{k+1} \otimes \cdot \cdot \cdot \otimes v_{2n})=-\Phi (\textbf{det})(v_1\otimes v_2 \otimes \cdot \cdot \cdot \otimes v_{k+1} \otimes v_k  \otimes \cdot \cdot \cdot \otimes v_{2n}).\ee
Denote $\sigma_{k,k+1}:={id_{\ObjOne}}^{\otimes(k-1)} \otimes \beta_{\ObjOne,\ObjOne} \otimes {id_{\ObjOne}}^{\otimes (n-k-1) } \in 
\Hom(\ObjOne^{\otimes 2n}, \ObjOne^{\otimes 2n})$.
On the one hand, 
\begin{align*}
    \Phi (\textbf{det} \circ \sigma_{k,k+1})(&v_1\otimes v_2 \otimes \cdot \cdot \cdot \otimes v_{k+1} \otimes v_k \otimes \cdot \cdot \cdot \otimes v_{2n})=\\
    &= (\Phi (\textbf{det}) \circ \Phi(\sigma_{k,k+1}))(v_1\otimes v_2 \otimes \cdot \cdot \cdot \otimes v_{k+1} \otimes v_k \otimes \cdot \cdot \cdot \otimes v_{2n})\\
    &=\Phi (\textbf{det}) (v_1\otimes v_2 \otimes \cdot \cdot \cdot \otimes v_k \otimes v_{k+1} \otimes \cdot \cdot \cdot \otimes v_{2n}).
\end{align*}
On the other hand, by Proposition \ref{ReideforCross}, 
\begin{align*}
    \textbf{det} \circ \sigma_{k,k+1} = \sum_{\alpha \in P_{2n}} (-1)^{K_\alpha+\frac{n(n+1)}{2}} (M_\alpha \circ \sigma_{k,k+1})
    =- \sum_{\tilde{\alpha} \in P_{2n}} (-1)^{K_{\tilde{\alpha}}+\frac{n(n+1)}{2}} M_{\tilde{\alpha}}.
\end{align*}
This tells us that $\textbf{det} \circ \sigma_{k,k+1}= - \textbf{det}$, which implies 
    $$\Phi (\textbf{det} \circ \sigma_{k,k+1})(v_1\otimes v_2 \otimes \cdot \cdot \cdot \otimes v_{k+1} \otimes v_k \otimes \cdot \cdot \cdot \otimes v_{2n})=-\Phi (\textbf{det})(v_1\otimes v_2 \otimes \cdot \cdot \cdot \otimes v_{k+1} \otimes v_k \otimes \cdot \cdot \cdot \otimes v_{2n}).$$ 
Combining (\ref{phidet1}) and (\ref{phidet2}) with linearity we have       
$\Phi (\textbf{det})( v_1\otimes v_2 \otimes \cdot \cdot \cdot \otimes v_{2n})= |v_1\wedge v_2 \wedge \cdot \cdot \cdot \wedge v_{2n}| $. 
\end{proof}

\subsection{$\Sp(2n)$ multiwebs on an annulus}

\def\Kloop{
\begin{trica}
   \draw [thick,black,fill=yellow!30] (0,0) circle (3cm);
   \draw [thick,black,fill=white] (0,0) circle (1cm);
   \draw [thick](0,0) circle (2cm);
      \filldraw[black] (0:2) circle (2pt) (180:2) circle (2pt);  
      \draw (0,2) node[above,black,scale=0.7]{$\mathbcal{k}$}
            (0,-2) node[below,black,scale=0.7]{$\mathbcal{2n-k}$}; 
\end{trica}
}

\def\loopcalc{
\begin{trics}
   \draw [thick,black,fill=yellow!30] (0,0) circle (3.3cm);
   \draw [thick,black,fill=white] (0,0) circle (0.6cm); 
      \draw (0,1.5) node[below,black,scale=0.7]{$\mathrm{k-1}$}
            (0,-1.5) node[below,black,scale=0.7]{$\mathrm{k}$}; 
      \draw    (2,0)..controls(1.5,1)and(0.7,1.5)..(0,1.5)..controls(-0.7,1.5)and(-1.5,1)..(-2,0);
      \draw    (2,0)..controls(1.5,-1)and(0.5,-1.5)..(0,-1.5)..controls(-0.5,-1.5)and(-1.5,-1)..(-2,0);

      \draw (-2,0)..controls(-2,2)and(0.5,2.5)..(0.8,2.7);
      \draw (-0.8,2.7)--(-0.2,2.5);
         \draw (2,0)..controls(2,1.5)and(0.5,2.1)..(0.2,2.3);
\end{trics}
}

\def\loopcalcA{
\begin{trics}
   \draw [thick,black,fill=yellow!30] (0,0) circle (3.3cm);
   \draw [thick,black,fill=white] (0,0) circle (0.6cm); 
      \draw (0,1.5) node[below,black,scale=0.7]{$\mathrm{k-1}$}
            (0,-1.5) node[below,black,scale=0.7]{$\mathrm{k}$}; 
      \draw    (2,0)..controls(1.5,1)and(0.7,1.5)..(0,1.5)..controls(-0.7,1.5)and(-1.5,1)..(-2,0);
      \draw    (2,0)..controls(1.5,-1)and(0.5,-1.5)..(0,-1.5)..controls(-0.5,-1.5)and(-1.5,-1)..(-2,0);
      
      \draw (-2,0)..controls(-2,1.5)and(-0.5,2.1)..(0,2.3);
         \draw (2,0)..controls(2,1.5)and(0.5,2.1)..(0,2.3);
    \draw (0,2.3)--(0,2.8) node[left,black,scale=0.7,midway]{$\mathrm{2}$};
    \draw (0,2.8)--(-0.8,3) (0,2.8)--(0.8,3);
\end{trics}
}

\def\loopcalcB{
\begin{trics}
   \draw [thick,black,fill=yellow!30] (0,0) circle (3.3cm);
   \draw [thick,black,fill=white] (0,0) circle (0.6cm); 
      \draw (0,1.5) node[below,black,scale=0.7]{$\mathrm{k-1}$}
            (0,-1.5) node[below,black,scale=0.7]{$\mathrm{k}$}; 
      \draw    (2,0)..controls(1.5,1)and(0.7,1.5)..(0,1.5)..controls(-0.7,1.5)and(-1.5,1)..(-2,0);
      \draw    (2,0)..controls(1.5,-1)and(0.5,-1.5)..(0,-1.5)..controls(-0.5,-1.5)and(-1.5,-1)..(-2,0);
      
      \draw (-2,0)..controls(-2,1.5)and(-0.5,2.1)..(-0.6,3);
         \draw (2,0)..controls(2,1.5)and(0.5,2.1)..(0.6,3);
\end{trics}
}

\def\loopcalcC{
\begin{trics}
   \draw [thick,black,fill=yellow!30] (0,0) circle (3.3cm);
   \draw [thick,black,fill=white] (0,0) circle (0.6cm); 
      \draw (0,1.5) node[below,black,scale=0.7]{$\mathrm{k-1}$}
            (0,-1.5) node[below,black,scale=0.7]{$\mathrm{k}$}; 
      \draw    (2,0)..controls(1.5,1)and(0.7,1.5)..(0,1.5)..controls(-0.7,1.5)and(-1.5,1)..(-2,0);
      \draw    (2,0)..controls(1.5,-1)and(0.5,-1.5)..(0,-1.5)..controls(-0.5,-1.5)and(-1.5,-1)..(-2,0);
      \draw (-0.7,2.9)..controls(-0.3,2.5)and(0.3,2.5)..(0.7,2.9);
      \draw    (2,0)..controls(2,1)and(1,2.4)..(0,2.4)..controls(-1,2.4)and(-2,1)..(-2,0);
\end{trics}
}

\def\loopcalcD{
\begin{trics}
   \draw [thick,black,fill=yellow!30] (0,0) circle (3.3cm);
   \draw [thick,black,fill=white] (0,0) circle (0.6cm); 
      \draw (0,1.5) node[below,black,scale=0.7]{$\mathrm{k+1}$}
            (0,-1.5) node[below,black,scale=0.7]{$\mathrm{k}$}; 
      \draw    (2,0)..controls(1.5,1)and(0.7,1.5)..(0,1.5)..controls(-0.7,1.5)and(-1.5,1)..(-2,0);
      \draw    (2,0)..controls(1.5,-1)and(0.5,-1.5)..(0,-1.5)..controls(-0.5,-1.5)and(-1.5,-1)..(-2,0);
      
      \draw (-2,0)..controls(-2,1.5)and(-0.5,2.1)..(-0.6,3);
         \draw (2,0)..controls(2,1.5)and(0.5,2.1)..(0.6,3);
\end{trics}
}

\def\loopcalcE{
\begin{trics}
   \draw [thick,black,fill=yellow!30] (0,0) circle (3.3cm);
   \draw [thick,black,fill=white] (0,0) circle (0.6cm); 
   \draw (-0.7,2.9)..controls(-0.3,2.5)and(0.3,2.5)..(0.7,2.9);
   \draw (0,0) circle (2);
   \draw  (0,-2) node[below,black,scale=0.7]{$\mathrm{k}$};
\end{trics}
}

\def\SReduceKnot{
\begin{trich}
   \draw [ultra thick,black,fill=yellow!30] (0,0) circle (5.2cm);
   \draw [ultra thick,black,fill=white] (0,0) circle (0.4cm);  
      \draw (0.8,0)..controls(0.8,-1)and(-0.8,-1)..(-0.8,0)
            (1.1,0)..controls(1.1,-1.5)and(-1.3,-1.5)..(-1.3,0)
            (1.7,0)..controls(1.7,-2.3)and(-2,-2.3)..(-2,0);
      \draw (-0.8,0)..controls(-0.8,1)and(1.1,1)..(1.1,0)
      (-1.3,0)..controls(-1.3,2)and(1.7,2)..(1.7,0)
      (-2,0)..controls(-2,3)and(2.5,3)..(2.5,0);

       \filldraw[black] (0,2.53) circle (1pt)
     (0,2.78) circle (1pt) (0,3.03)circle (1pt);

     \draw 
     (-4.5,0)..controls(-4.5,-5.8)and(4.5,-5.8)..(4.5,0);

      \draw  (4.5,0)..controls(4.5,4.5)and(-3.7,5)..(-3.7,0)
      (-3.7,0)..controls(-3.7,-4.7)and(3.7,-4.7)..(3.7,0);

      \draw (3.7,0)..controls(3.7,0.7)and(3.5,1.2)..(3.3,1.5)
            (2.5,0)..controls(2.5,-0.7)and(2.3,-1.2)..(2,-1.5);

     \filldraw[black] (2.8,0) circle (1pt) (3.05,0) circle (1pt)
     (3.3,0) circle (1pt);

  \draw [double=darkblue,double distance=0.8pt,ultra thick,yellow!30,line width=3pt] 
 (0.8,0)--(0.8,3.75)..controls(0.8,5)and(-4.5,4.5)..(-4.5,0);

\end{trich}
}

Consider link diagrams on an annulus (arising from knots and links embedded in a thickened annulus). Use $\mathcal{K}_k$ to denote a knot winding around the puncture of the annulus k times as the following:  
$$ \mathcal{K}_k : = \SReduceKnot.$$

\begin{lemma} \label{LinktoWinding}
When over and under crossings are identified, any link diagram on an annulus is isotopic to a disjoint union of contractible circles and elements in $\{ \mathcal{K}_k: k\ge 1 \}$. Suppose that there is a path from one boundary circle to the other, which intersects the link $k$ times. 
Then the link is a disjoint union of contractible circles and 
elements $\mathcal{K}_{k_i}, k_i \ge 1$, such that $\sum k_i \le k$.
\end{lemma}

\begin{proof}
An elementary application of the Reidemeister theorem. 
\end{proof}

\begin{lemma} \label{WindingToColor}
  
  $$   \loopcalc= - \loopcalcA  + \loopcalcB + \frac{1}{n}  \loopcalcC= $$
  $$ - \loopcalcD + \frac{1}{n-k+1} \loopcalcB + \frac{1}{n}  \loopcalcC + \frac{n-k}{n} \loopcalcE $$
\end{lemma}

\begin{proof}
 Resolve the crossing by Equation \eqref{GeneralCross} to deduce the first equation, then apply Equation \eqref{SpHighIH} to deduce the second equation. 
\end{proof}

\def\SReduceKnotA{
\begin{trich}
   \draw [ultra thick,black,fill=yellow!30] (0,0) circle (5.2cm);
   \draw [ultra thick,black,fill=white] (0,0) circle (0.4cm);  
     
      \draw (0.8,0)..controls(0.8,-1)and(-0.8,-1)..(-0.8,0)
            (0.8,0)..controls(1.1,-1.5)and(-1.3,-1.5)..(-1.3,0)
            (1.7,0)..controls(1.7,-2.3)and(-2,-2.3)..(-2,0)
             (-0.8,0)..controls(-0.8,0.8)and(0.5,0.8)..(0.8,0.5) ;
             
      \draw(0.75,0.2) node[right,scale=0.7,black] {$\mathrm{2}$};
            
      \draw 
      (-1.3,0)..controls(-1.3,1.8)and(1.7,1.8)..(1.7,0)
      (-2,0)..controls(-2,3)and(2.5,3)..(2.5,0);

\filldraw[black] (0,2.53) circle (1pt)
     (0,2.78) circle (1pt) (0,3.03)circle (1pt);
 \draw 
     (-4.5,0)..controls(-4.5,-5.8)and(4.5,-5.8)..(4.5,0);
 \draw  (4.5,0)..controls(4.5,4.5)and(-3.7,5)..(-3.7,0)
      (-3.7,0)..controls(-3.7,-4.7)and(3.7,-4.7)..(3.7,0);

 \draw (3.7,0)..controls(3.7,0.7)and(3.5,1.2)..(3.3,1.5)
            (2.5,0)..controls(2.5,-0.7)and(2.3,-1.2)..(2,-1.5);

     \filldraw[black] (2.8,0) circle (1pt) (3.05,0) circle (1pt)
     (3.3,0) circle (1pt);

\draw (0.8,0)--(0.8,0.9); 
\draw [double=darkblue,double distance=0.8pt,ultra thick,yellow!30,line width=3pt] 
 (0.8,0.8)--(0.8,3.75)..controls(0.8,5)and(-4.5,4.5)..(-4.5,0); 
     
\end{trich}
}

\def\SReduceKnotB{
\begin{trich}
   \draw [ultra thick,black,fill=yellow!30] (0,0) circle (5.2cm);
   \draw [ultra thick,black,fill=white] (0,0) circle (0.4cm);  
  
      \draw (0.8,0)..controls(0.8,-1)and(-0.8,-1)..(-0.8,0)
            (0.8,0)..controls(1.1,-1.5)and(-1.3,-1.5)..(-1.3,0)
            (1.7,0)..controls(1.7,-2.3)and(-2,-2.3)..(-2,0)
             (-0.8,0)..controls(-0.8,0.8)and(0.5,0.8)..(0.8,0.5) ;
             
      \draw(0.75,0.2) node[right,scale=0.7,black] {$\mathrm{2}$}
           (-0.70,0) node[left,scale=0.7,black] {$\mathrm{3}$};
            
      \draw
      (-1.3,0)..controls(-1.3,1)and(0.8,1)..(0.8,1.5) 
      (0.8,0.8)..controls(0.8,1.3)and(1.7,1.3)..(1.7,0)
      (-2,0)..controls(-2,3)and(2.5,3)..(2.5,0);

   \filldraw[black] (0,2.53) circle (1pt)
     (0,2.78) circle (1pt) (0,3.03)circle (1pt);

     \draw 
     (-4.5,0)..controls(-4.5,-5.8)and(4.5,-5.8)..(4.5,0);
      \draw  (4.5,0)..controls(4.5,4.5)and(-3.7,5)..(-3.7,0)
      (-3.7,0)..controls(-3.7,-4.7)and(3.7,-4.7)..(3.7,0);

      \draw (3.7,0)..controls(3.7,0.7)and(3.5,1.2)..(3.3,1.5)
            (2.5,0)..controls(2.5,-0.7)and(2.3,-1.2)..(2,-1.5);

     \filldraw[black] (2.8,0) circle (1pt) (3.05,0) circle (1pt)
     (3.3,0) circle (1pt);
\draw (0.8,0)--(0.8,0.8); 
\draw [double=darkblue,double distance=0.8pt,ultra thick,yellow!30,line width=3pt] 
 (0.8,1.5)--(0.8,3.75)..controls(0.8,5)and(-4.5,4.5)..(-4.5,0); 
\end{trich}
}

\def\SReduceKnotC{
\begin{trich}
   \draw [ultra thick,black,fill=yellow!30] (0,0) circle (5.2cm);
   \draw [ultra thick,black,fill=white] (0,0) circle (0.4cm);  
      
      \draw (0.8,0)..controls(0.8,-1)and(-0.8,-1)..(-0.8,0)
            (0.8,0)..controls(1.1,-1.5)and(-1.3,-1.5)..(-1.3,0)
            (1.7,0)..controls(1.7,-2.3)and(-2,-2.3)..(-2,0)
             (-0.8,0)..controls(-0.8,0.8)and(0.5,0.8)..(0.8,0.5) ;
             
      \draw(0.75,0.2) node[right,scale=0.7,black] {$\mathrm{4}$}
           (-0.7,0) node[left,scale=0.7,black] {$\mathrm{3}$};
            
      \draw 
      (-1.3,0)..controls(-1.3,1)and(0.8,1)..(0.8,1.5) 
      (0.8,0.8)..controls(0.8,1.3)and(1.7,1.3)..(1.7,0);

\draw (-2,0)..controls(-2,2)and(0.8,2)..(0.8,2.5)
       (0.8,1.8) ..controls(0.8,2.5) and(2.5,2.5)..(2.5,0);

   \filldraw[black] (0,2.53) circle (1pt)
     (0,2.78) circle (1pt) (0,3.03)circle (1pt);

     \draw 
     (-4.5,0)..controls(-4.5,-5.8)and(4.5,-5.8)..(4.5,0);

      \draw  (4.5,0)..controls(4.5,4.5)and(-3.7,5)..(-3.7,0)
      (-3.7,0)..controls(-3.7,-4.7)and(3.7,-4.7)..(3.7,0);

      \draw (3.7,0)..controls(3.7,0.7)and(3.5,1.2)..(3.3,1.5)
            (2.5,0)..controls(2.5,-0.7)and(2.3,-1.2)..(2,-1.5);

     \filldraw[black] (2.8,0) circle (1pt) (3.05,0) circle (1pt)
     (3.3,0) circle (1pt);
    
    \draw (0.8,0)--(0.8,0.8) (0.8,1.5)--(0.8,1.8) 
            (0.8,2.5)--(0.8,2.8); 
\draw [double=darkblue,double distance=0.8pt,ultra thick,yellow!30,line width=3pt] 
 (0.8,2.7)--(0.8,3.75)..controls(0.8,5)and(-4.5,4.5)..(-4.5,0); 
\end{trich}
}

\begin{thm}\label{Sp2nAnnulus}
Any $\Sp(2n)$ web on an annulus can be written as a linear combination of disjoint unions of elements in $\{ \mathcal{K}_k : 1\le k \le n \}$.
\end{thm}

\begin{proof}

\old{
The idea of the proof is as follows: 
First we show that any knot with more than $n$ crossings can be reduced to a linear combination of knots each with strictly fewer crossings.
Then we only need to consider knots with less or equal to $n$ crossings, and prove that there is a unitriangular change of basis from knots to loops.
}

First, for any $\Sp(2n)$ web $\mathcal{W}$, one can apply Relation \eqref{SpHighIH} to write $\mathcal{W}$ as a linear combination of webs with edges labeled by $1$ and $2$. Furthermore, edges labeled $2$ can be replaced by linear combinations of planar matchings and crossings by Relation \eqref{GeneralCross}. In conclusion, $\mathcal{W}$ is equal to a linear combination of webs with only edges 
labeled $1$ and crossings.
    
By Proposition \ref{ReideforCross}, an $\Sp(2n)$ web with edges labeled $1$ and crossings can be seen as a link diagram, where over and under crossings are identified. Then by Lemma \ref{LinktoWinding}, $\mathcal{W}$ is a linear combination of disjoint unions of contractible circles and elements in $\{ \mathcal{K}_k: k\ge 1 \}$.

Use $\mathbf{D}_{< m}$ to denote a linear combination of diagrams which are disjoint unions of contractible circles and elements in $\{\mathcal{K}_k: 1\le k \le m-1  \}$. We now show that for $m>n$, $\mathcal{K}_m$ can be written as $\mathbf{D}_{< m}$ by Definition \ref{CnSpider} and Lemma \ref{WindingToColor}. 
Apply Relation \eqref{GeneralCross} to resolve the crossing of $\mathcal{K}_m$ that is closest to the puncture of the annulus, we know that $\mathcal{K}_m$ can be written as the following: 
$$ \SReduceKnot=\SReduceKnotA +\mathbf{D}_{< m} $$

By Lemma \ref{WindingToColor}, we can further write the expression as: 
       
       \begin{align*}
           \SReduceKnotB +\mathbf{D}_{< m} 
           =\SReduceKnotC +\mathbf{D}_{< m} 
       \end{align*}

Since $m>n$, we can apply Lemma \ref{WindingToColor} consecutively $n$ times, and write $\mathcal{K}_m$ as a sum of $\mathbf{D}_{< m}$ and an $\Sp(2n)$ web with an edge labelled by $(n+1)$.

By Definition \ref{CnSpider}, we know that any $\Sp(2n)$ web with an edge labelled by $(n+1)$ is equal to $0$. So $\mathcal{K}_m$ is equal to $\mathbf{D}_{< m}$ when $m>n$. By induction, any $\Sp(2n)$ web $\mathcal{W}$ can be written as $\mathbf{D}_{< (n+1)}$. Then by Definition \ref{CnSpider}, a contractible circle can be evaluated as a scalar, so $\mathcal{W}$ can be written as a linear combination of disjoint unions of elements in $\{ \mathcal{K}_k : 1\le k \le n \}$.  
\end{proof}

\begin{definition} \label{defLoopVertN}
An $\Sp(2n)$ k-loop is defined as $\mathcal{L}_k=\Kloop$. 
\end{definition}

\begin{thm}
Different disjoint unions of elements in $\{\mathcal{L}_k: 1\le k \le n \}$ are linearly independent with each other. Furthermore, disjoint unions of elements in $\{ \mathcal{L}_k : 1\le k \le n \}$ form a basis for the space of $\Sp(2n)$ webs on an annulus.  
\end{thm}

\begin{proof}
Consider $\mathcal{L}_k$ for $1\le k\le n$ on an annulus with flat connection having monodromy $A \in Sp(2n)$. One can verify that $\Tr(\mathcal{L}_k)= \tr(\bigwedge^k A)$ by vector calculations. Since $\prod_{i=1}^{N} \tr(\bigwedge^{k_i} A)$ are linearly independent for different multi-sets $\{k_i: 1 \le i \le N\}$, we know that different disjoint unions of elements in $\{\mathcal{L}_k: 1\le k\le n \}$ are linearly independent.

On the other hand, $\# \{\mathcal{L}_k: 1\le k\le n \}= \# \{\mathcal{K}_k: 1\le k\le n \}=n$. 
We show that $\{\mathcal{L}_k: 1\le k\le n \}$ and $\{\mathcal{K}_k: 1\le k\le n \}$ are related by the following relations: 
\[ \mathcal{L}_k= \lambda_k\mathcal{K}_k+ \mathbf{D}_{<k} \]
where $\lambda_k \ne 0$ and, as in the previous proof, $\mathbf{D}_{<k}$ represents linear combinations of unions of $ \mathcal{K}_j$ for $j<k$.

By Theorem \ref{Sp2nAnnulus}, for $1\le k\le n$ we have $\mathcal{L}_k=\mathbf{D}_{<n+1}$. Then by the second half of Lemma \ref{LinktoWinding}, $ \mathcal{L}_k= \lambda_k\mathcal{K}_k+\mathbf{D}_{<k}.$

Now we show that $\lambda_k \ne 0$. This is by induction on $k$; the case $k=1$ follows by converting the degree-$2n$ vertices into 
linear combinations of crossings as in Example \ref{hexavalentex}. Suppose $\lambda_{k_0}=0$ and $\lambda_{\ell} \ne 0$ for $\ell<k_0$. Then $\mathcal{L}_{k_0}=\mathbf{D}_{<k_0}$. Since $\lambda_{\ell} \ne 0$ for $\ell<k_0$, we can inductively prove that 
$\mathcal{K}_\ell= \lambda_\ell^{-1}\mathcal{L}_\ell+ \mathbf{E}_{<\ell}$
where $\mathbf{E}_{<\ell}$ represents linear combinations of unions of $ \mathcal{L}_j$ for $j<\ell$. This gives a linear 
dependence between $\mathcal{L}_{k_0}$ and $\mathbf{E}_{<k_0}$, a contradiction. 
\end{proof}

So any $\Sp(2n)$ web on an annulus can be written as a unique linear combination of disjoint unions of elements in $\{ \mathcal{L}_k : 1\le k \le n \}$.  

\appendix

\section{Vertices and quantum determinants}

Web categories when $q\ne 1$ are equivalent to the fundamental representation categories of the corresponding quantum groups instead of the classical Lie groups.

\subsection{Degree-$n$ vertex in the $\SL(n)$ web category}\label{quantSLn}

In the $\SL(n)$ web category \cite{CKM} when $q\ne 1$, we can define the following degree-$n$ vertex, which corresponds to the quantum determinant \cite{QDetR}:

$$\NTetraIsDetA:=\BigIk,$$
where the trivalent vertices correspond to taking exterior product in the quantum exterior algebra, and the edge labelled $n$ corresponds to the isomorphism from the determinant representation to the trivial representation 
and therefore is a ``dangling'' edge which corresponds to a scalar output.

We know that 
$$
\QuantumDet=-q\QuantumDetA ,  1\le i < j\le n, 
 \QuantumDetB=0,  1\le i \le n.  
$$

In other words, for $1\le i < j\le n,$
$$
 y_1\wedge y_2\wedge...\wedge e_j \wedge e_i \wedge...\wedge y_n
= -q \cdot y_1\wedge y_2\wedge...\wedge e_i \wedge e_j \wedge...\wedge y_n,
$$
and
$$y_1\wedge y_2\wedge...\wedge e_i \wedge e_i \wedge...\wedge y_n= 0.$$

This implies that if $y_i=\sum_{j=1}^{n}a_{i,j}e_j$ then
\[ y_1\wedge y_2\wedge...\wedge y_n
 =\text{det}_q(A) e_1\wedge e_2\wedge...\wedge e_n,\]
where 
$$A=(a_{i,j}), \quad \text{det}_q(A)= \sum_{\sigma \in S_n} (-q)^{l(\sigma)} a_{1,\sigma(1)}a_{2,\sigma(2)}...a_{n,\sigma(n)}.$$

\subsection{Degree-$4$ vertex in the $\mathfrak{sp}_4$ web category}\label{quantSp4}

In the $\Sp(4)$ web category when $q\ne 1$, consider the following tetravalent vertex: 
\begin{align*}
   \tetravalent := \Skeineadouble\ \   - \  \frac{1}{[2]} \Skeinec \ \    =  \  \Skeinebdouble \ - \  \frac{1}{[2]}\Skeined  ,
\end{align*}
 where $[2]=q+q^{-1}$.

By vector calculations, one can see that this vertex corresponds to a quantum determinant different from the $SL(n)$ case.
For example one can use the vector assignments in \cite{Sp4tilt} to deduce the following relations:
$$\TetraIsDetSwap=Q\TetraIsDetSwapA,$$
 where $Q=-q$ when $(i,j)=(1,2),(1,4),(2,3)$ or $(4,3)$, and $Q=-q^2$ when $(i,j)=(1,3)$ or $(2,4)$.

The above quantum determinants given by the $\SL(n)$ degree-n vertex and the $\Sp(4)$ degree-4 vertex can both be seen as quantizations of the classical determinant. In other words, when $q=1$, the above quantum determinants agree with the classical determinant. See Theorems \ref{Sp4det} and \ref{Sp2ndet}.

\bibliographystyle{alpha}
\bibliography{C2webrefs.bib}

\end{document}